\documentclass[twoside,a4paper]{article}
\usepackage{fleqn,epsf,graphicx}
\usepackage{authblk}

\usepackage{epsfig}
\usepackage{graphicx}
\usepackage{caption}
\usepackage{subcaption}

\begin{document}

\title{Multiple~scattering~and~accidental~coincidences in~the~J-PET~detector simulated~using~GATE~package}

\author{P.~Kowalski$^{a}$, P.~Moskal$^{b}$, W.~Wi\'slicki$^{a}$, L.~Raczy\'nski$^{a}$, T.~Bednarski$^{b}$, P.~Bia\l as$^{b}$, J.~Bu\l ka$^{c}$, E.~Czerwi\'nski$^{b}$, A.~Gajos$^{b}$, A.~Gruntowski$^{b}$, D.~Kami\'nska$^{b}$, \L .~Kap\l on$^{b,d}$, A.~Kochanowski$^{e}$, G.~Korcyl$^{b}$, J.~Kowal$^{b}$, T.~Kozik$^{b}$, W.~Krzemie\'n$^{f}$, E.~Kubicz$^{b}$, Sz.~Nied\'zwiecki$^{b}$,  M.~Pa\l ka$^{b}$, Z.~Rudy$^{b}$, P.~Salabura$^{b}$, N.G.~Sharma$^{b}$, M.~Silarski$^{b}$, A.~S\l omski$^{b}$, J.~Smyrski$^{b}$, A.~Strzelecki$^{b}$, A.~Wieczorek$^{b,d}$, I.~Wochlik$^{c}$,  M.~Zieli\'nski$^{b}$, N.~Zo\'n$^{b}$}

\affil{

       $^{a}$\'Swierk Computing Centre, National Centre for Nuclear Research, 05-400 Otwock-\'Swierk, Poland\\       
       $^{b}$
  Faculty of Physics, Astronomy and Applied Computer Science, Jagiellonian University, 30-348 Cracow, Poland\\
       $^{c}$Department of Automatics and Bioengineering, AGH University of Science and Technology, Cracow, Poland\\
       $^{d}$Institute of Metallurgy and Materials Science of Polish Academy of Sciences, 30-059 Cracow, Poland\\
       $^{e}$Faculty of Chemistry, Jagiellonian University, 30-060 Cracow, Poland\\
       $^{f}$High Energy Physics Division, National Centre for Nuclear Research, 05-400 Otwock-\'Swierk, Poland\\
     }

\maketitle                   

{PACS:
  29.40.Mc, 
  87.57.uk, 
  87.10.Rt, 
  34.50.-s 
}

\begin{abstract}
Novel Positron Emission Tomography system, based on plastic scintillators, is developed by the J-PET collaboration. In order to optimize geometrical configuration of built device, advanced computer simulations are performed. Detailed study is presented of background given by accidental coincidences and multiple scattering of gamma quanta. 
\end{abstract}

\section{Introduction}

The GEANT4 Application for Tomographic Emission (GATE \cite{gate_paper}) represents one of the most advanced specialized software packages for simulations of PET scanners. 
Despite the complexity of the simulated system, GATE is easily configurable and facilitates convenient use of the powerful GEANT4 simulation toolkit.

Thanks to the fact, that the software was widely verified, it may be used for simulations of such a prototype devices as Strip-PET scanner \cite{jpet_paper_1}-\cite{jpet_paper_3}, build by the J-PET collaboration. The scanner is based on the plastic scintillators representing innovative approach in the field of PET tomography. 
Another important feature of the scanner is large axial field-of-view (AFOV). PET scanners with large AFOV are also developed by other collaborations \cite{insaini}-\cite{blanco}.

\section{Setting parameteres of the simulations in the GATE software}

Properties of the scintillating material and the detecting surface, were set using three GATE-specific files: \textit{GateMaterials.db}, \textit{Materials.xml} and \textit{Surfaces.xml}. Some of them, could be fixed using data from documentation prepared by the producers of the equipment.

\begin{figure}[!h]
\hspace{0.25\textwidth}
\parbox{0.50\textwidth}{
   \centerline{
   \includegraphics[width=0.5\textwidth]{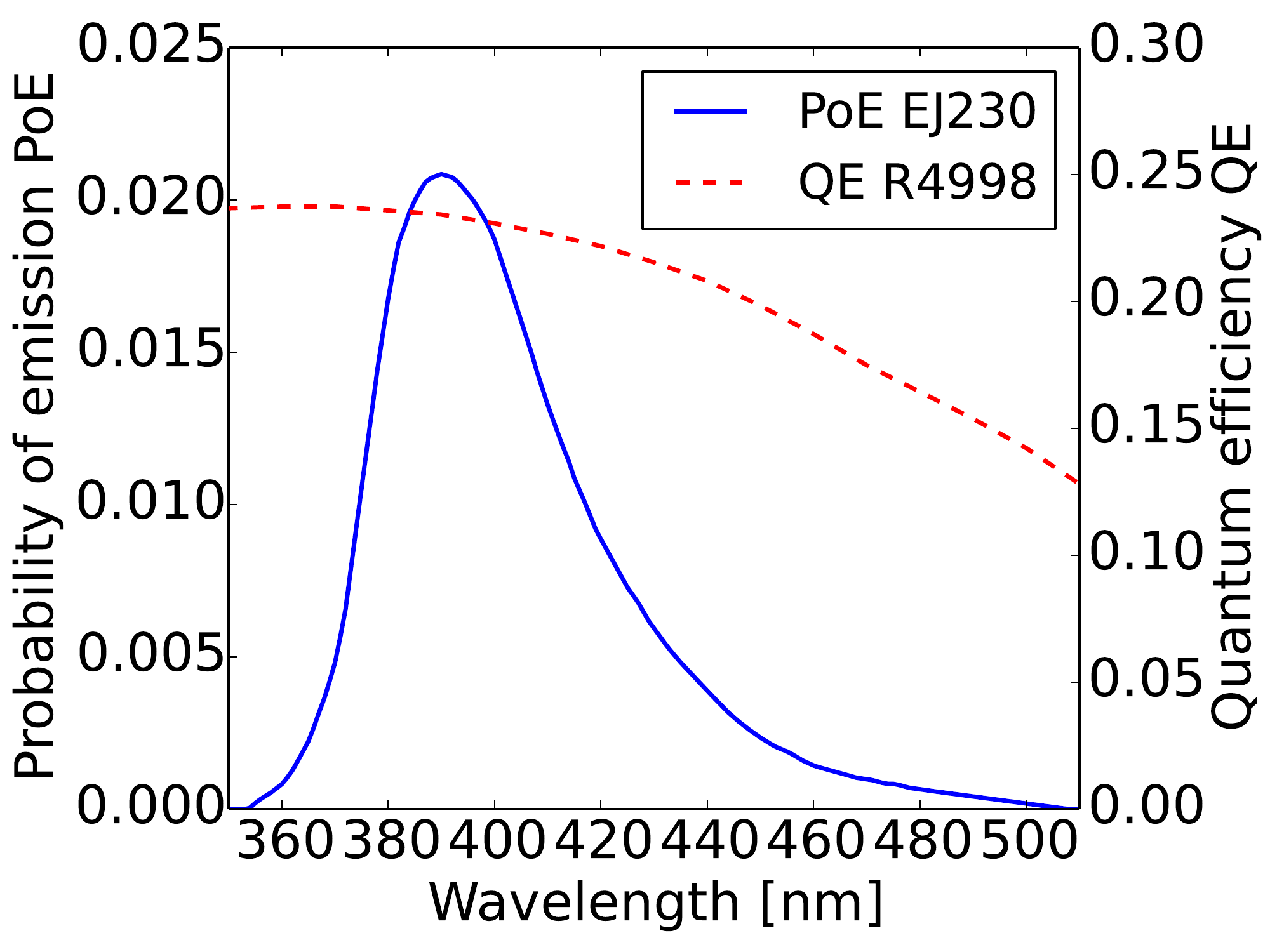}
}}
\caption{Emission spectrum of the EJ230 material \cite{ej230datasheet} and quantum efficiency of the R4998 photomultiplier \cite{hamamatsu_private}}
\label{emission_pmt}
\end{figure}

For example the properties of the scintillating material EJ230 \cite{ej230datasheet}, that is used by the collaboration in real-life experiments are:
\begin{itemize}
  \item scintillation yield - 9,700 1/MeV
  \item refraction index - 1.58
  \item density 1.023 g/cm$^3$
  \item emission spectrum - Fig. \ref{emission_pmt}; maximum of emission at 391 nm
\end{itemize}

The only property of the detecting surface (which immitates the photomultiplier Hamamatsu R4998 \cite{hamamatsu_R4998}), that has to be set by the user is the dependence of quantum efficiency on the wavelength of optical photons \cite{hamamatsu_private} (Fig. \ref{emission_pmt}).

Some important properties, however, are not given by the producer. One of them is the absorption length dependence on the light wavelength. Therefore, we adopted and tested this dependence from another similar material as described below.

\subsection{Simulations of the single strip}

The dependence of absorption length on light wavelength for plastic scintillator may be found in Ref. \cite{UPS2006}. Similar borrowing has been applied by the authors simulating the NEMO detector \cite{UPS2011}. The dependence taken from publication was read out from the picture, smoothed using line interpolation and implemented in GATE software. After that some simulations of the single strip were performed and their results were compared with the experiment.

\begin{figure}[h!]
\centering

\begin{subfigure}{0.49\textwidth}
    \centering
    \includegraphics[width=\textwidth]{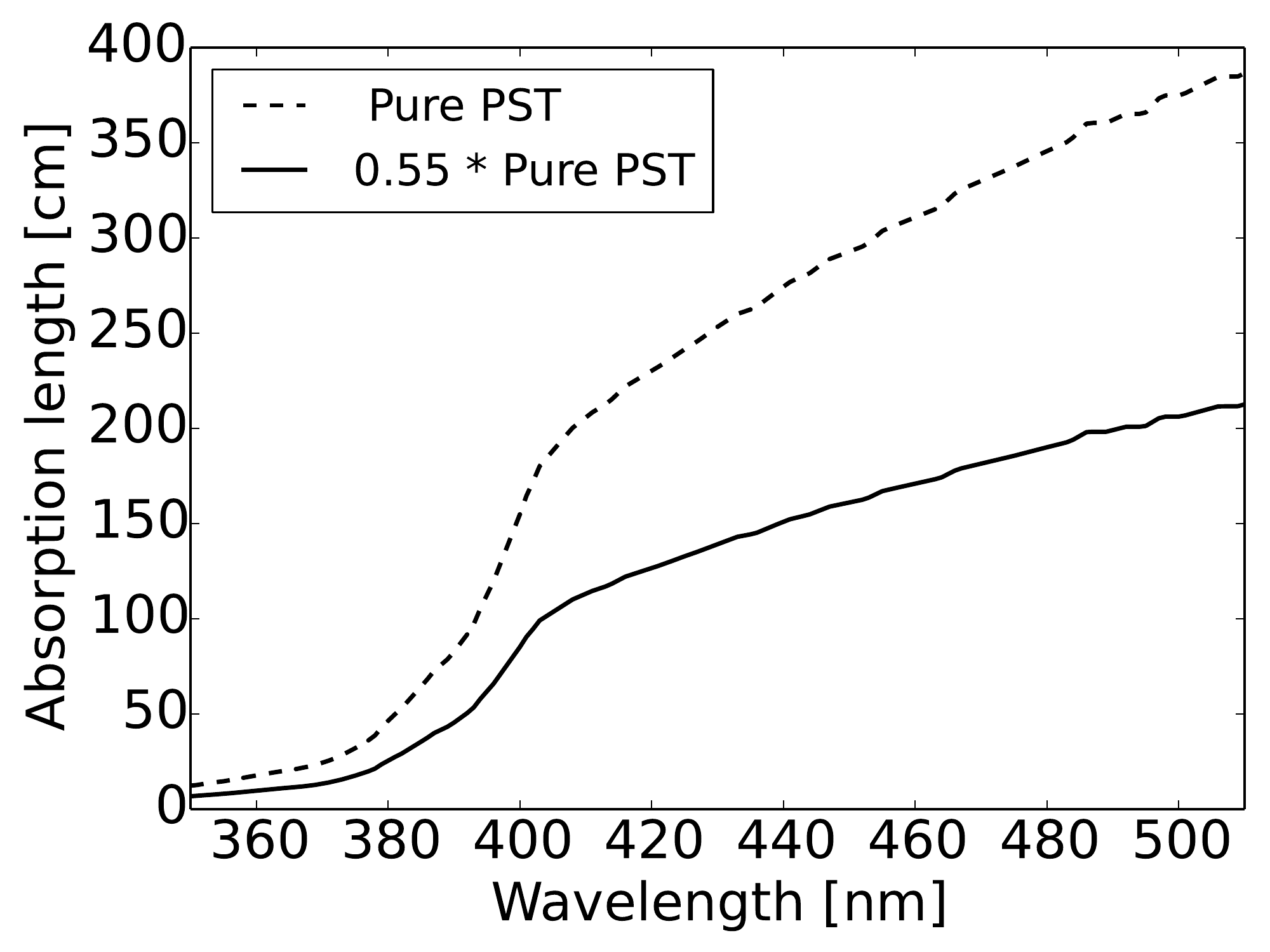}
\end{subfigure}
\begin{subfigure}{0.49\textwidth}
    \centering
    \includegraphics[width=\textwidth]{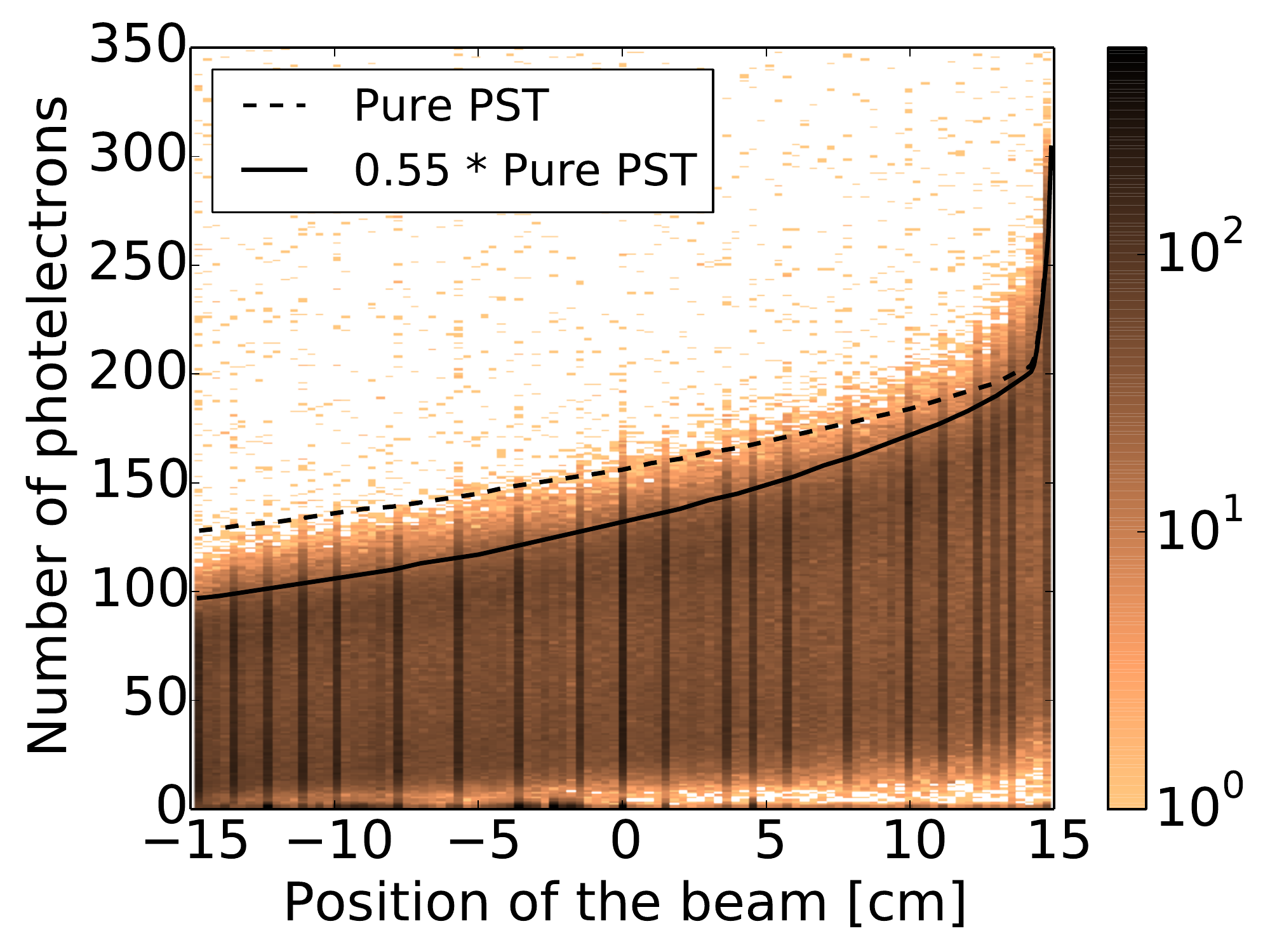}
\end{subfigure}

\caption{The left part of the figure presents the dependence of the absorption length on the light wavelength. Upper line shows result obtained for pure polystyrene (PST) \cite{UPS2006} and the lower line is scaled by factor of 0.55. The right part of the figure shows the spectra of number of photoelectrons as a position of the beam of gamma quanta. Dashed line indicates maximum number of  photoelectrons produced by 511 keV gamma quanta as a function of position of irradiation assuming absorption as measured for PST (dashed line) and  PST absorption scaled by factor of 0.55 (solid line).}
\label{sim_vs_exp_abs_len}
\end{figure}

In the experiment, the collimated source of gamma quanta (Na-22) was moved along the scintillator EJ230 (5 mm x 19 mm x 30 cm) with step 3~mm and the beam was directed perpendicularly into the scintillator. For each position of the beam, 1-dimensional histogram of the number of photoelectrons was created. The histogram was put into the single column of two-dimensional histogram presented in the background of the Fig. \ref{sim_vs_exp_abs_len}b. In this figure, one can see the dependence between the number of photons detected by the photomultiplier and the position of the beam of gamma quanta. Experimental data are available for positions between -14.7~cm and 14.7~cm and the width of bins is 3~mm. In this figure results for R4998 photomultiplier attached to the scintillator at the end (position 15~cm) are shown.

\begin{figure}[h!]
\centering

\begin{subfigure}{0.325\textwidth}
    \centering
    \includegraphics[width=\textwidth]{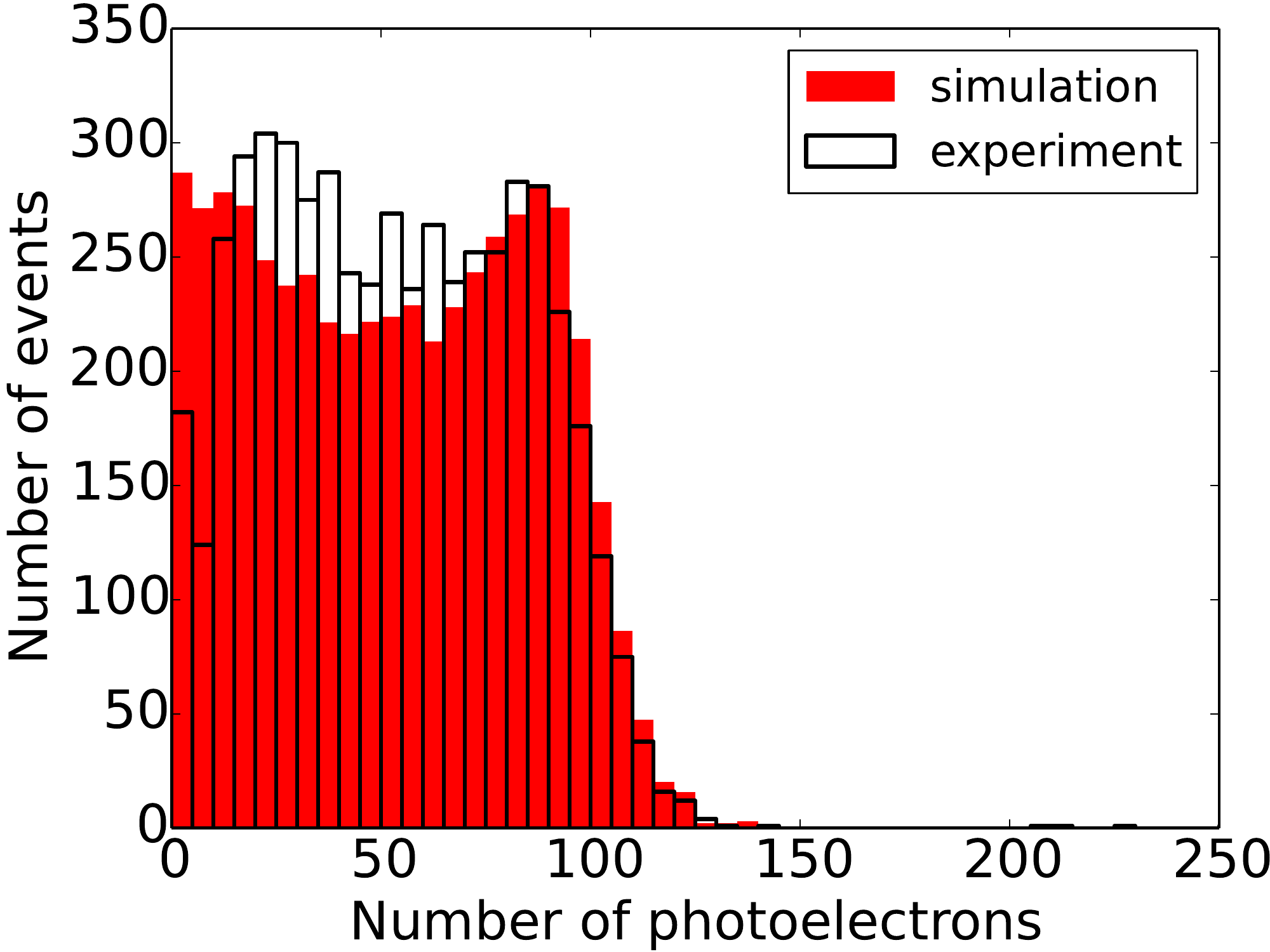}
\end{subfigure}
\begin{subfigure}{0.325\textwidth}
    \centering
    \includegraphics[width=\textwidth]{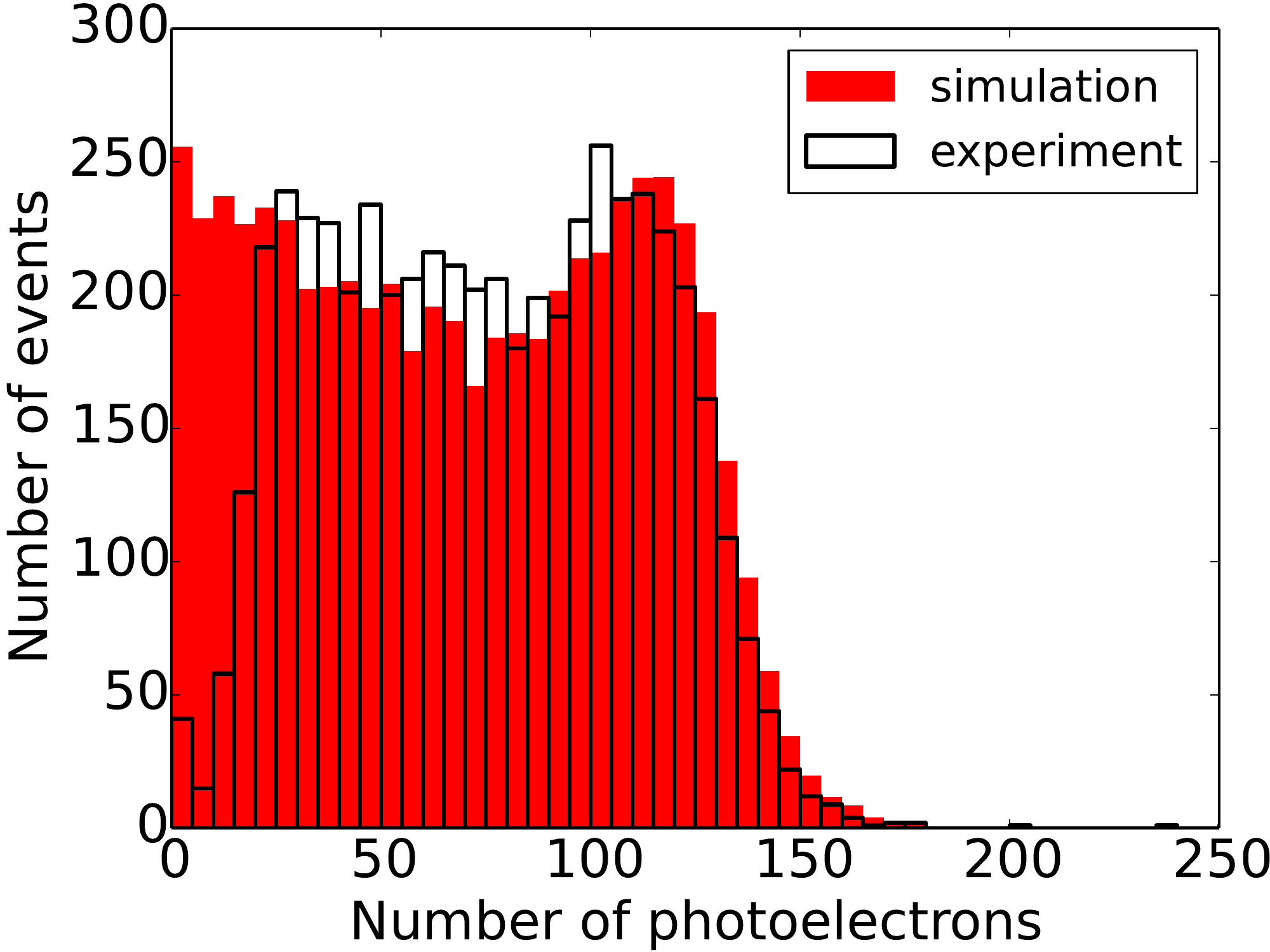}
\end{subfigure}
\begin{subfigure}{0.325\textwidth}
    \centering
    \includegraphics[width=\textwidth]{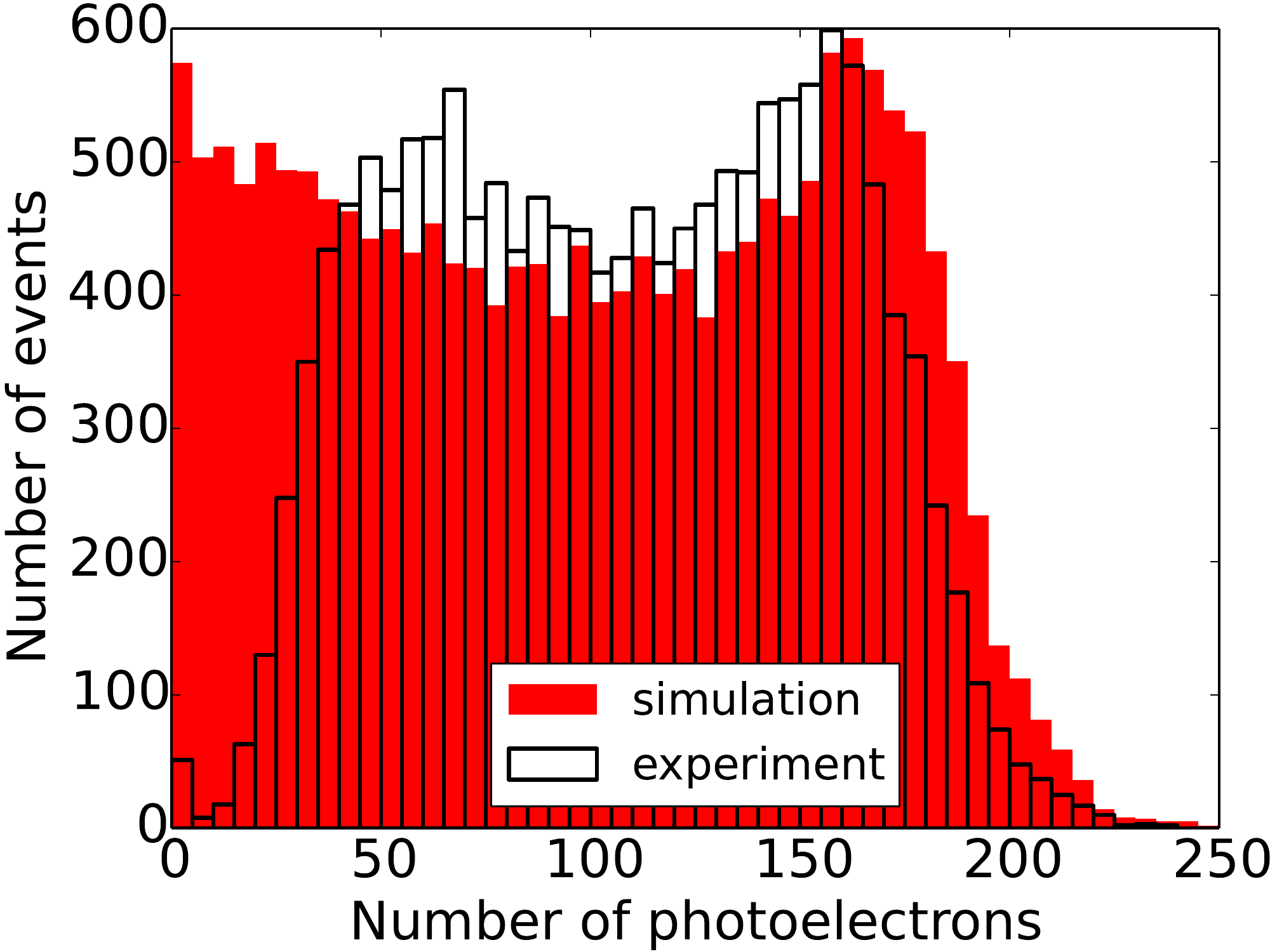}
\end{subfigure}

\caption{Comparison of the simulated and experimental histograms of energy deposited by 511~keV gamma quanta (in number of photoelectrons) for the beam positions -12~cm (left), 0~cm (middle), 12~cm (right); experimental spectrum is suppressed at low values due to the triggering conditions \cite{jpet_paper_1}.}
\label{three_beams_positions}
\end{figure}

Fig. \ref{three_beams_positions} shows comparison of simulated and experimental distributions of number of photoelectrons for three exemplary positions. A good agreement was obtained when scaling the absorption length of pure polystyrene \cite{UPS2006} by factor of 0.55 (Fig. \ref{sim_vs_exp_abs_len}).
The scaling factor accounts effectively for the absorption due to the primary and secondary admixture in the scintillator material, imperfections of surfaces and reflectivity of the foil.
Dashed and solid line in right part of Fig.~\ref{sim_vs_exp_abs_len} presents results of simulations performed for energy loss of 341 keV  corresponding to the maximum energy of the electron scattered by the 511 keV gamma quantum via Compton effect. 
Dashed line was obtained assuming the absorption length as determined for the  pure PST, whereas solid line shows result after scaling the absorption by a factor of 0.55. The scaling factor was optimised to the experimental results. 


\section{Simulations of the single layer J-PET scanner}

\begin{figure}[h!]
\centering

\begin{subfigure}{0.325\textwidth}
    \centering
	\includegraphics[width=\textwidth]{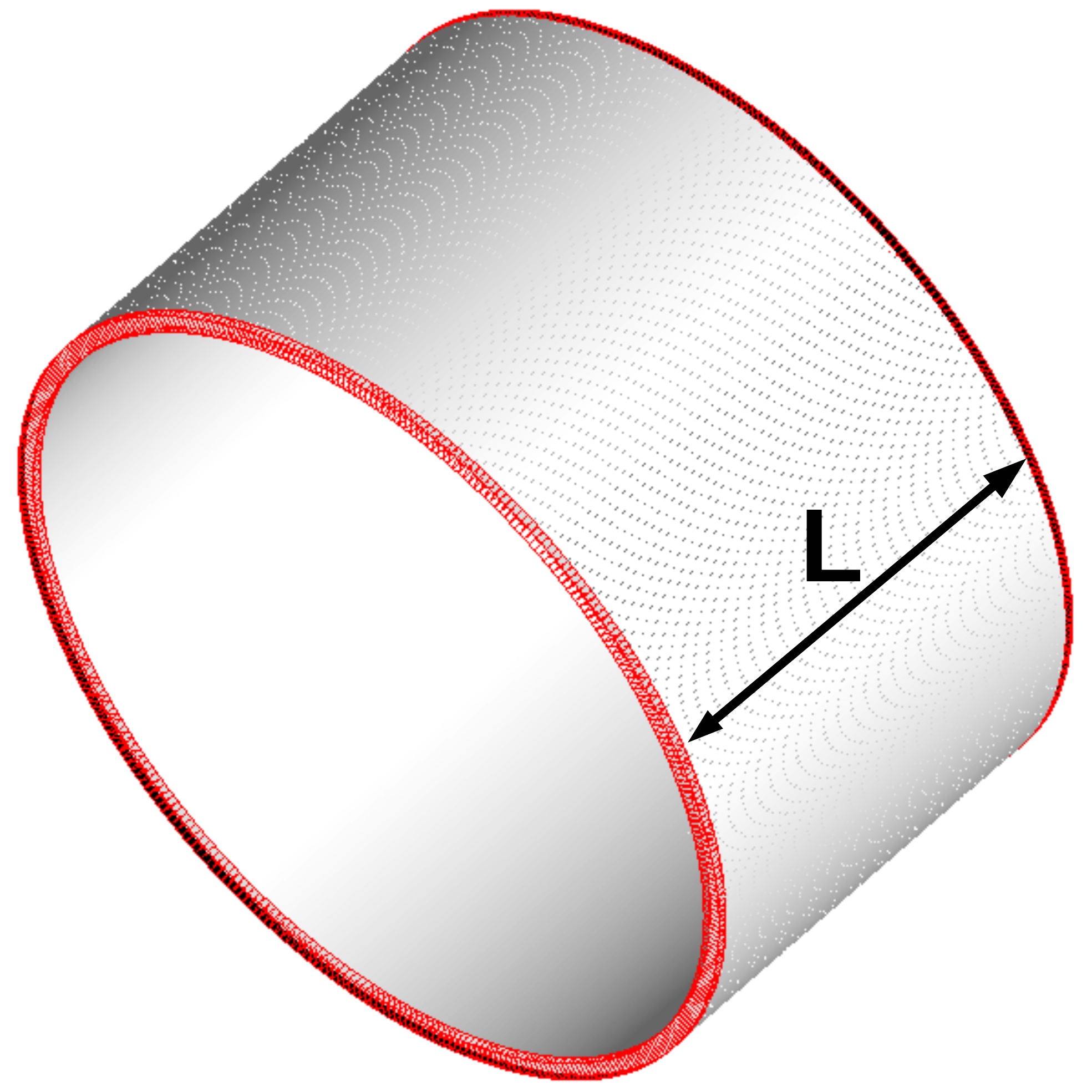}
\end{subfigure}
\begin{subfigure}{0.325\textwidth}
    \centering
	\includegraphics[width=\textwidth]{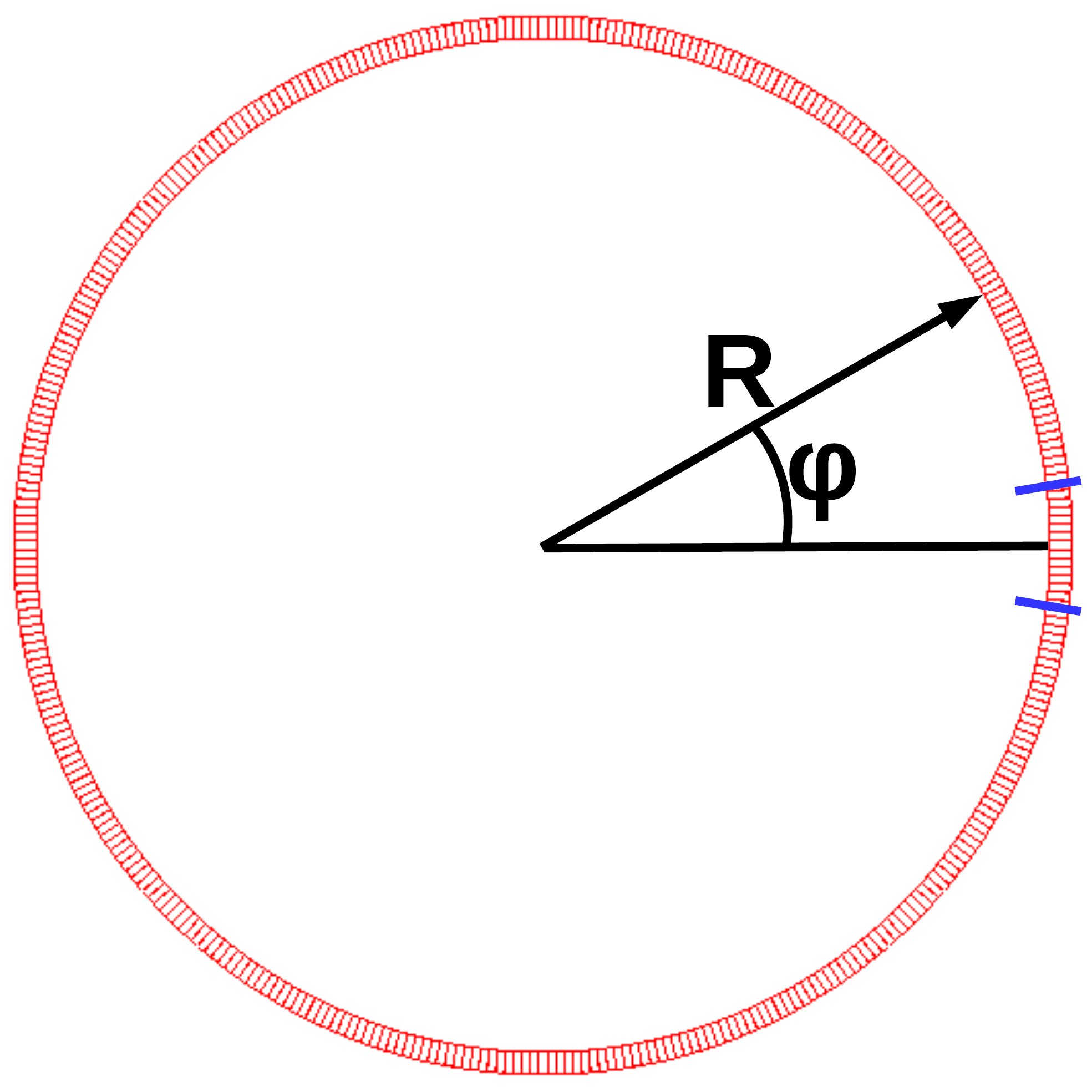}
\end{subfigure}
\begin{subfigure}{0.325\textwidth}
    \centering
	\includegraphics[width=\textwidth]{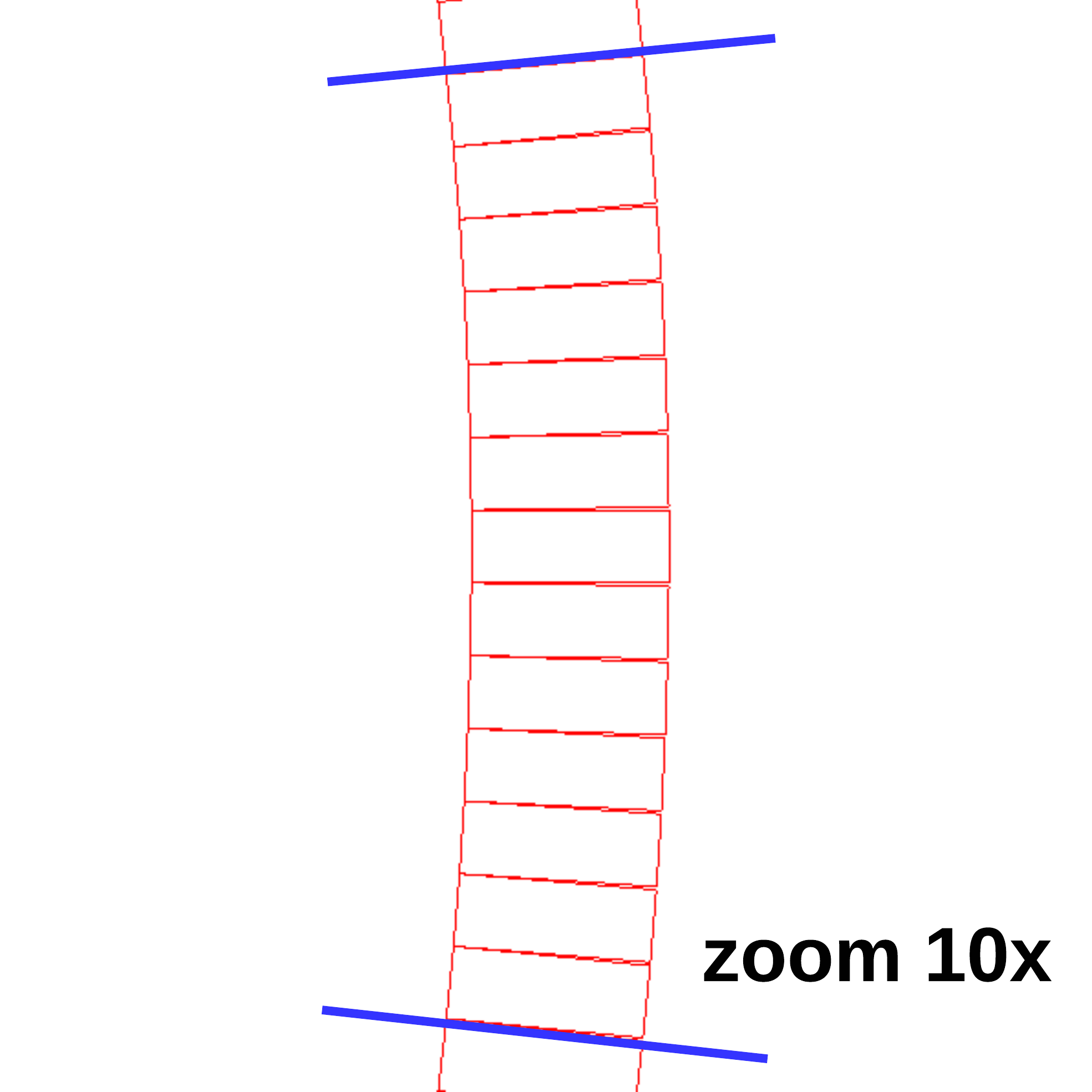}
\end{subfigure}

\caption{Visualisation of the geometry of the single-layer J-PET scanner with radius of the cylinder R and the length of the scintillators L.}
\label{geometry}
\end{figure}

A diagnostic chamber of the J-PET detector will form a cylinder which will be constructed from the plastic scintillator strips \cite{med_rev1}-\cite{med_rev3}. In this article we present simulations for the detector with the inner radius of R=427.8~mm (radius similar to commercially available PET systems \cite{commercial}, \cite{commercial2}). We assume that the detector possesses one layer build out of 384 EJ230 scintillator strips with dimensions of 7~mm x 19~mm x L (L = 20~cm, 50~cm, 100~cm or 200~cm). Geometry of the simulated scanner is visualised in Fig. \ref{geometry}.

\subsection{Scattered coincidences}
\label{scattered_coincidences}

\begin{figure}[h!]
\centering

\begin{subfigure}{0.3\textwidth}
    \centering
	\includegraphics[width=\textwidth]{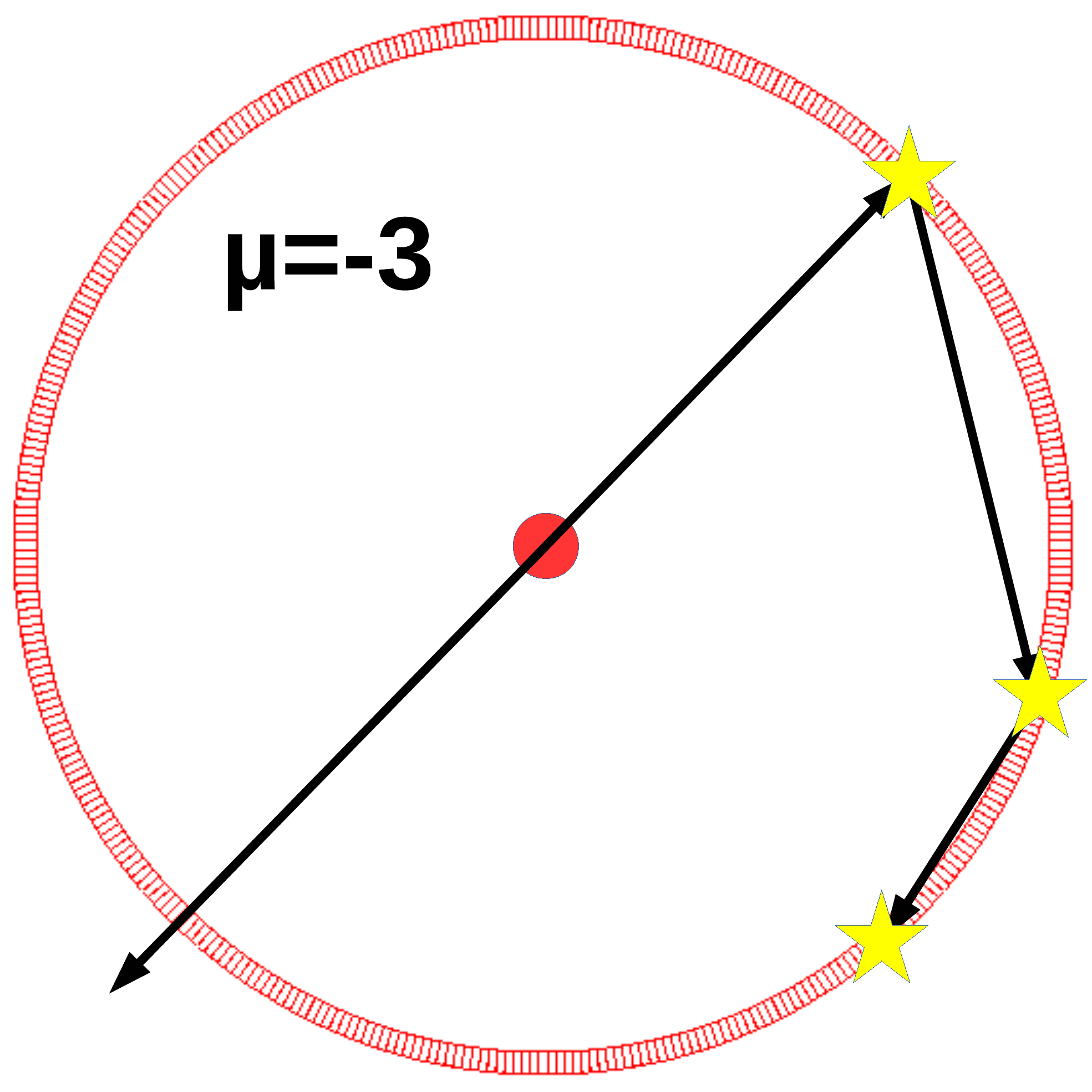}
\end{subfigure}
\begin{subfigure}{0.3\textwidth}
    \centering
	\includegraphics[width=\textwidth]{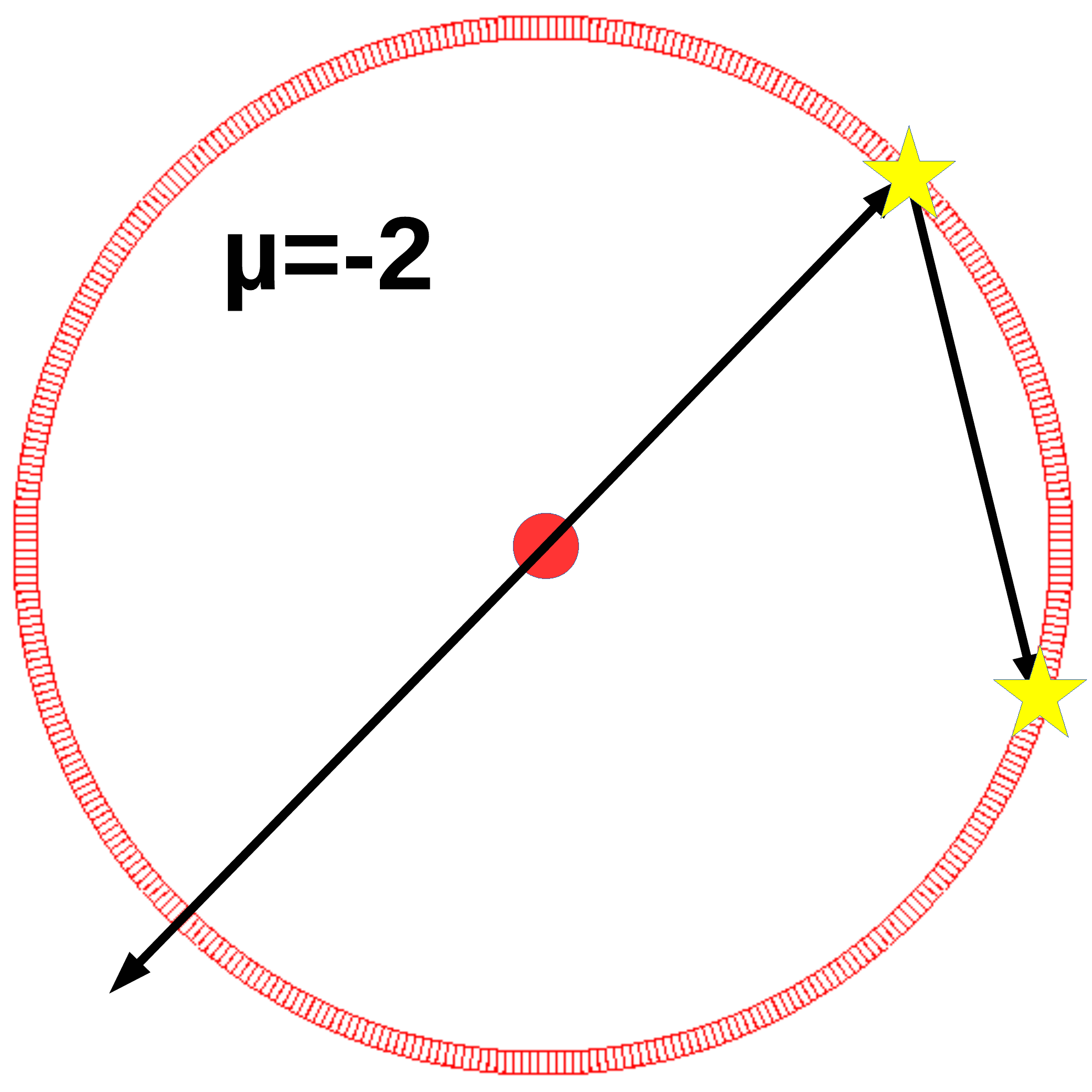}
\end{subfigure}
\begin{subfigure}{0.3\textwidth}
    \centering
	\includegraphics[width=\textwidth]{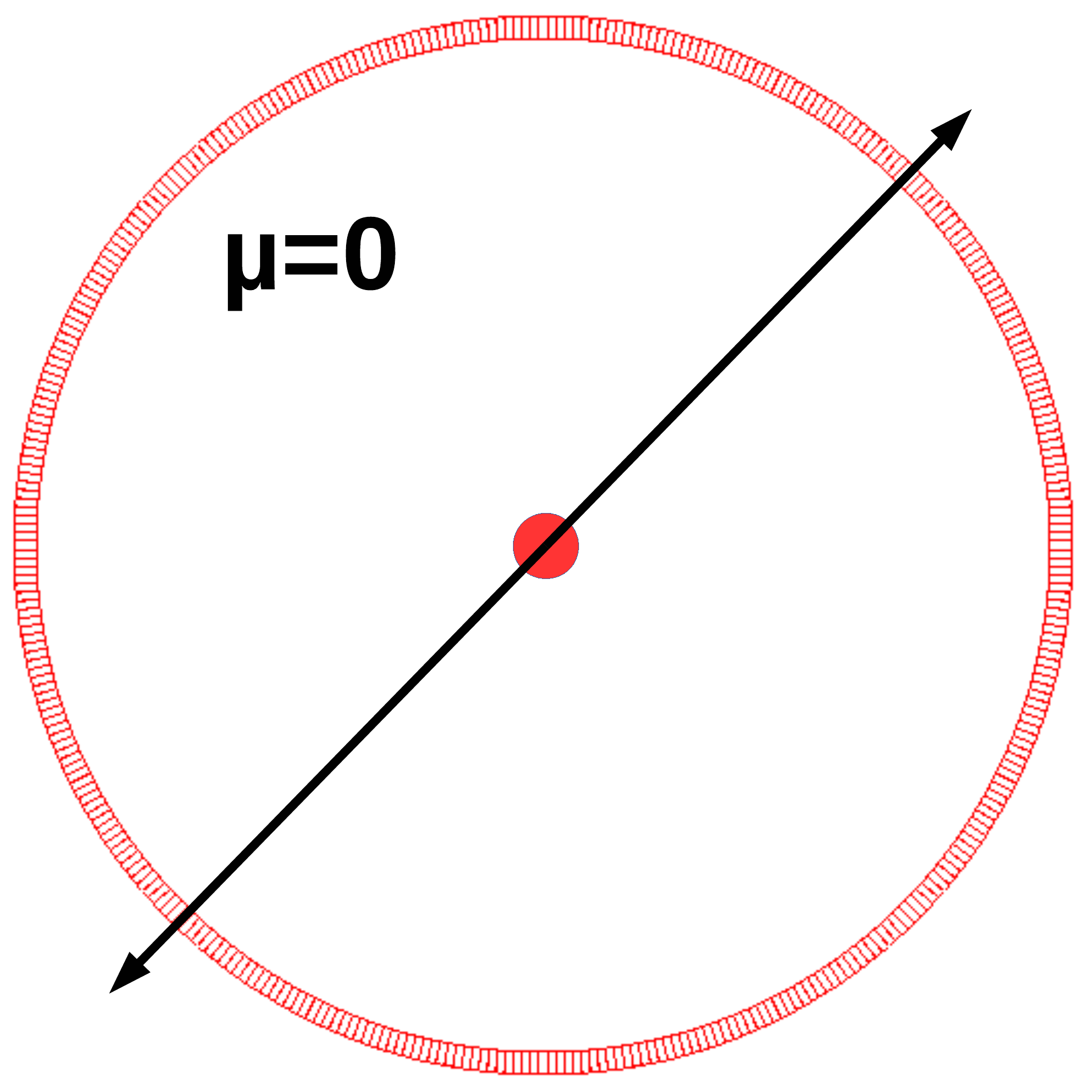}
\end{subfigure}

\begin{subfigure}{0.3\textwidth}
    \centering
	\includegraphics[width=\textwidth]{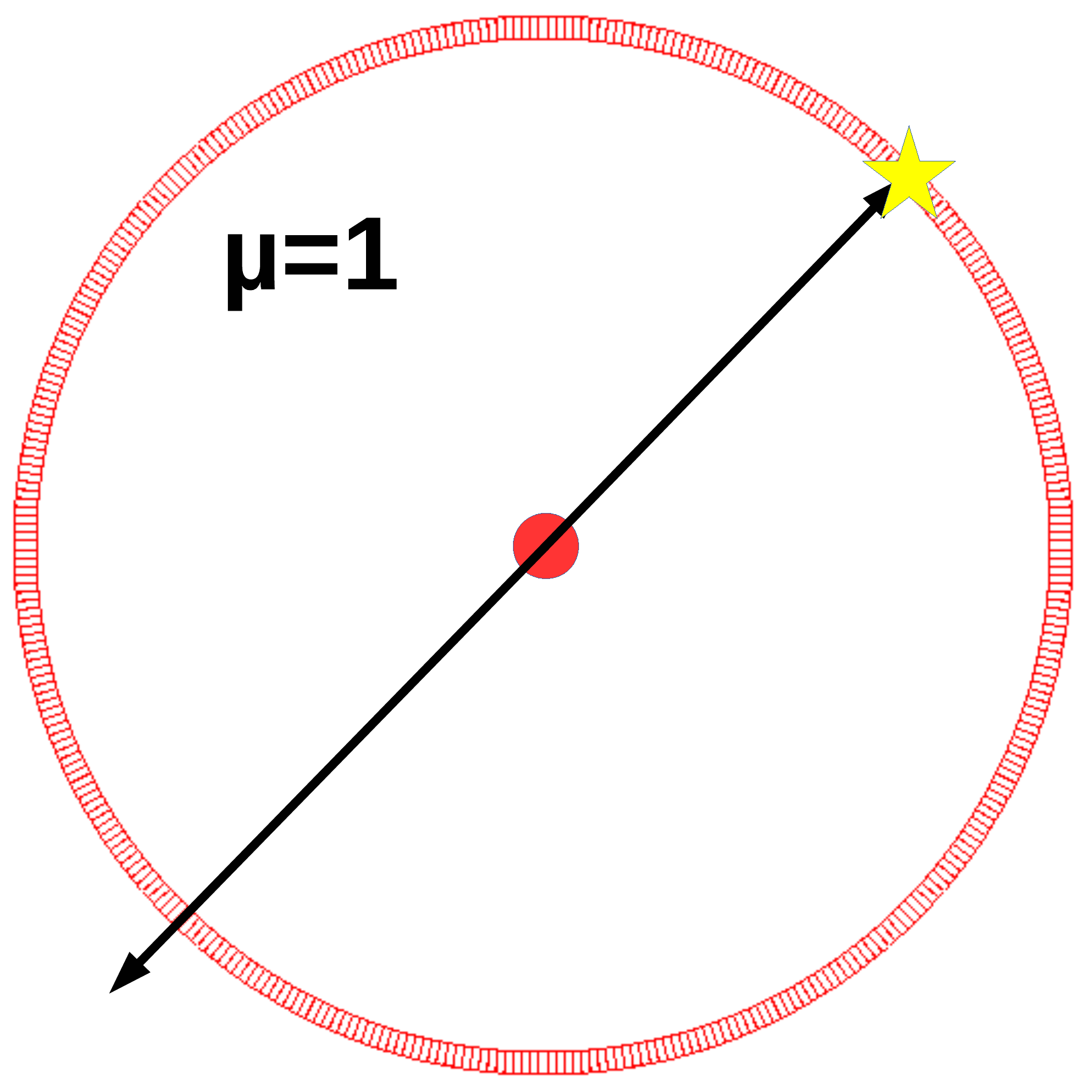}
\end{subfigure}
\begin{subfigure}{0.3\textwidth}
    \centering
	\includegraphics[width=\textwidth]{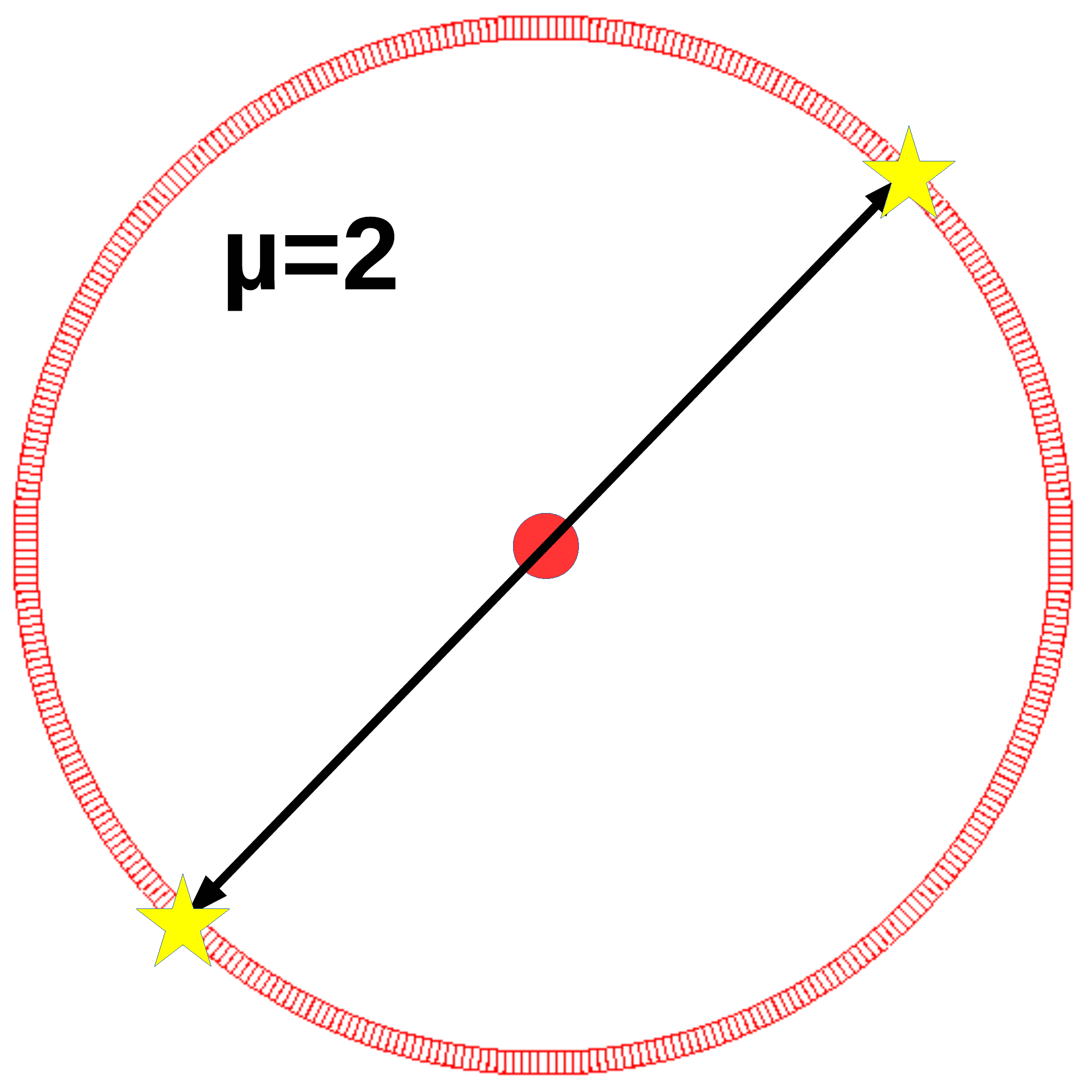}
\end{subfigure}
\begin{subfigure}{0.3\textwidth}
    \centering
	\includegraphics[width=\textwidth]{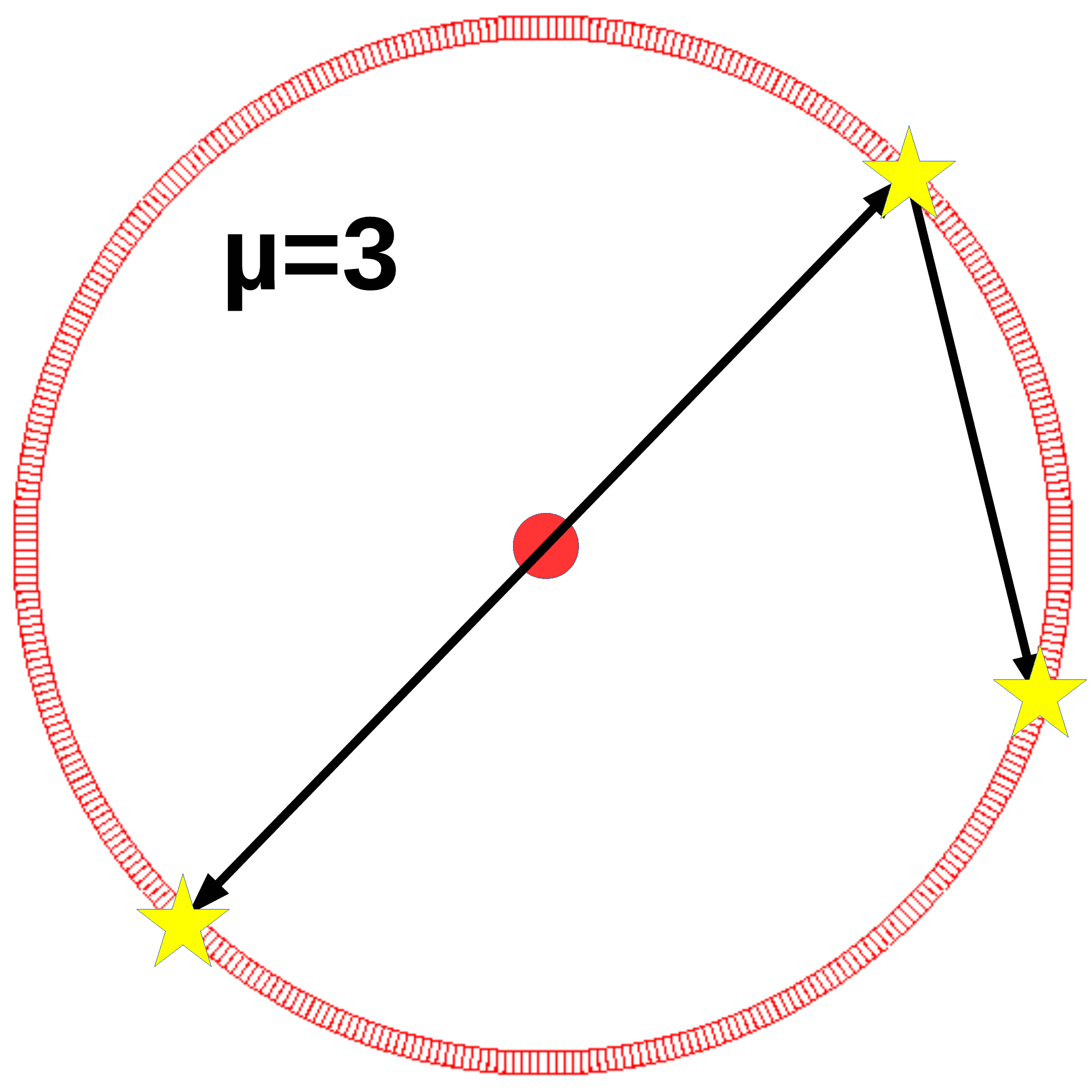}
\end{subfigure}

\caption{Pictorial definitions of the value of multiplicity $\mu$  used further in the following figures.}
\label{mi}
\end{figure}

In order to estimate secondary scattering of gamma quanta in the detector material, we have simulated annihilations homogeneously in the 2~m long line placed along the central axis of the scanner. In the following, we consider few most probable responses of the detector system (see Fig. \ref{mi}). In the most probable case both gamma quanta will escape detection and no signal will be observed ($N_{strips}=0$). The second frequent category corresponds to events when only one strip was hit ($N_{strips}=1$). Further on for the multiplicity of strips $N_{strips}>=2$ we can distinguish different cases for the same value of $N_{strips}$. Therefore for the univocal description we introduce one more parameter $\mu$. Various possibilities which may occur are listed below and depicted in Fig. \ref{mi}:

\begin{itemize}
  \item $N_{strips}=3$, $\mu=-3$ \newline 3 quanta in 3 different strips with two secondary scatterings
  \item $N_{strips}=2$, $\mu=-2$ \newline 2 quanta in 2 different strips with one secondary scattering
  \item $N_{strips}=0$, $\mu=0$ \newline no gamma quanta registered
  \item $N_{strips}=1$, $\mu=1$ \newline interaction in only one strip
  \item $N_{strips}=2$, $\mu=2$ \newline 2 interactions in 2 different strips
  \item $N_{strips}=3$, $\mu=3$ \newline 3 scatterings in 3 different strips; 2 primary and 1 secondary scattering
\end{itemize}

It is also possible that there are 4, 5 or even more scatterings, depending on the energy threshold applied to each hit. 

\begin{figure}[h!]
\centering

\begin{subfigure}{0.49\textwidth}
    \centering	\includegraphics[width=\textwidth]{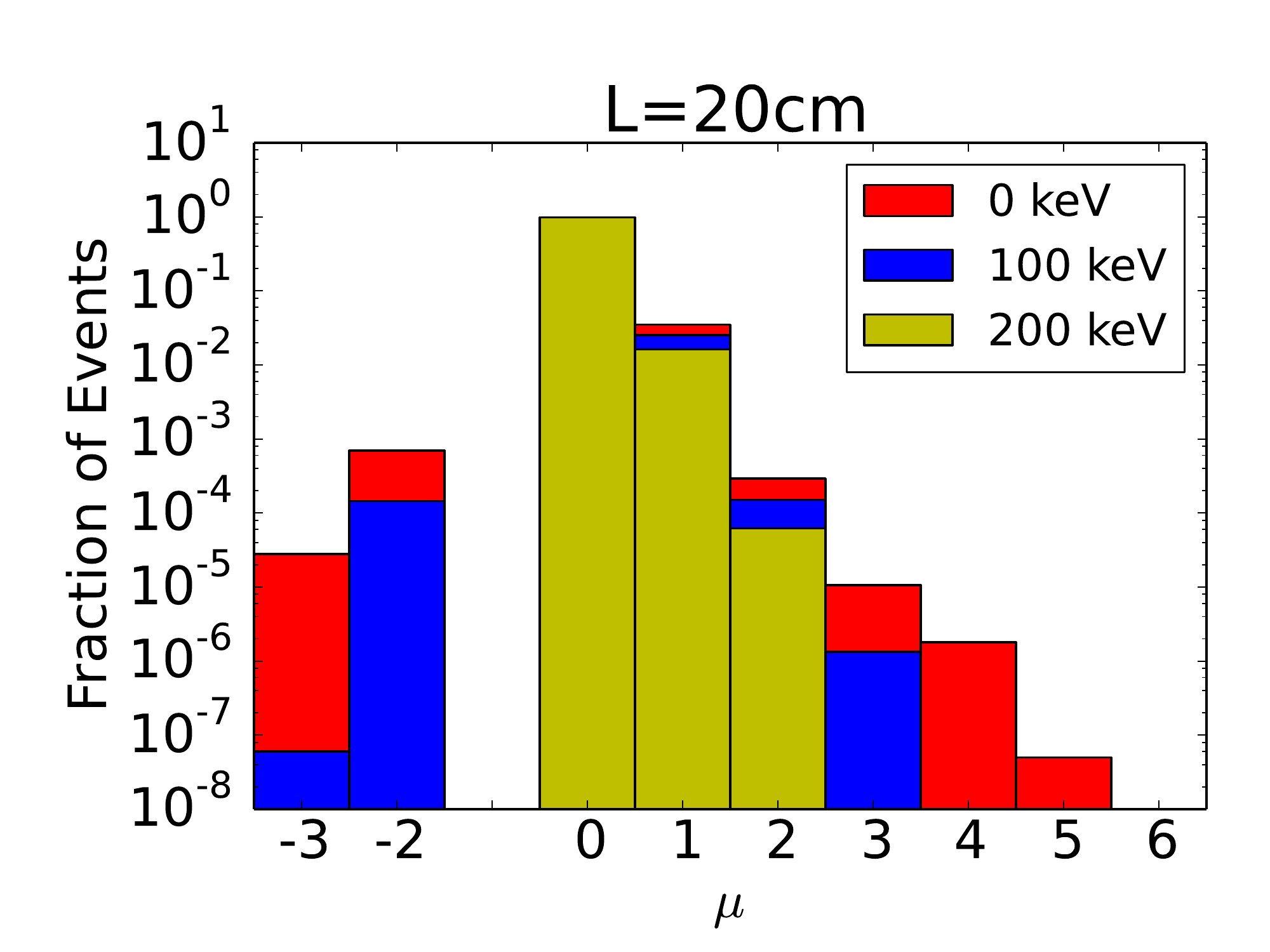}
\end{subfigure}
\begin{subfigure}{0.49\textwidth}
    \centering	\includegraphics[width=\textwidth]{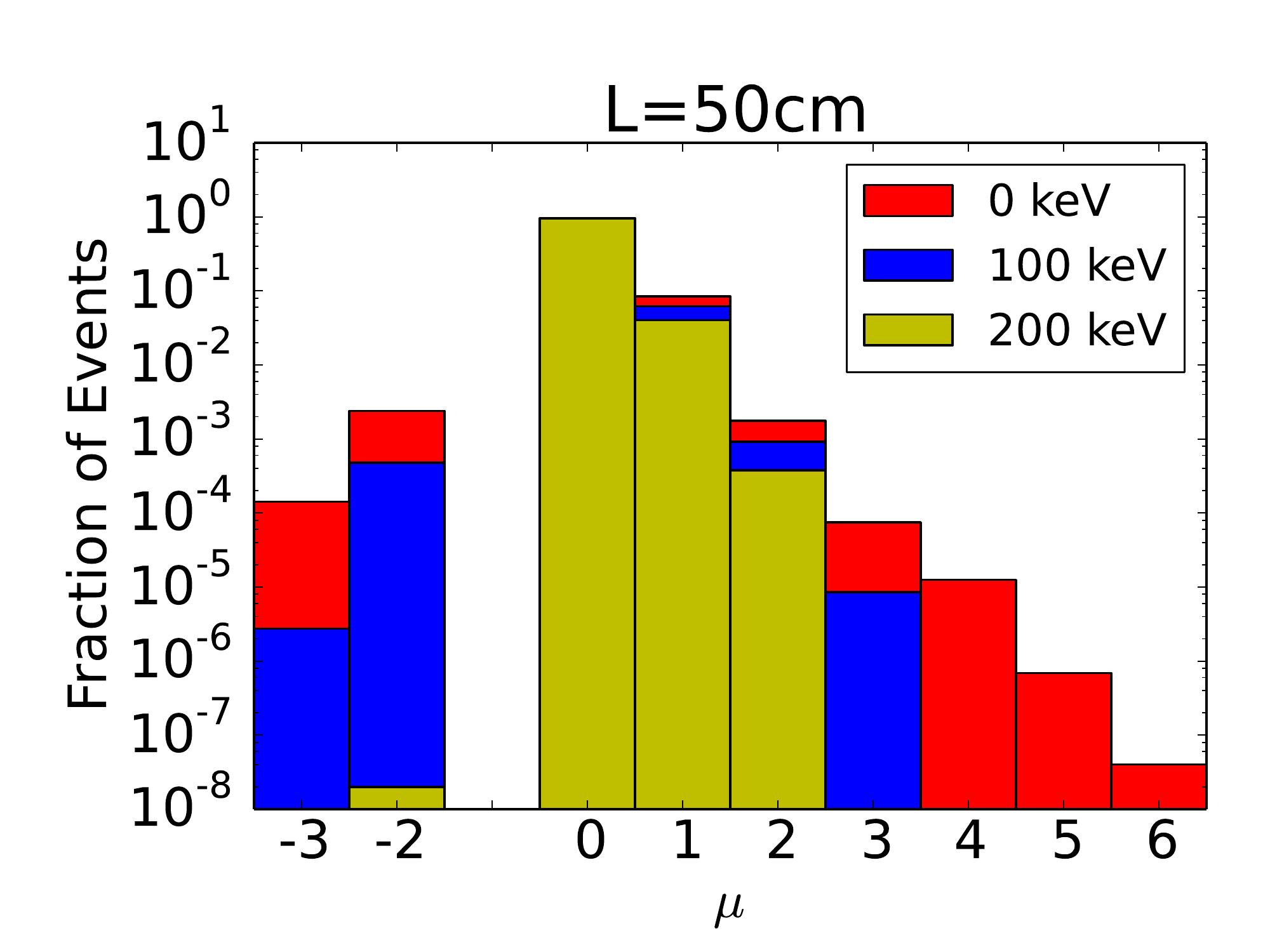}
\end{subfigure}

\begin{subfigure}{0.49\textwidth}
    \centering	\includegraphics[width=\textwidth]{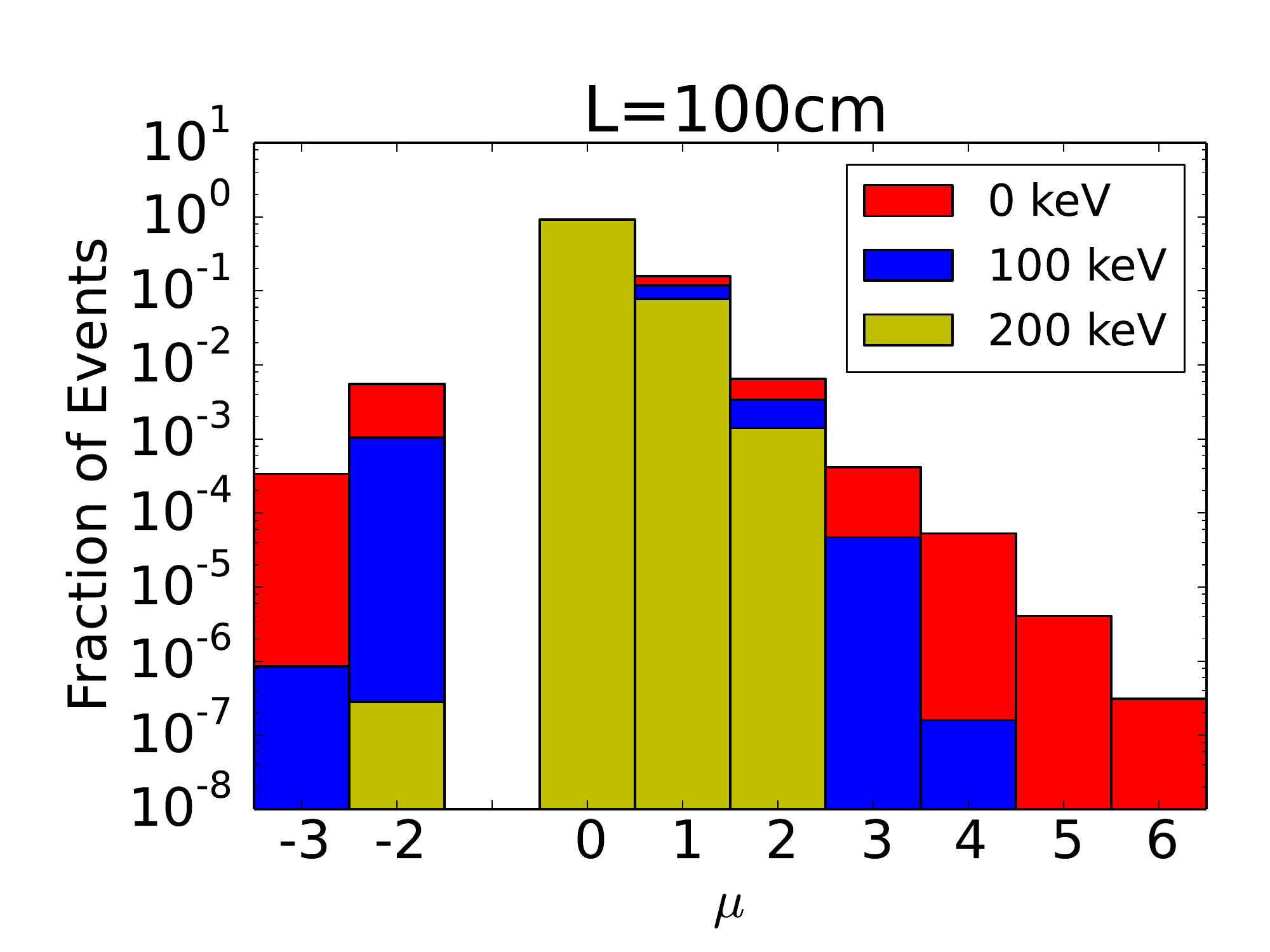}
\end{subfigure}
\begin{subfigure}{0.49\textwidth}
    \centering	\includegraphics[width=\textwidth]{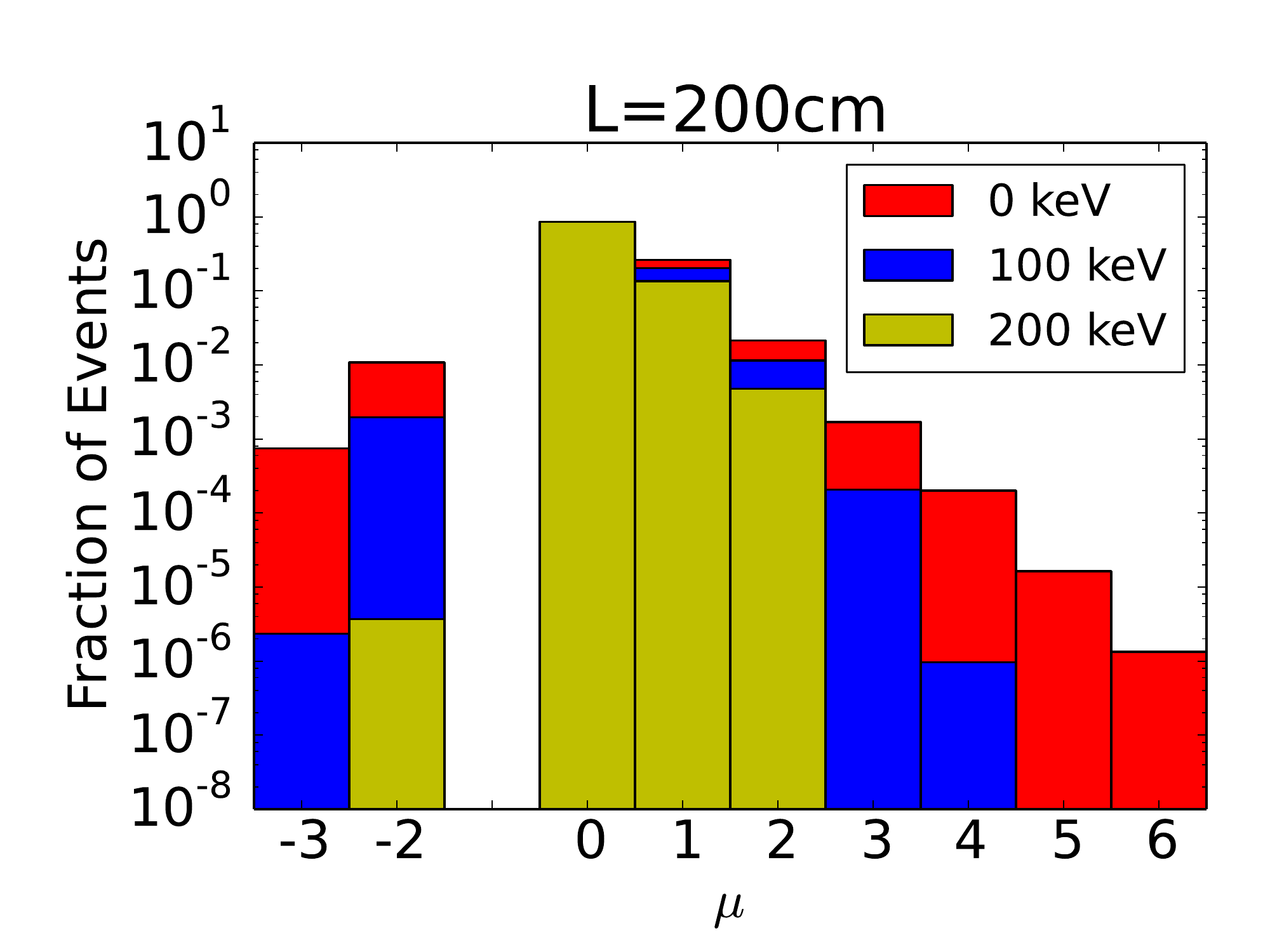}
\end{subfigure}

\caption{Histogram of the multiplicity $\mu$ for different lengths L of the diagnostic chamber. Meanings of different values of the multiplicity $\mu$ are described in the text and defined in Fig. \ref{mi}.}
\label{hist_mi}
\end{figure}

Histograms of the multiplicity for three different energy thresholds (0 keV, 100 keV and 200 keV) and for four different lengths of scintillators (20 cm, 50 cm, 100 cm and 200 cm) are presented in the Fig. \ref{hist_mi}. Results of the simulations show that if energy threshold is set to 200 keV, there are no events where number of hits is bigger than 2. Most of scattered coincidences (with multiplicity -2) is also eliminated with this energy threshold. If the energy threshold is set to 100 keV, for lengths of scitnillators 100 cm and 200 cm, there would be even events with four scatterings, which may negatively influence the quality of reconstructed images. 

\begin{figure}[h!]
\centering

\begin{subfigure}{0.49\textwidth}
    \centering
	\includegraphics[width=\textwidth]{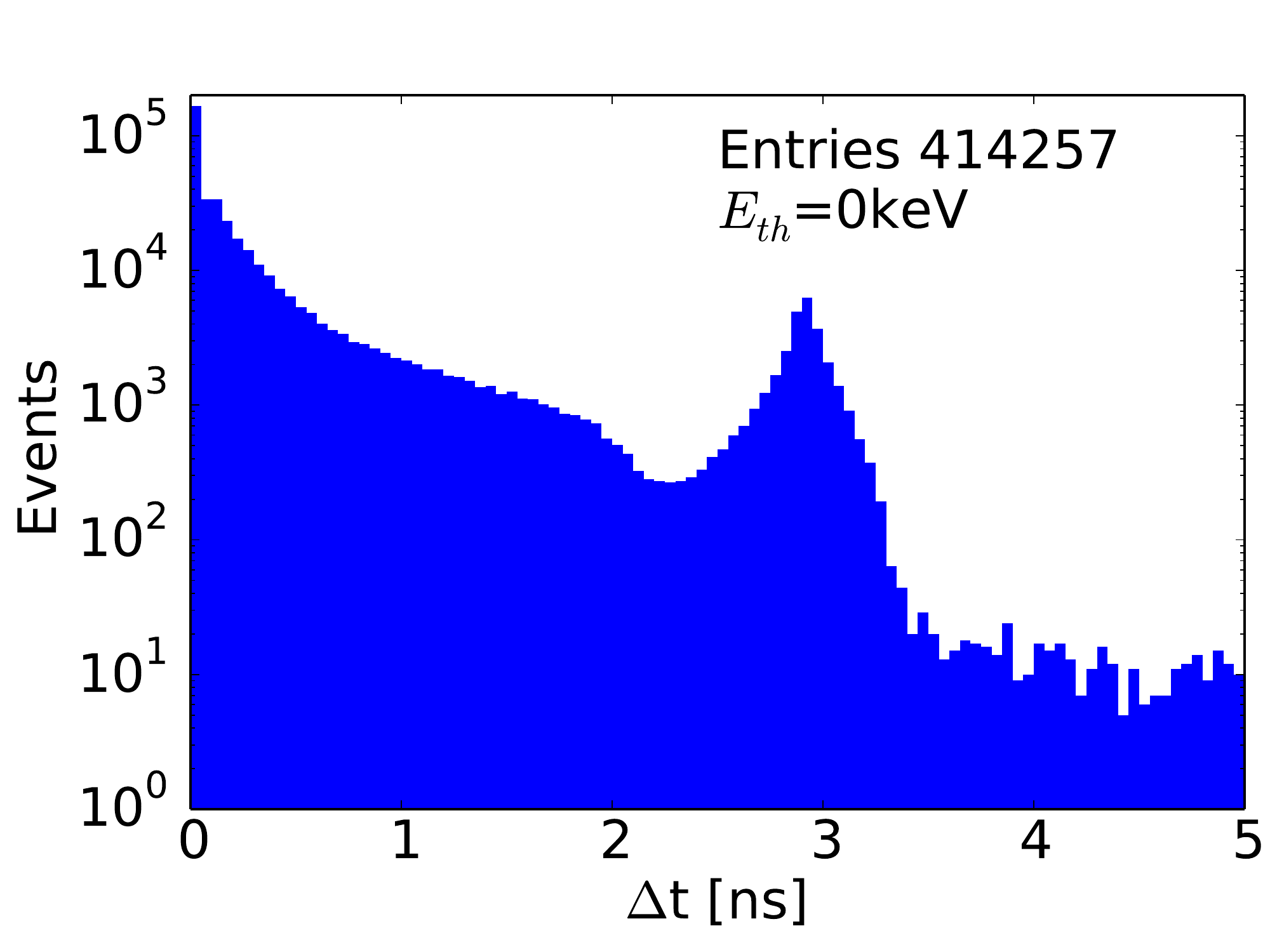}
\end{subfigure}
\begin{subfigure}{0.49\textwidth}
    \centering	
	\includegraphics[width=\textwidth]{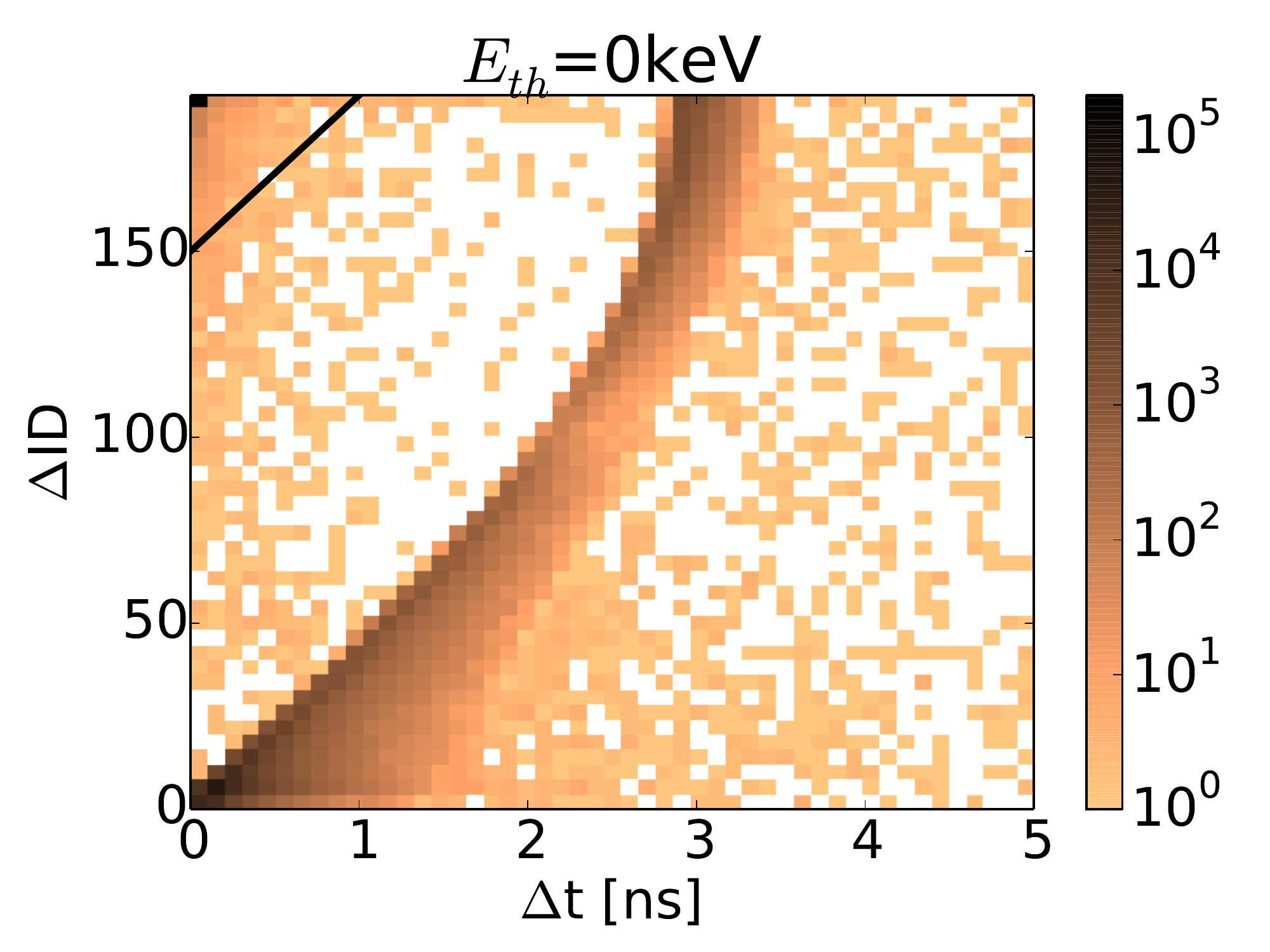}
\end{subfigure}

\begin{subfigure}{0.49\textwidth}
    \centering	\includegraphics[width=\textwidth]{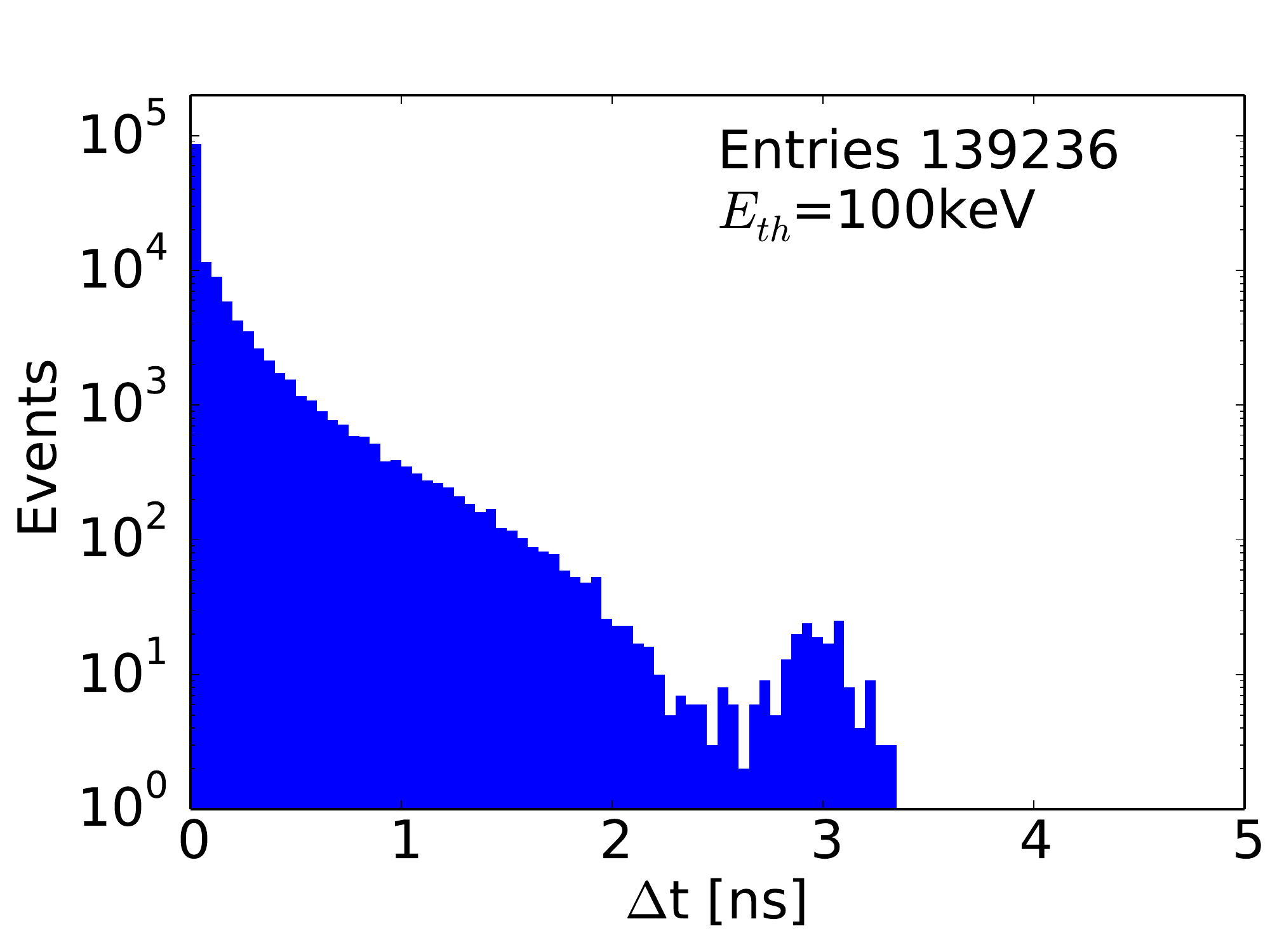}
\end{subfigure}
\begin{subfigure}{0.49\textwidth}
    \centering	
	\includegraphics[width=\textwidth]{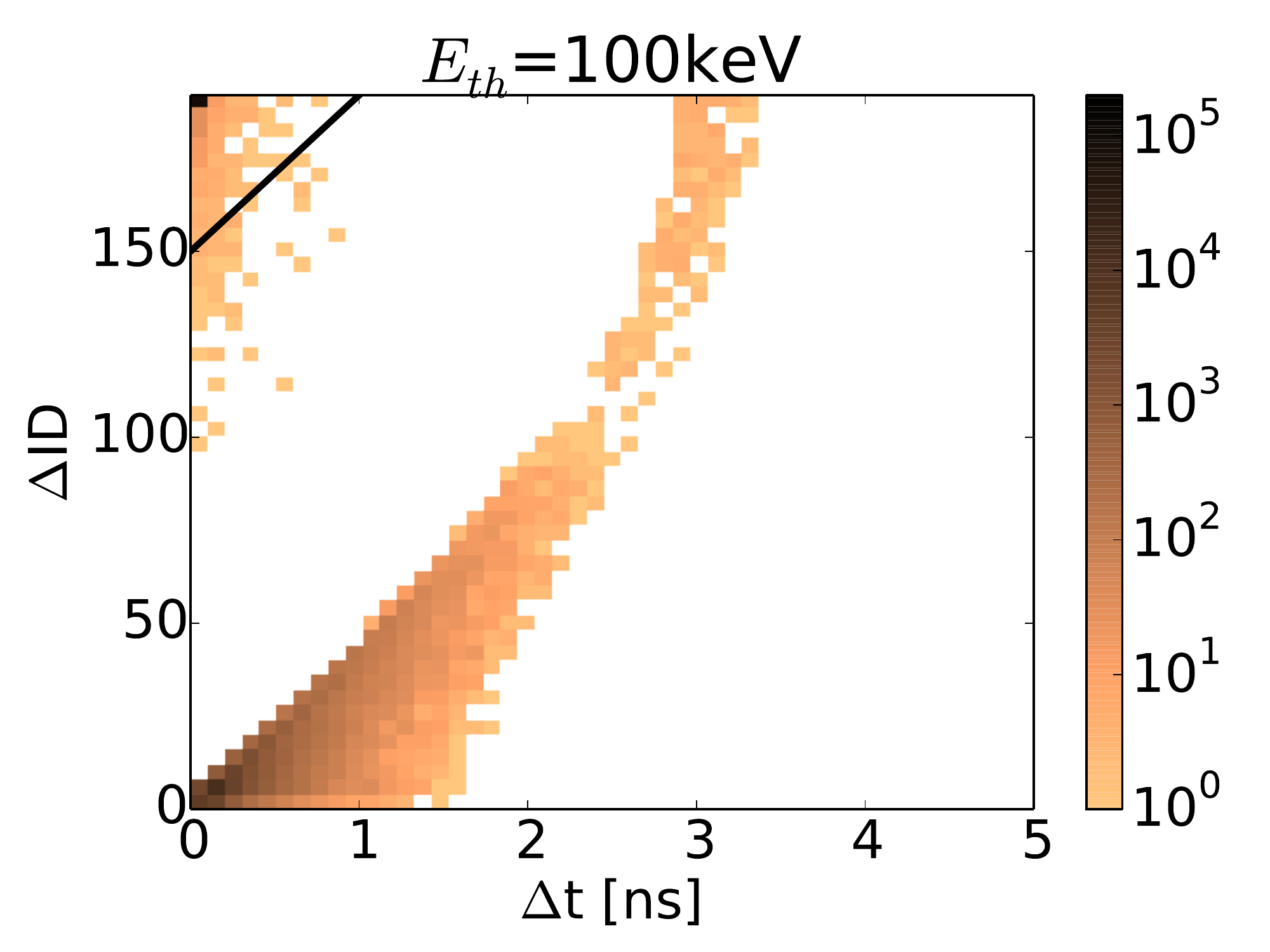}
\end{subfigure}

\begin{subfigure}{0.49\textwidth}
    \centering	\includegraphics[width=\textwidth]{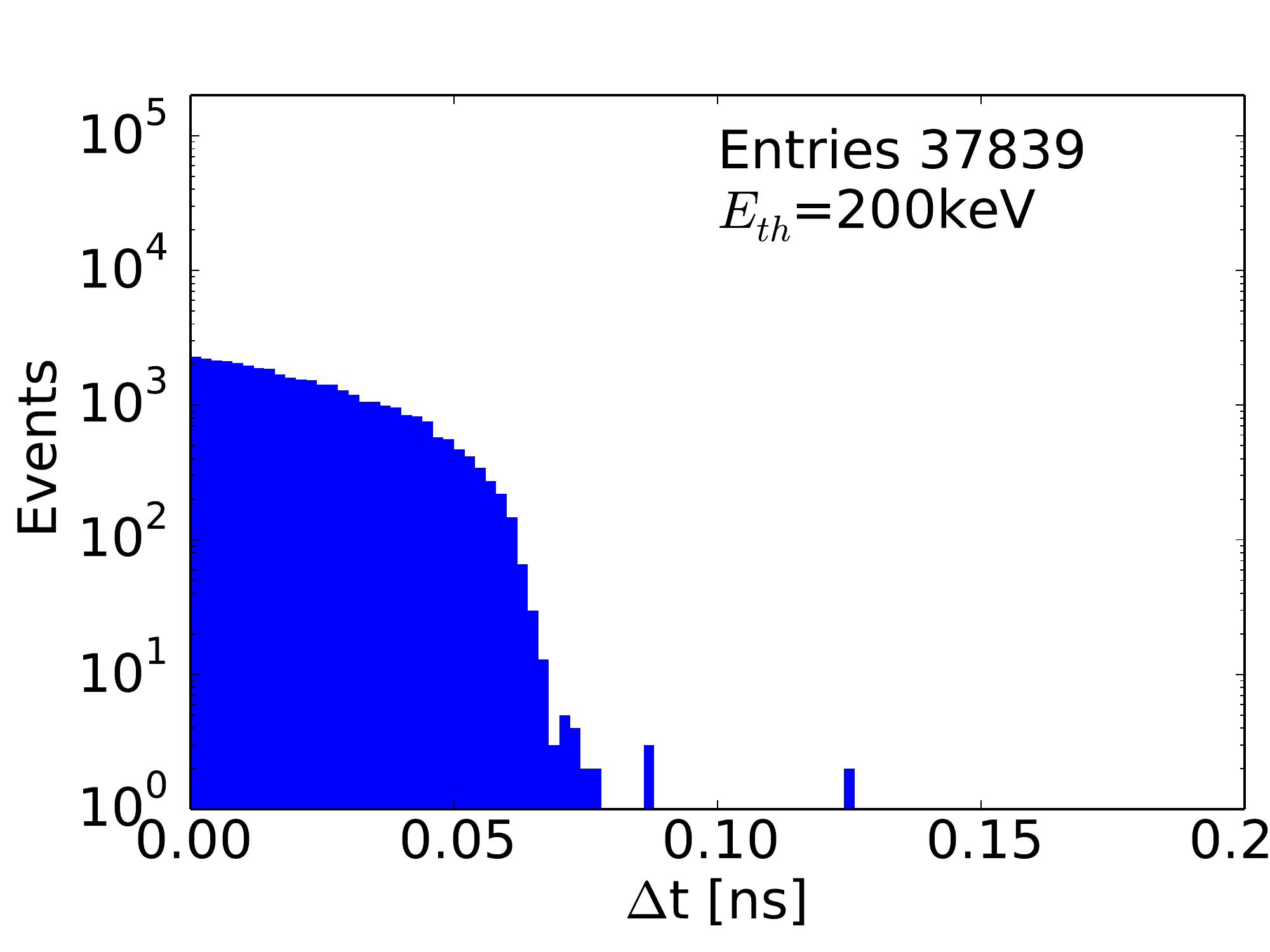}
\end{subfigure}
\begin{subfigure}{0.49\textwidth}
    \centering	
	\includegraphics[width=\textwidth]{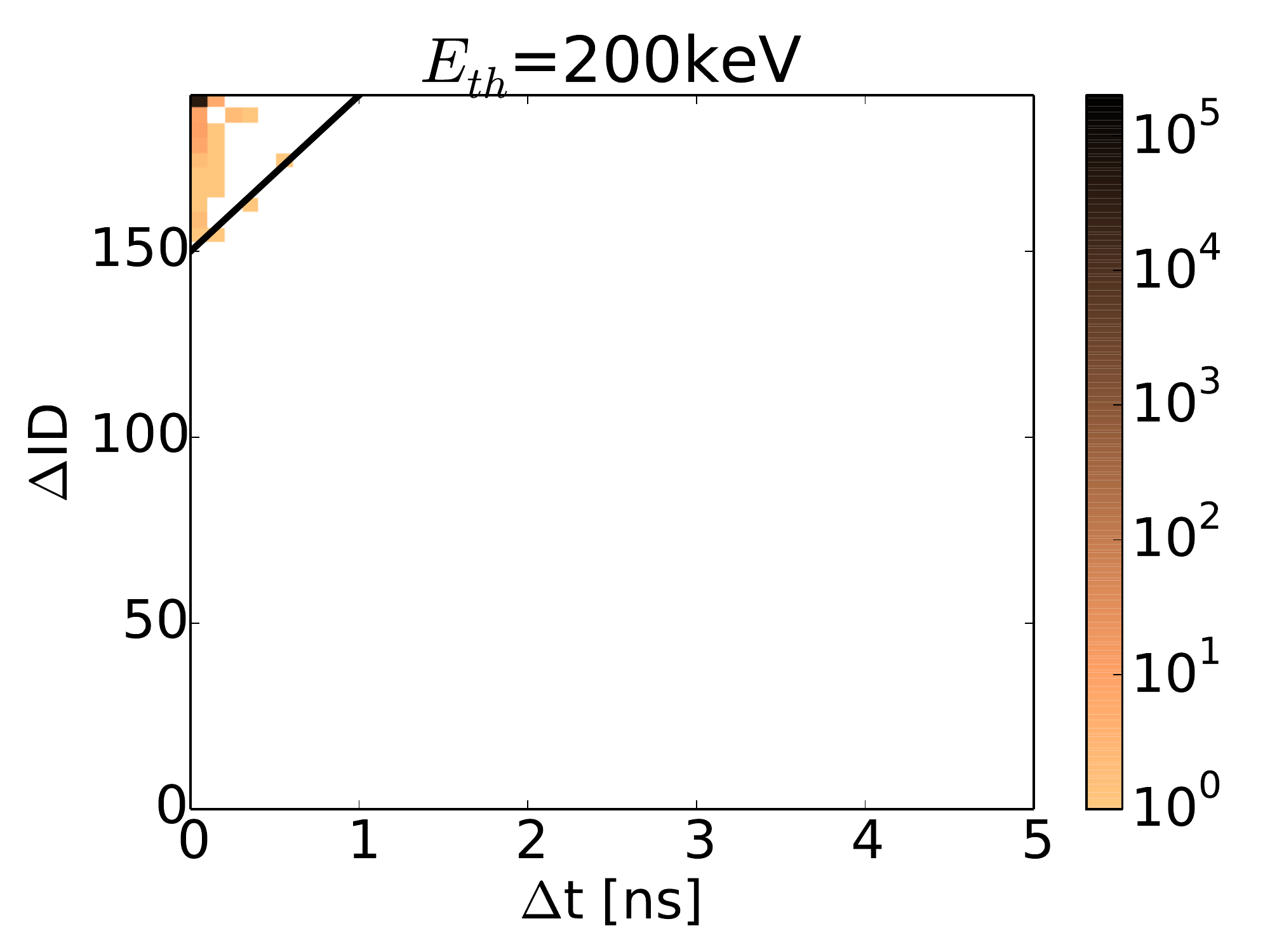}
\end{subfigure}

\caption{Distributions of differences of hit times; $N_{strips}=2$, $\mu=-2$ or $\mu=2$; black line in the two-dimensional histogram shows the boundary between events treated as originating from primary interactions only (above the line) and events including secondary interactions (below the line). Figure presents results of simulations for L=50~cm. The time differences are calculated only for interactions originating from the same annihilation process.}
\label{Nstrips2}
\end{figure}

\begin{figure}[h!]
\centering

\begin{subfigure}{0.49\textwidth}
    \centering	\includegraphics[width=\textwidth]{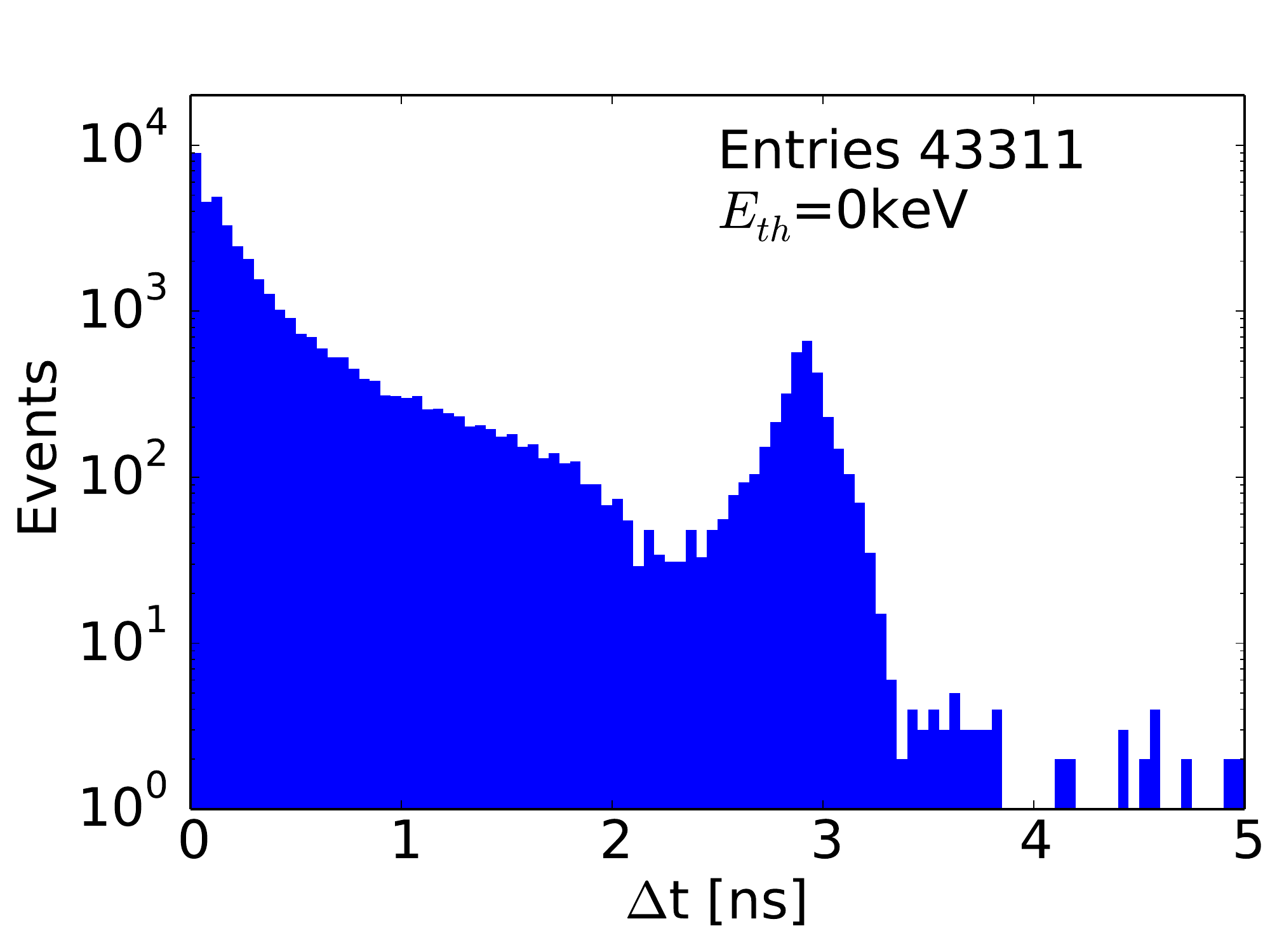}
\end{subfigure}
\begin{subfigure}{0.49\textwidth}
    \centering	
	\includegraphics[width=\textwidth]{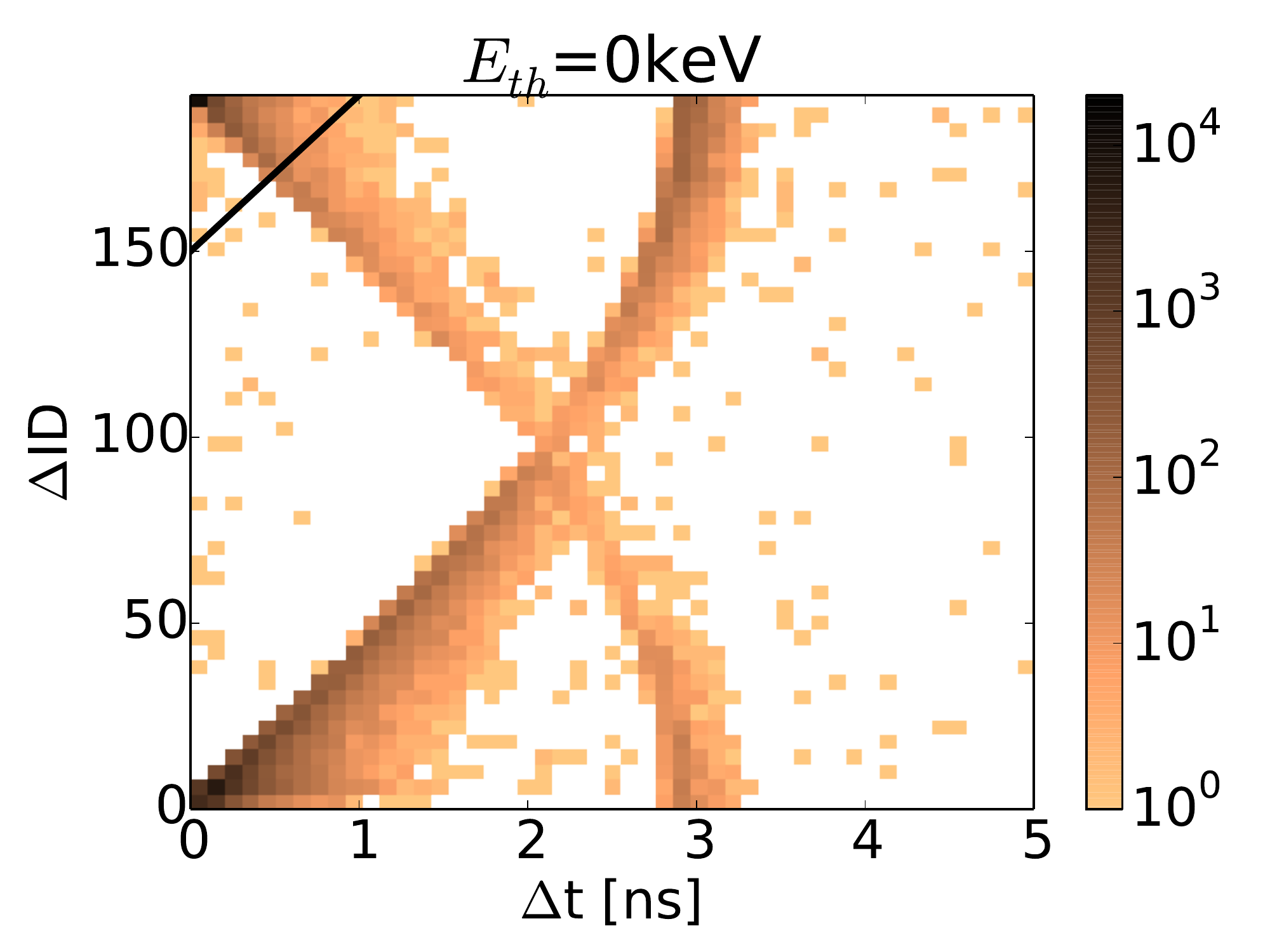}
\end{subfigure}

\begin{subfigure}{0.49\textwidth}
    \centering	\includegraphics[width=\textwidth]{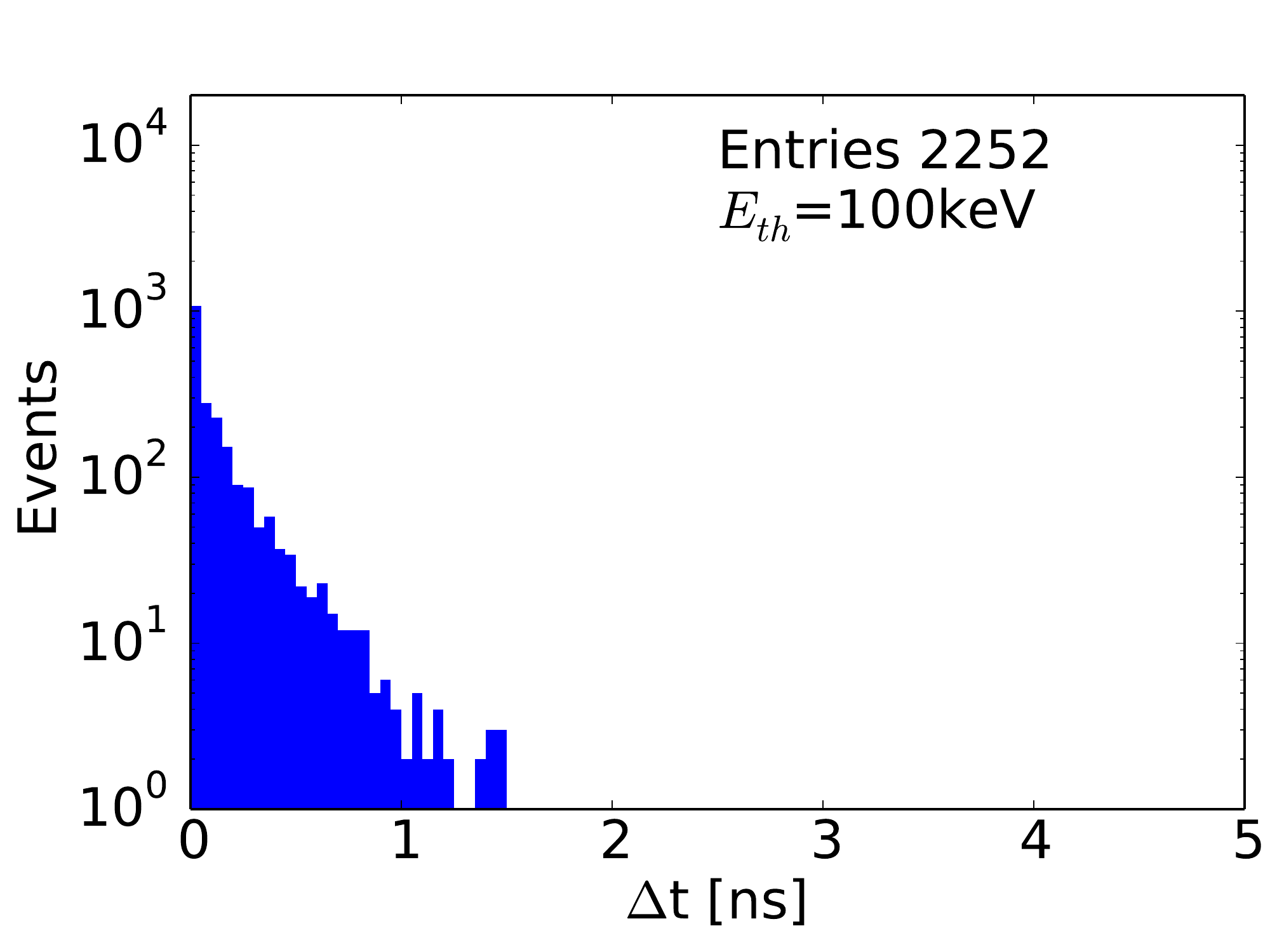}
\end{subfigure}
\begin{subfigure}{0.49\textwidth}
    \centering	
    \includegraphics[width=\textwidth]{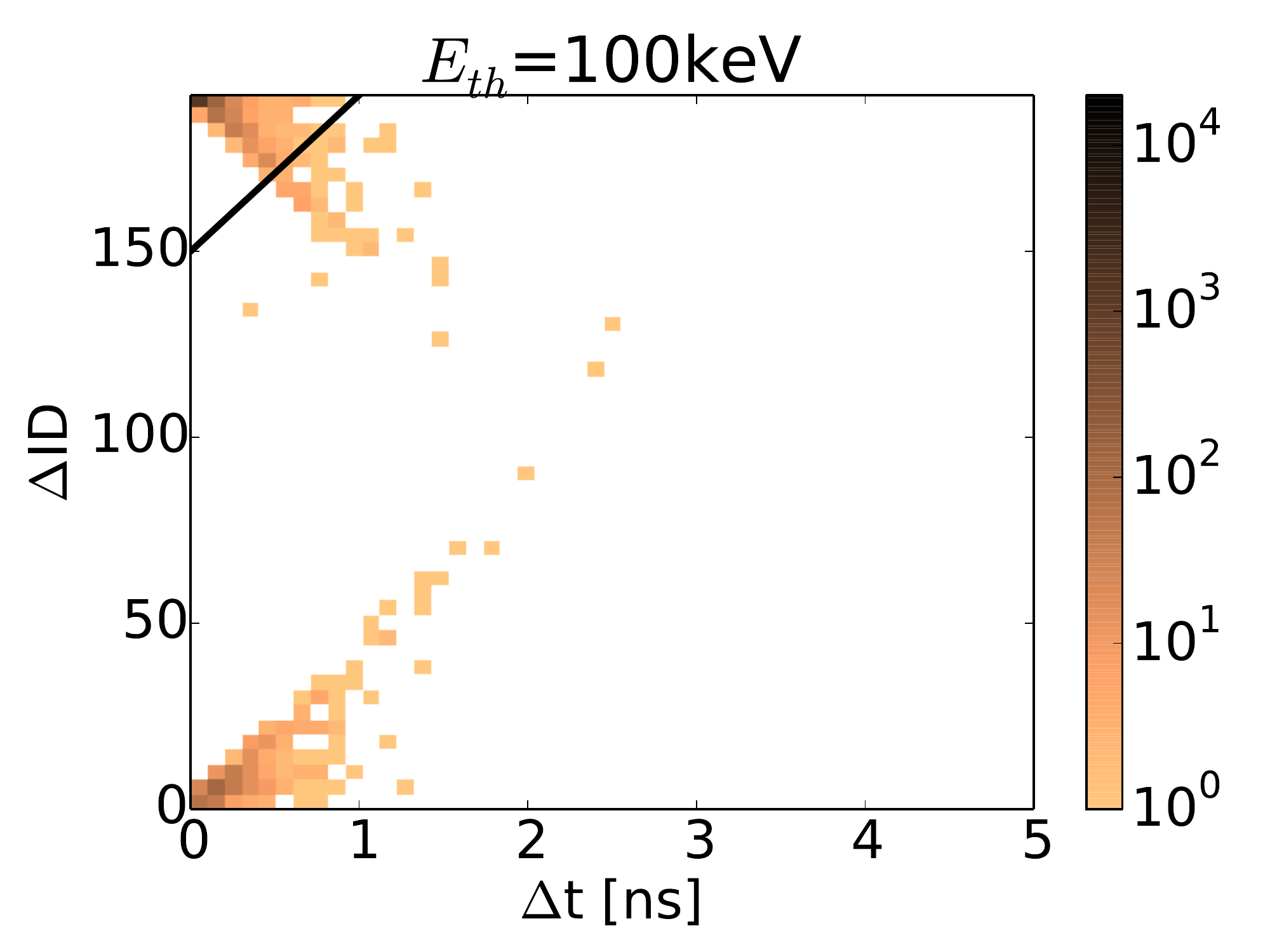}
\end{subfigure}

\caption{Distributions of differences of hit times; $N_{strips}$=3, $\mu=-3$ or $\mu=3$, both time differences are taken into account. Figure presents results of simulations for L=50~cm. The time differences are calculated only for interactions originating from the same annihilation process. Figure is described with details in the text.}
\label{Nstrips3_2diff}
\end{figure}

\begin{figure}[h!]
\centering

\begin{subfigure}{0.49\textwidth}
    \centering	\includegraphics[width=\textwidth]{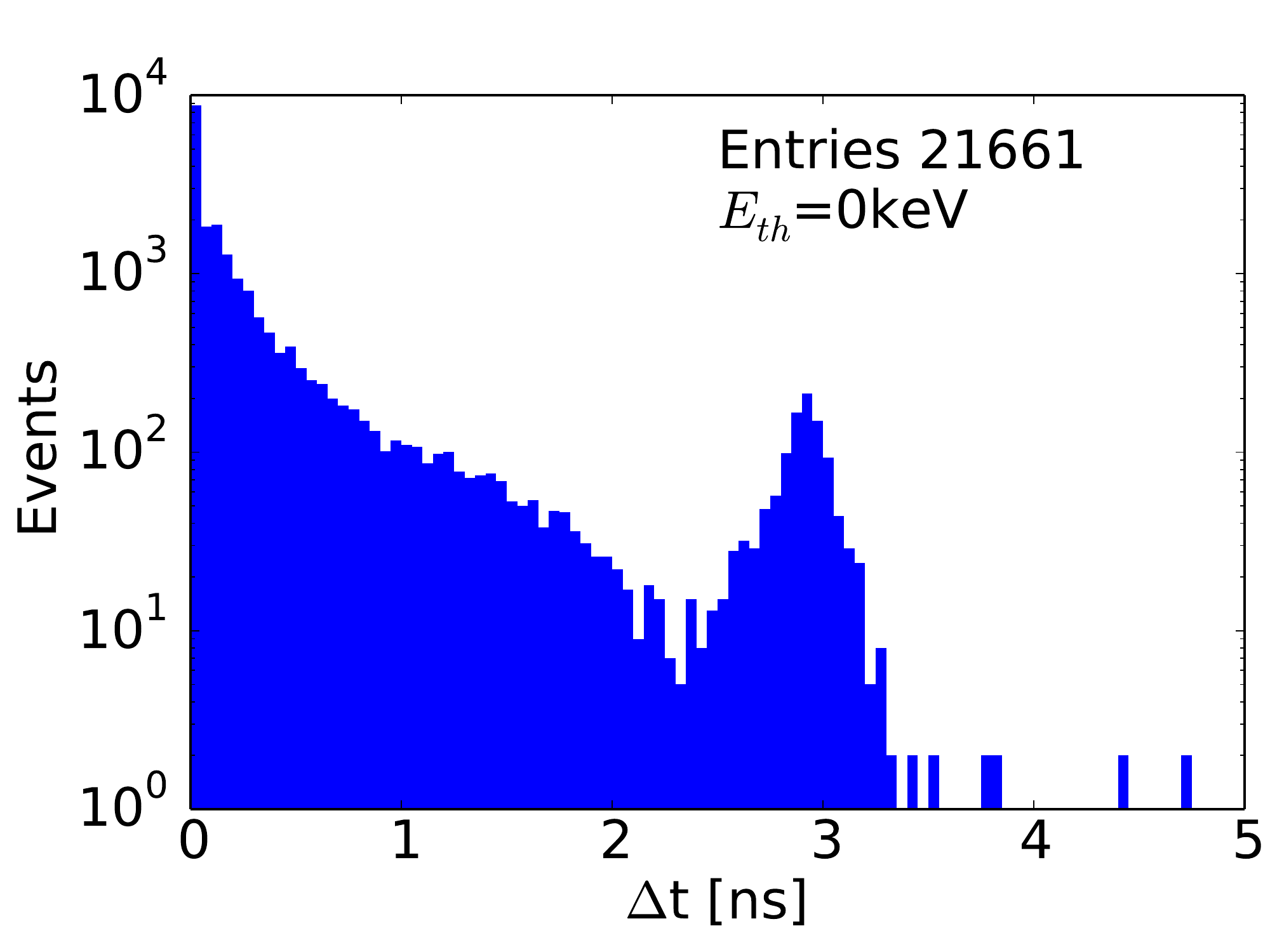}
\end{subfigure}
\begin{subfigure}{0.49\textwidth}
    \centering	
    \includegraphics[width=\textwidth]{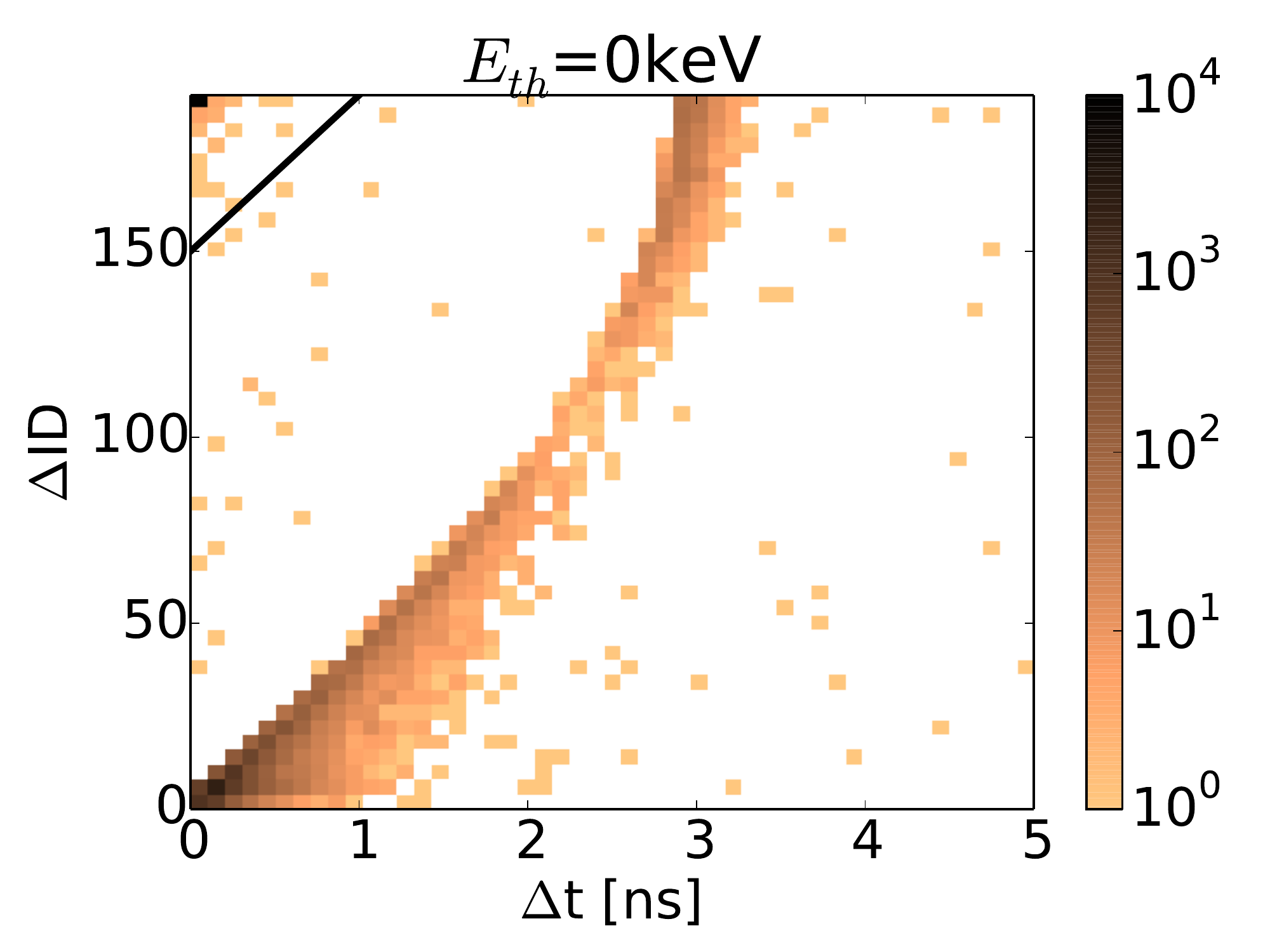}
\end{subfigure}

\begin{subfigure}{0.49\textwidth}
    \centering	\includegraphics[width=\textwidth]{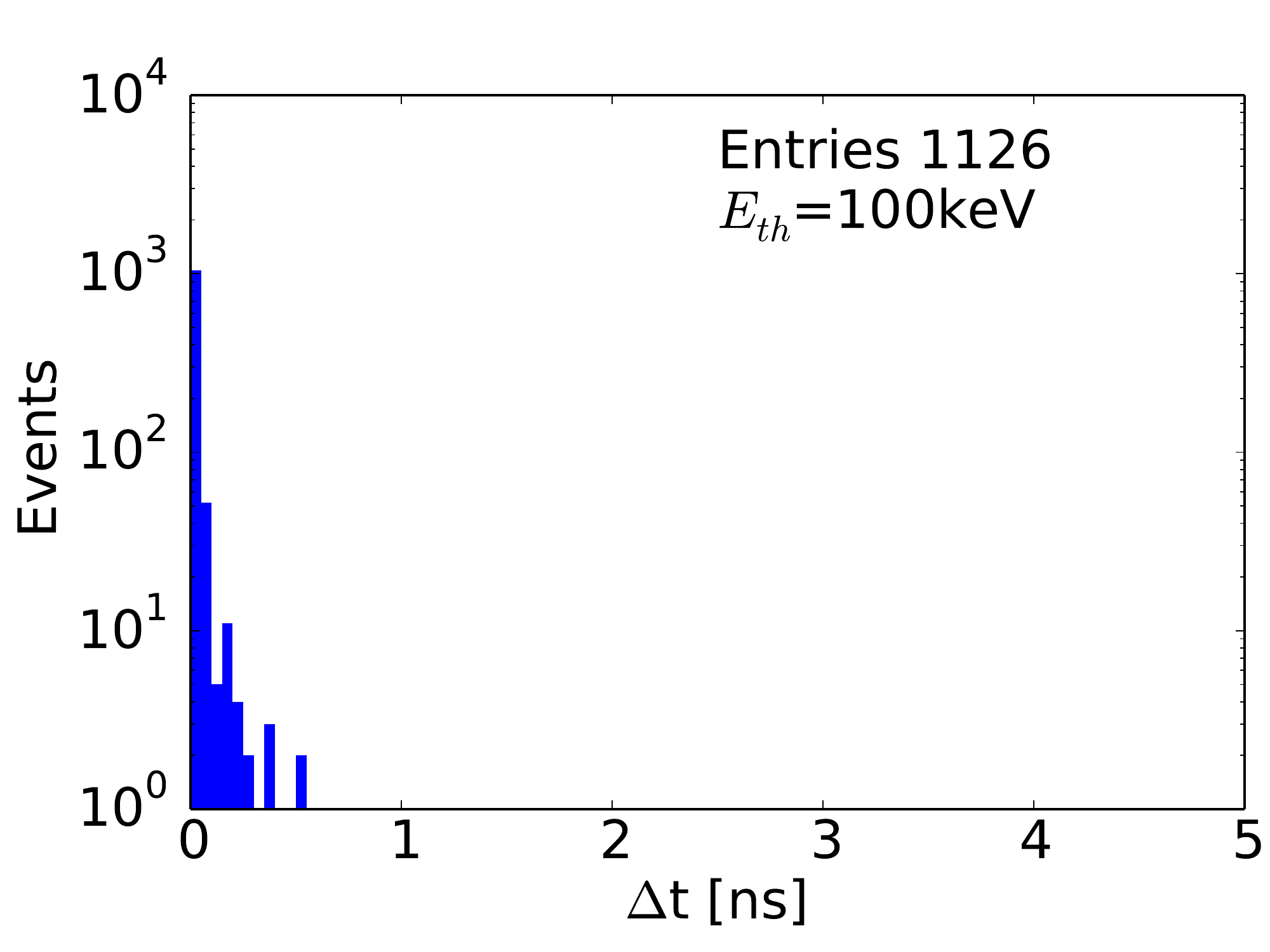}
\end{subfigure}
\begin{subfigure}{0.49\textwidth}
    \centering	
    \includegraphics[width=\textwidth]{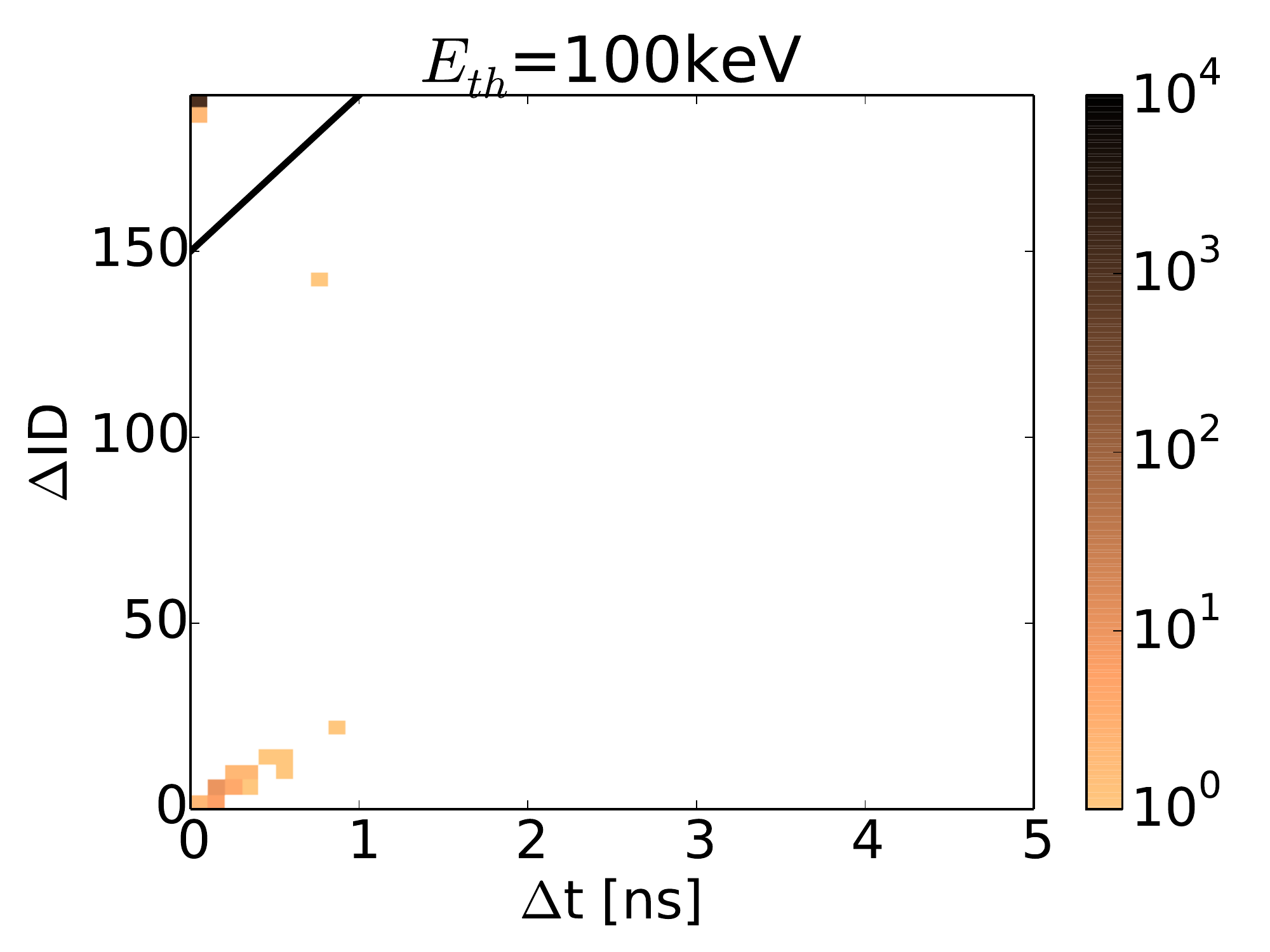}
\end{subfigure}

\caption{Distributions of differences of hit times; $N_{strips}$=3, $\mu=-3$ or $\mu=3$, only first time difference is taken into account. Figure presents results of simulations for L=50~cm. The time differences are calculated only for interactions originating from the same annihilation process. Figure is described with details in the text.}
\label{Nstrips3_1diff}
\end{figure}

\begin{figure}[h!]
\centering

\begin{subfigure}{0.49\textwidth}
    \centering	
    \includegraphics[width=\textwidth]{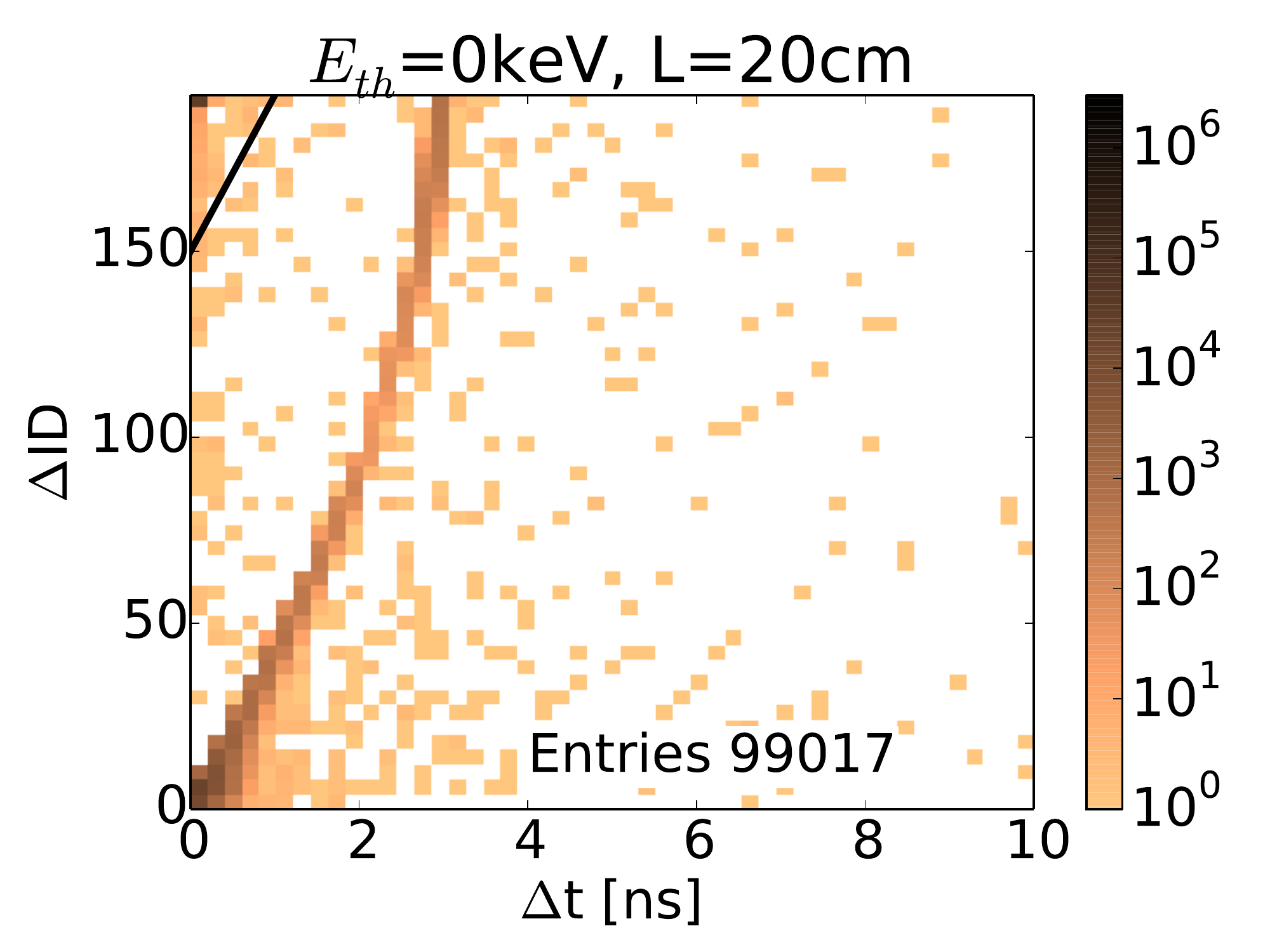}
\end{subfigure}
\begin{subfigure}{0.49\textwidth}
    \centering	
    \includegraphics[width=\textwidth]{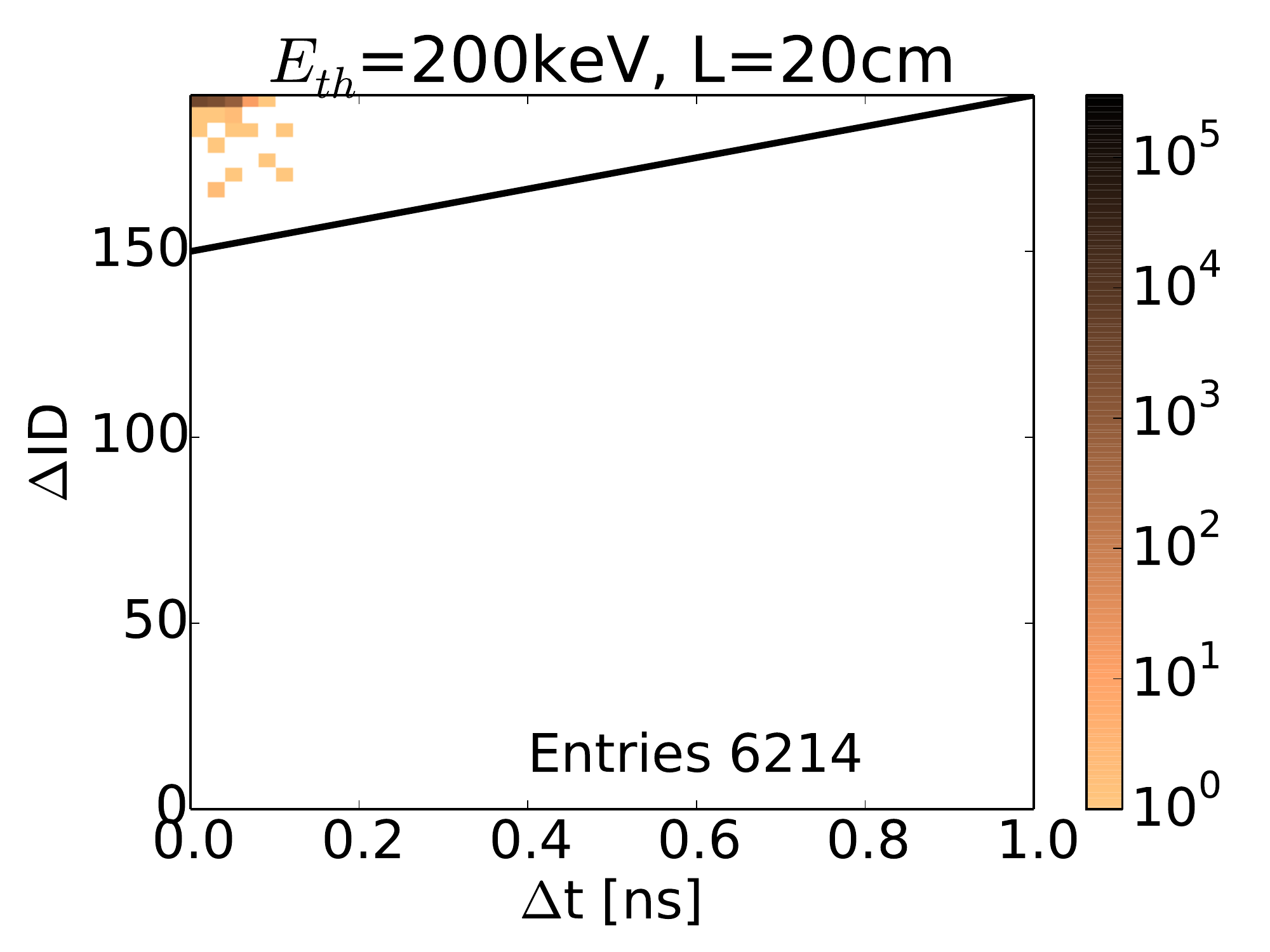}
\end{subfigure}

\begin{subfigure}{0.49\textwidth}
    \centering	
    \includegraphics[width=\textwidth]{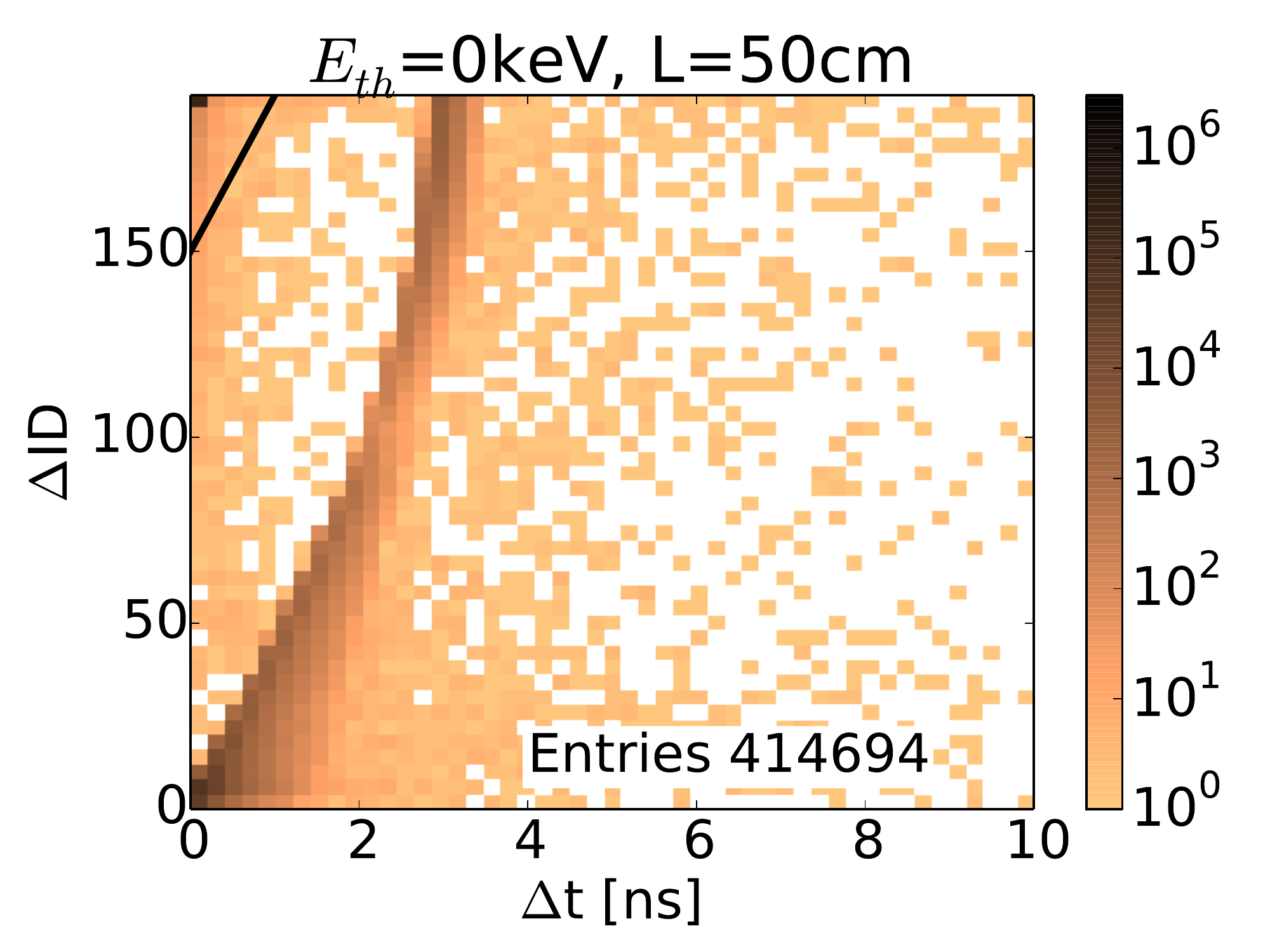}
\end{subfigure}
\begin{subfigure}{0.49\textwidth}
    \centering	
    \includegraphics[width=\textwidth]{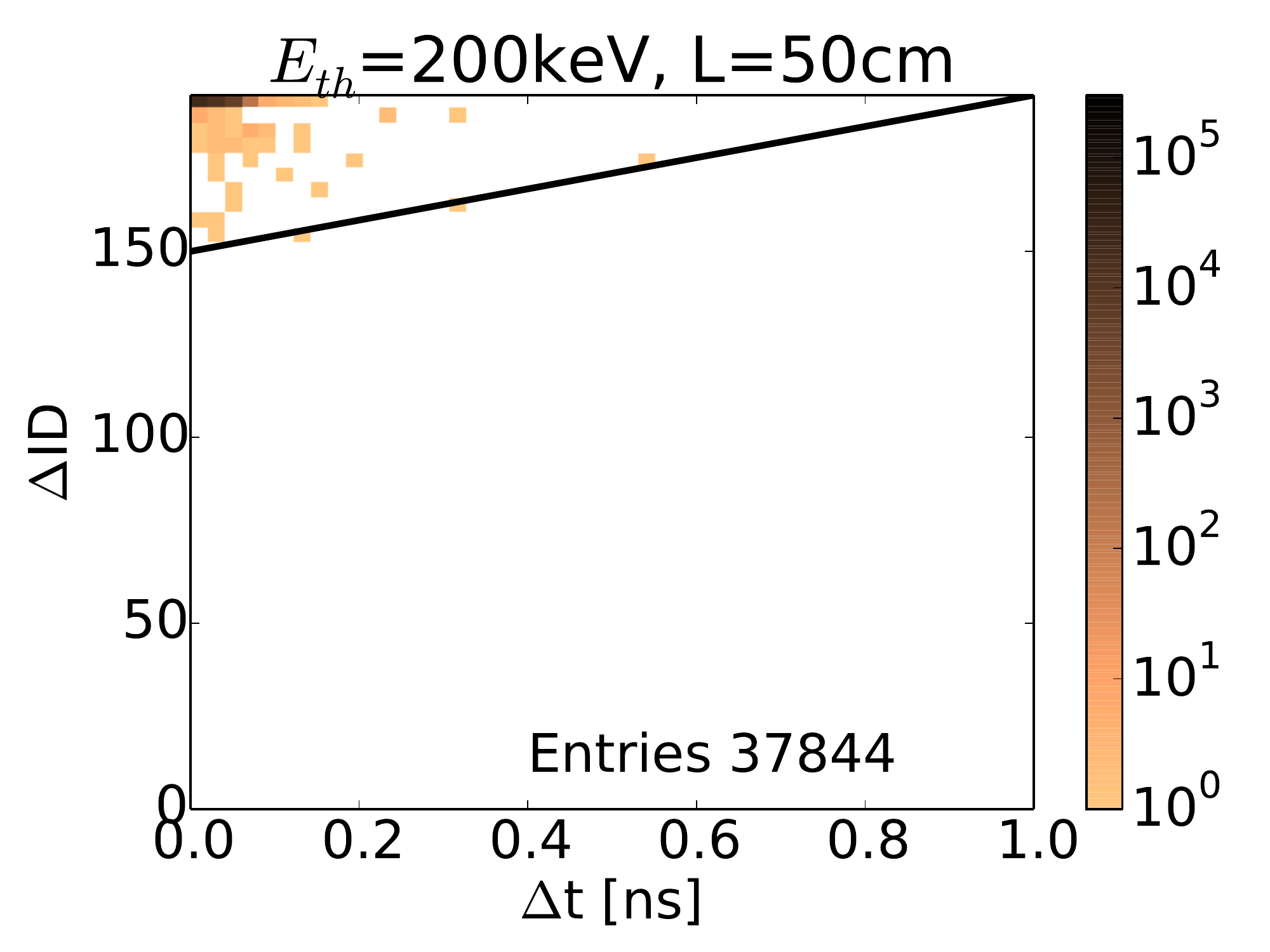}
\end{subfigure}

\begin{subfigure}{0.49\textwidth}
    \centering	
    \includegraphics[width=\textwidth]{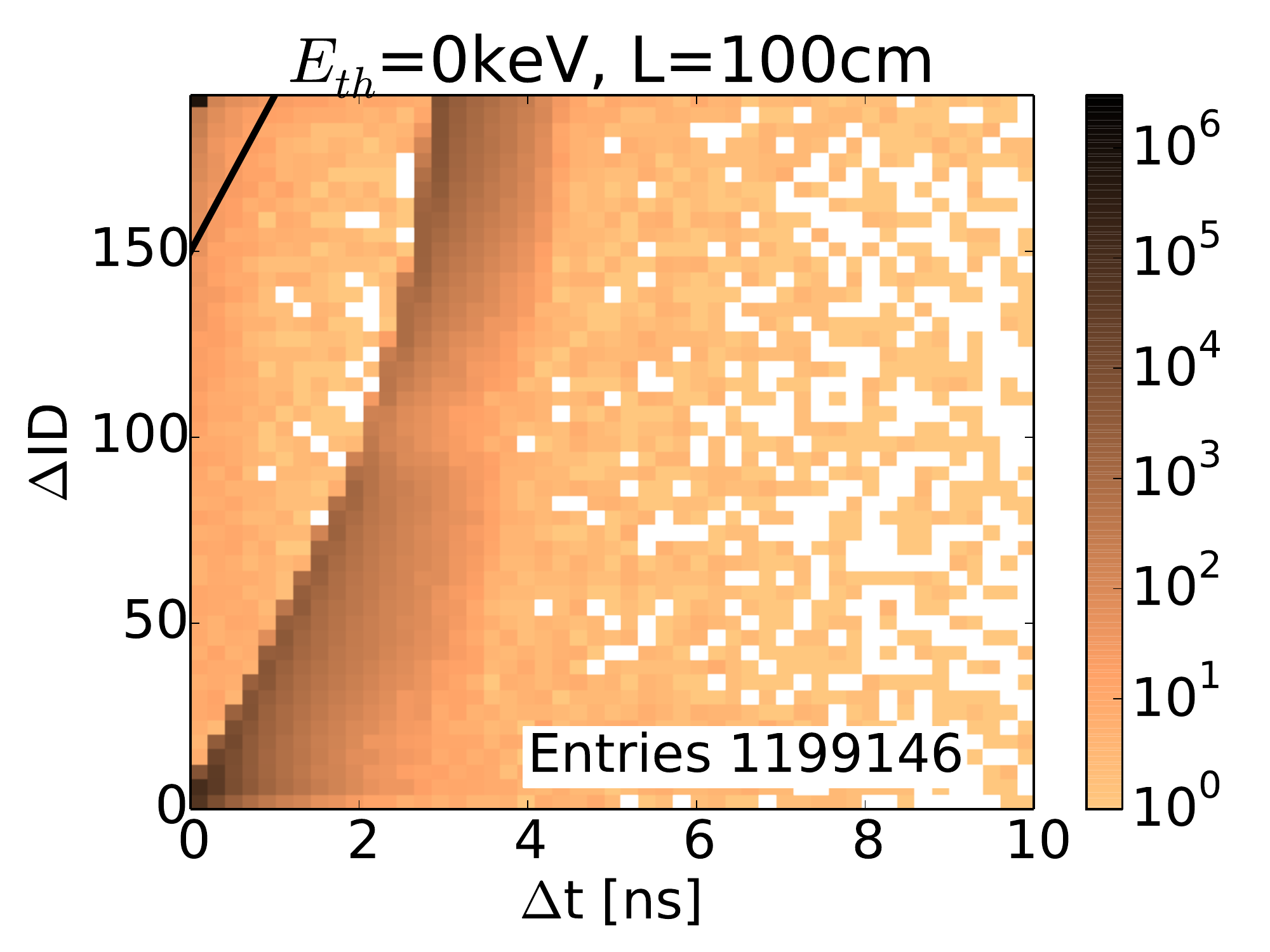}
\end{subfigure}
\begin{subfigure}{0.49\textwidth}
    \centering	
    \includegraphics[width=\textwidth]{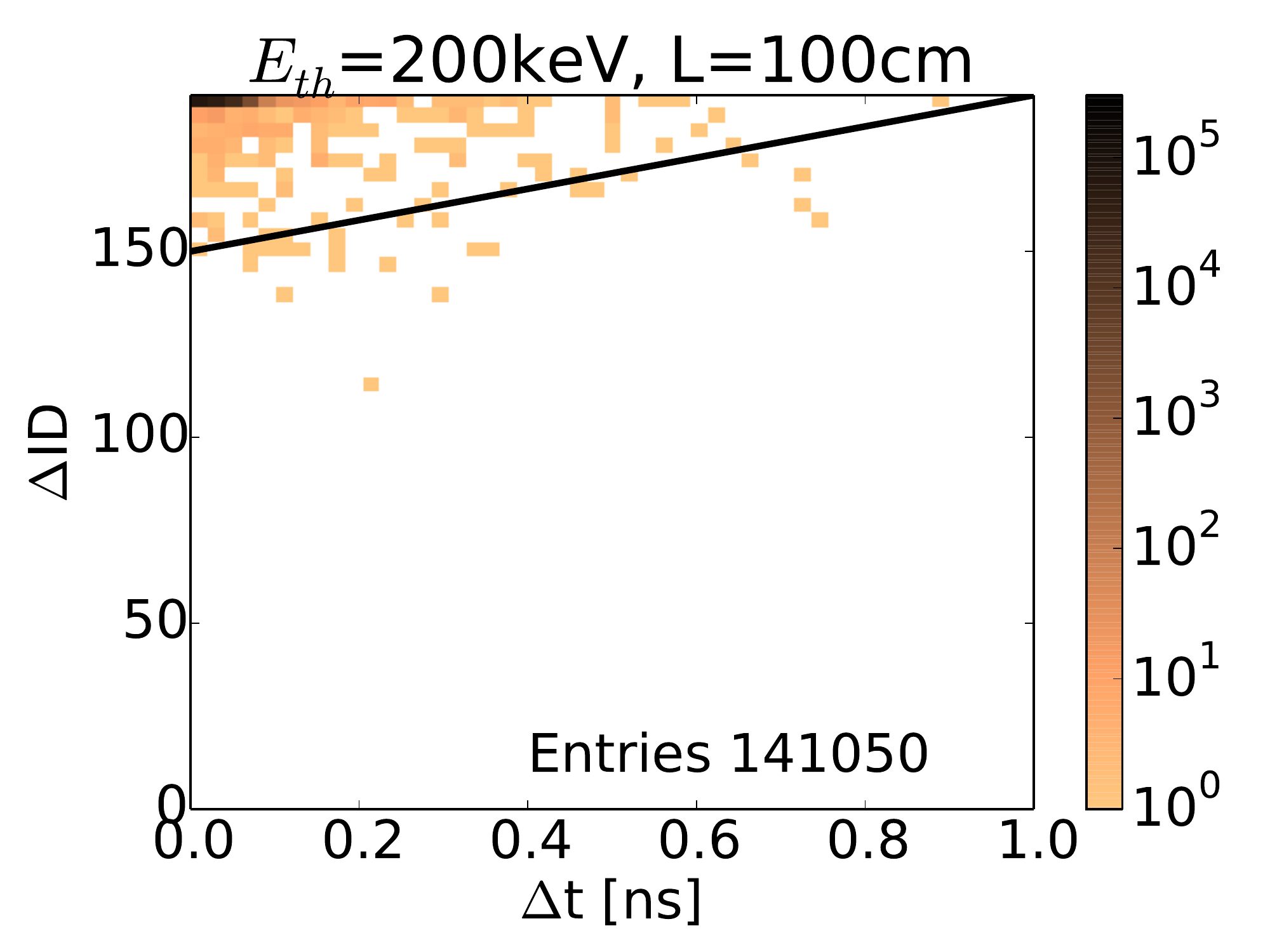}
\end{subfigure}

\begin{subfigure}{0.49\textwidth}
    \centering	
    \includegraphics[width=\textwidth]{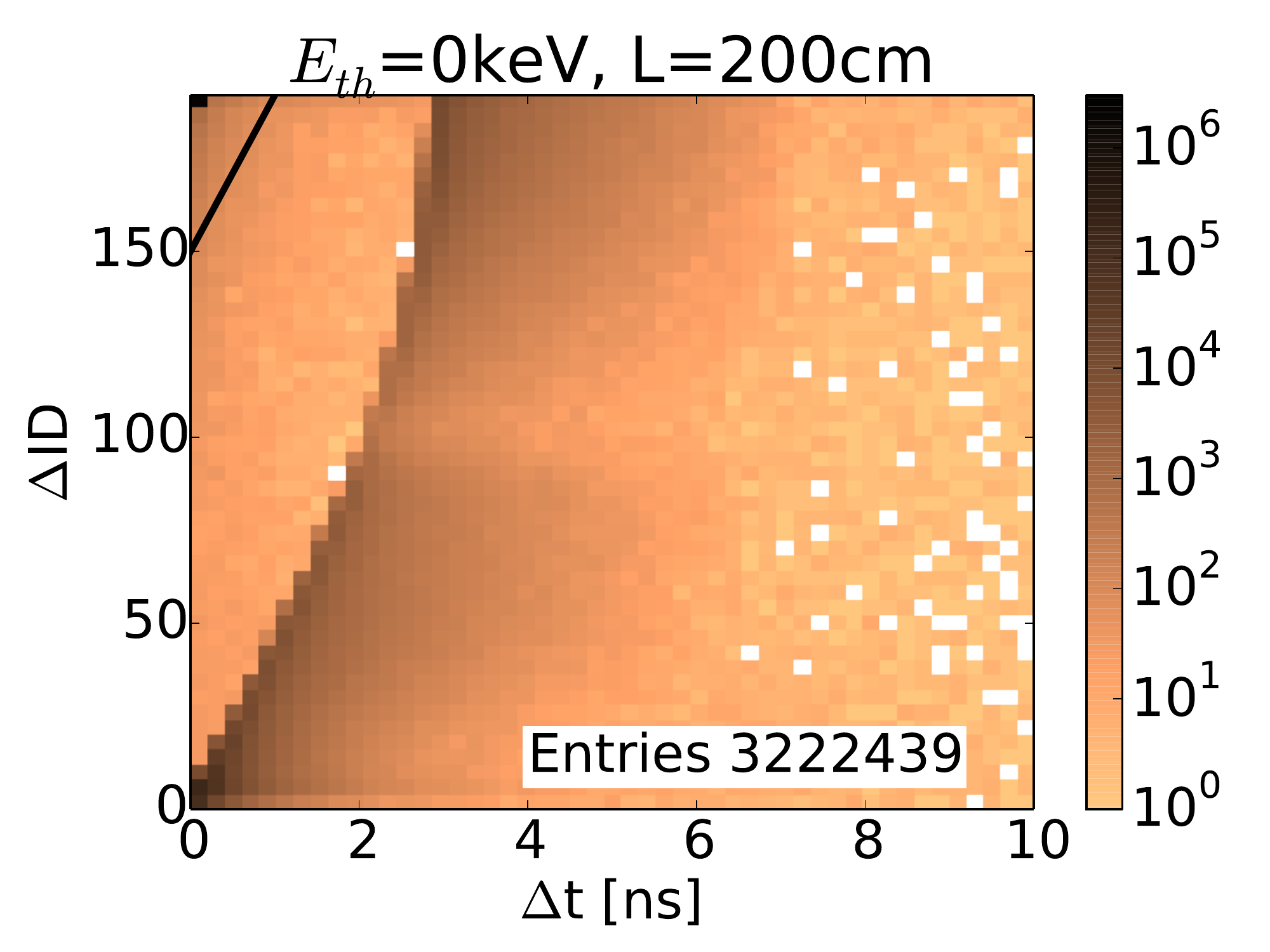}
\end{subfigure}
\begin{subfigure}{0.49\textwidth}
    \centering	
    \includegraphics[width=\textwidth]{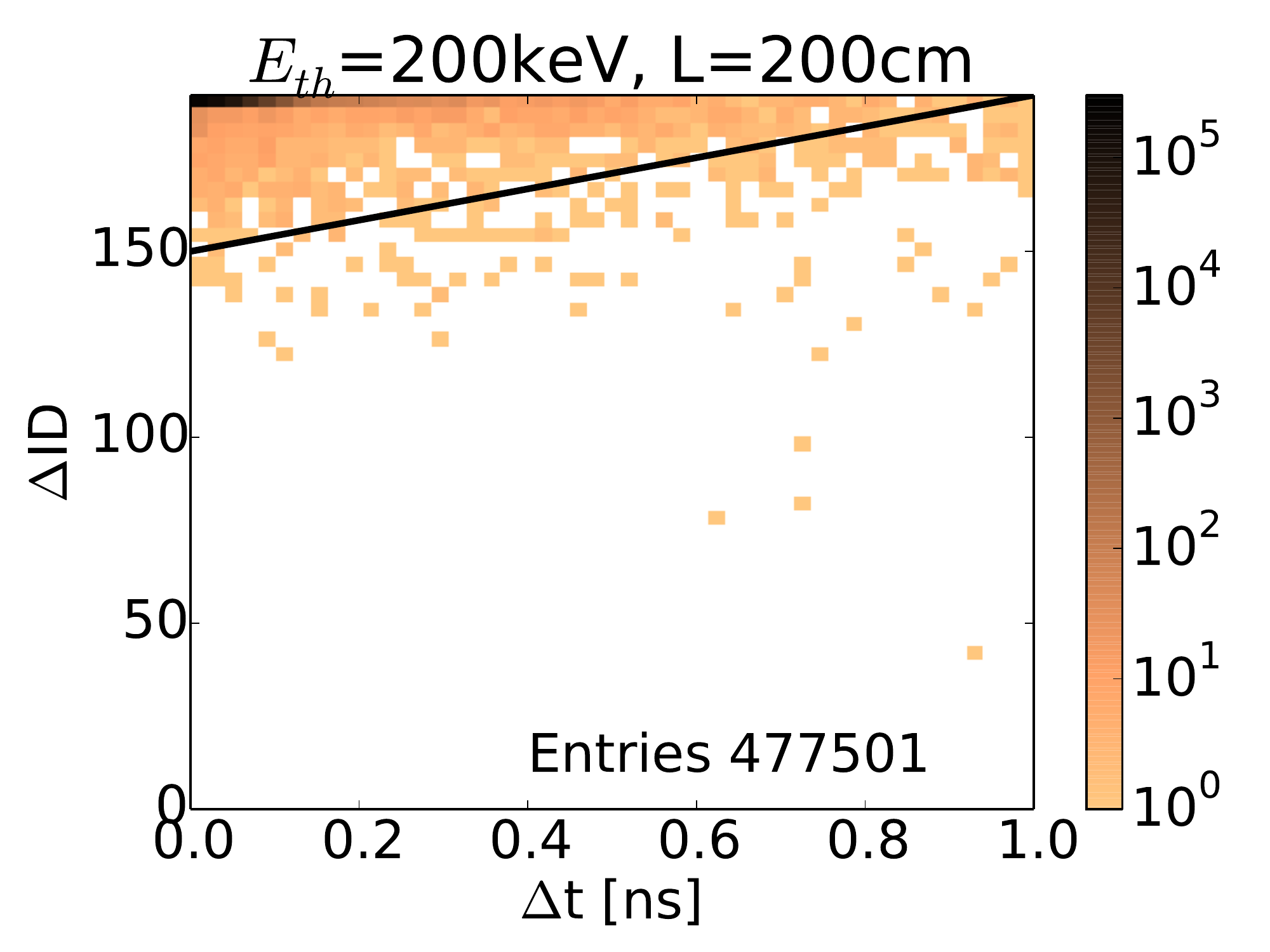}
\end{subfigure}

\caption{Distributions of differences of hit times ($\Delta t$) as a function of ID difference between hit modules ($\Delta ID$). The length L of the detector and the energy thresholds are indicated above the figures. Only interactions originating from the same annihilation process are taken into account.}
\label{histograms_diff_lengths}
\end{figure}

For $N_{strips}$=2 and $N_{strips}$=3 and length of scintillators equal to L~=~50~cm, histograms of time differences between subsequent hits were calculated. These histograms are presented in Figs \ref{Nstrips2} - \ref{Nstrips3_1diff}. 
Right panel of these figures show distribution of difference between ID of hit modules ($\Delta ID$) as a function of hit time difference.\footnote{If IDs of hit modules are $ID_1$ and $ID_2$ then $\Delta ID = min(|ID_1-ID_2|, 384-|ID_1-ID_2|)$.} The module ID increases monotonically with the grows of the azimuthal angle $\varphi$ (see Fig. \ref{geometry}).
Black lines in the two-dimensional histograms (Figs \ref{Nstrips2} - \ref{histograms_diff_lengths}) show the boundaries between events treated as useful coincidences and events treated as background coincidences due to the secondary scatterings. A positive value of $\mu$ (2 or 3) is assigned to events above the line, which are treated in further analysis as true conicidences. Whereas to events below the line a negative value of $\mu$ (-3 or -2) is assigned since these events include secondary scattering of gamma quanta. This boundary was used to separate events with different multiplicities for preparation of histograms presented in Fig. \ref{hist_mi}.

In Fig. \ref{Nstrips2}, for energy thresholds 0 keV and 100 keV, in two-dimensional histograms there is longitudinal structure extending between points (0 ns, 0) and (3 ns, 192). These events correspond to difference between time of primary reaction of the gamma quantum in a given scintillator and a time of the secondary scattering. The larger is the angle of the primary scattered gamma quantum the larger will be the $\Delta ID$ value and also a $\Delta t$. For example bin with coordinates (2.9 ns, 192) corresponds to the backscattering - primary particle is backscattered and it is registered in the strip on the opposite side of the scintillator (2.9 ns is the time needed by the gamma quanta to travel between opposite strips with speed of light).

If the energy threshold is set to 200 keV, nearly all scattered coincidences are eliminated. In the lower panel of this figure there are results for this threshold. In ideal situation, time difference for this simulation for true coincidences would be always 0 and we would have only one bin for 0 ns. Because of the fact that gamma quanta interact with matter in different depths (Depth of Interaction), time difference is changing from 0 to about 80 ps. This picture show, what is the time limit for time-of-flight determination with scintillator strips of 19 mm thickness.

In Fig. \ref{Nstrips3_2diff}, for energy thresholds 0 keV and 100 keV, in two-dimensional histograms there is symmetrical butterfly-shape structure extending between points (0 ns, 0) and (3 ns, 192) and between points (3 ns, 0) and (0 ns, 192). 
Each event with three hits and deposited energy above the energy thresold, gives two inputs to these histograms. 
An additional structure (for $N_{strips}=3$) which is spanned between points (3 ns, 0) and (0 ns, 192) originates from the time differences between the primary interaction of one of the gamma quantum and a secondary interaction of the other or from the time difference between two secondary interactions. 
Pictorial definitions of these situations are presented in Fig.~\ref{mi}. If only the first time difference is taken into account, histograms for 3~hits (Fig. \ref{Nstrips3_1diff}) look like histograms for 2~hits (Fig. \ref{Nstrips2}).

Response of the detector to the annihilations in the 2~m long line placed along the detector axis was simulated also for other lengths of scintillators L~=~20~cm, 100~cm and 200~cm. Results of these simulations for two energy thresholds (0~keV and 200~keV) are presented in Fig. \ref{histograms_diff_lengths}. One can see that, the longer the scintillators, the wider the longitudal structure described above. It is caused by the fact that, the longer the scintillators, the longer the possible distance between places of the primary and secondary interactions. For the scanner with 20 cm scintillators, the longest possible path along the diagonal of the longitudinal cross-section of the scanner has length of 88 cm ($\sim$ 2.9~ns) and for the scanner with 200 cm scintillators, the longest possible path is equal to 218~cm (7.3~ns). 

\clearpage
\subsection{Accidental coincidences}

An accidental coincidence is the coincidence, in which two events occur simultaneously in a fixed time window but in fact they are independent, they come from different annihilations. 
Because of that, number of accidental coincidences depends on the width of the time window, the size of the detector and in contrast to the secondary scattering, the accidental coincidences depend on the activity of the source. 

\subsubsection{Accidental coincidences as a function of the source activity}

\begin{figure}[h!]
\centering
\begin{subfigure}{0.49\textwidth}
    \centering	\includegraphics[width=\textwidth]{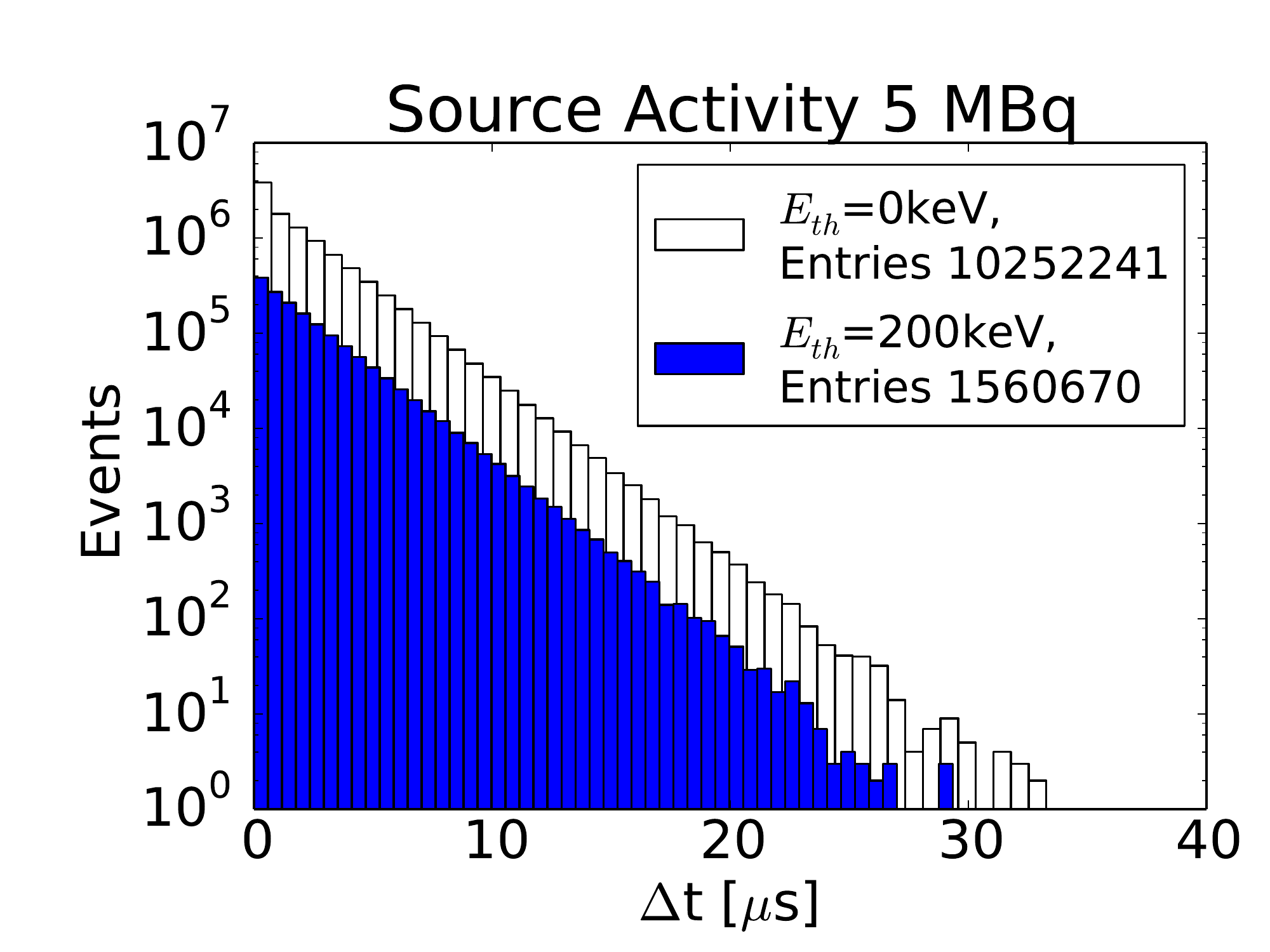}
\end{subfigure}
\begin{subfigure}{0.49\textwidth}
    \centering	\includegraphics[width=\textwidth]{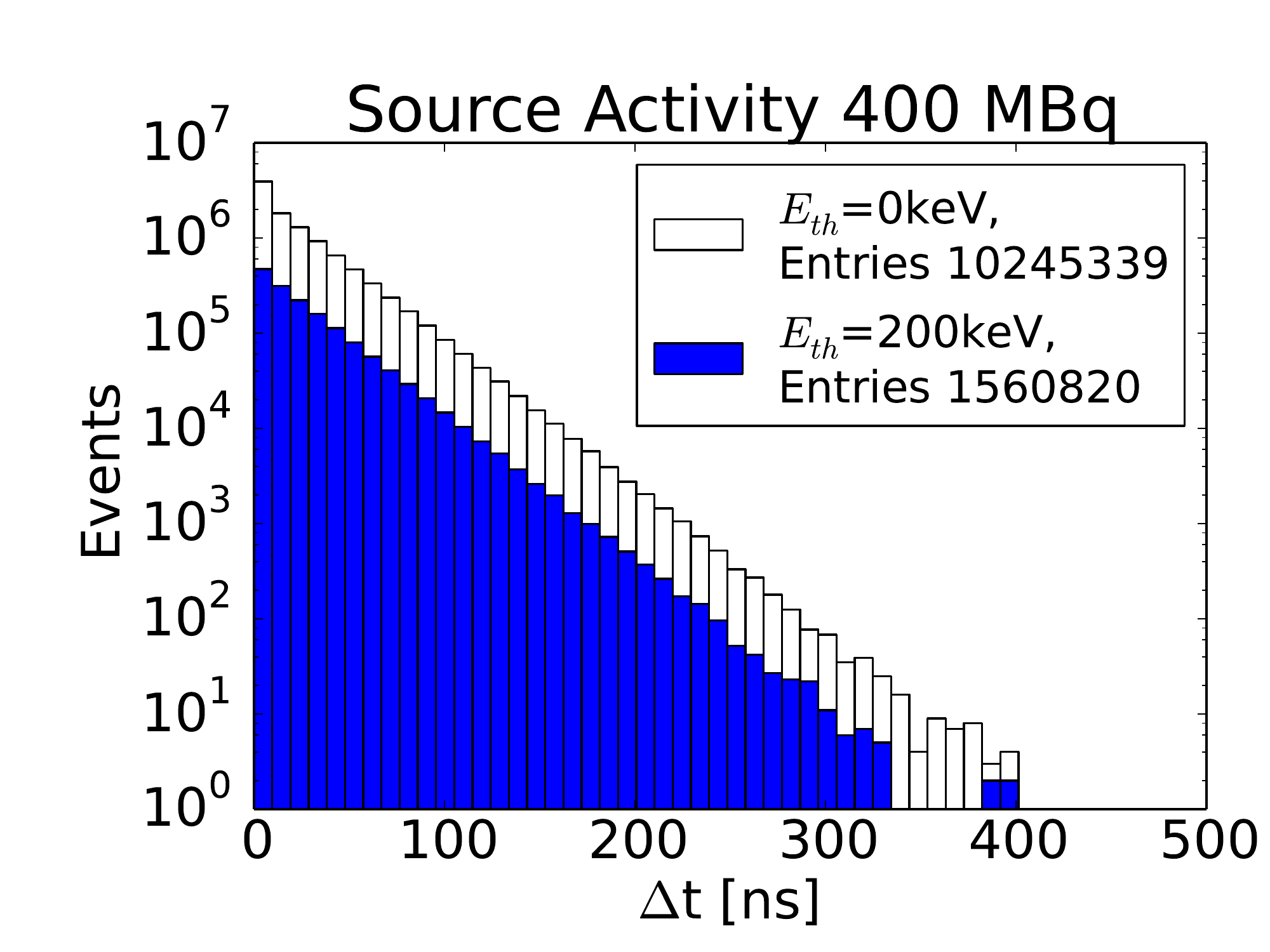}
\end{subfigure}
\caption{Histograms of numbers of all differences of hit times for length of scintillators equal to 50~cm, activities of 5 MBq and 400 MBq and for two different energy thresholds as indicated in the legends.}
\label{time_differences_all}
\end{figure}

\begin{figure}[h!]
\centering

\begin{subfigure}{0.49\textwidth}
    \centering	\includegraphics[width=\textwidth]{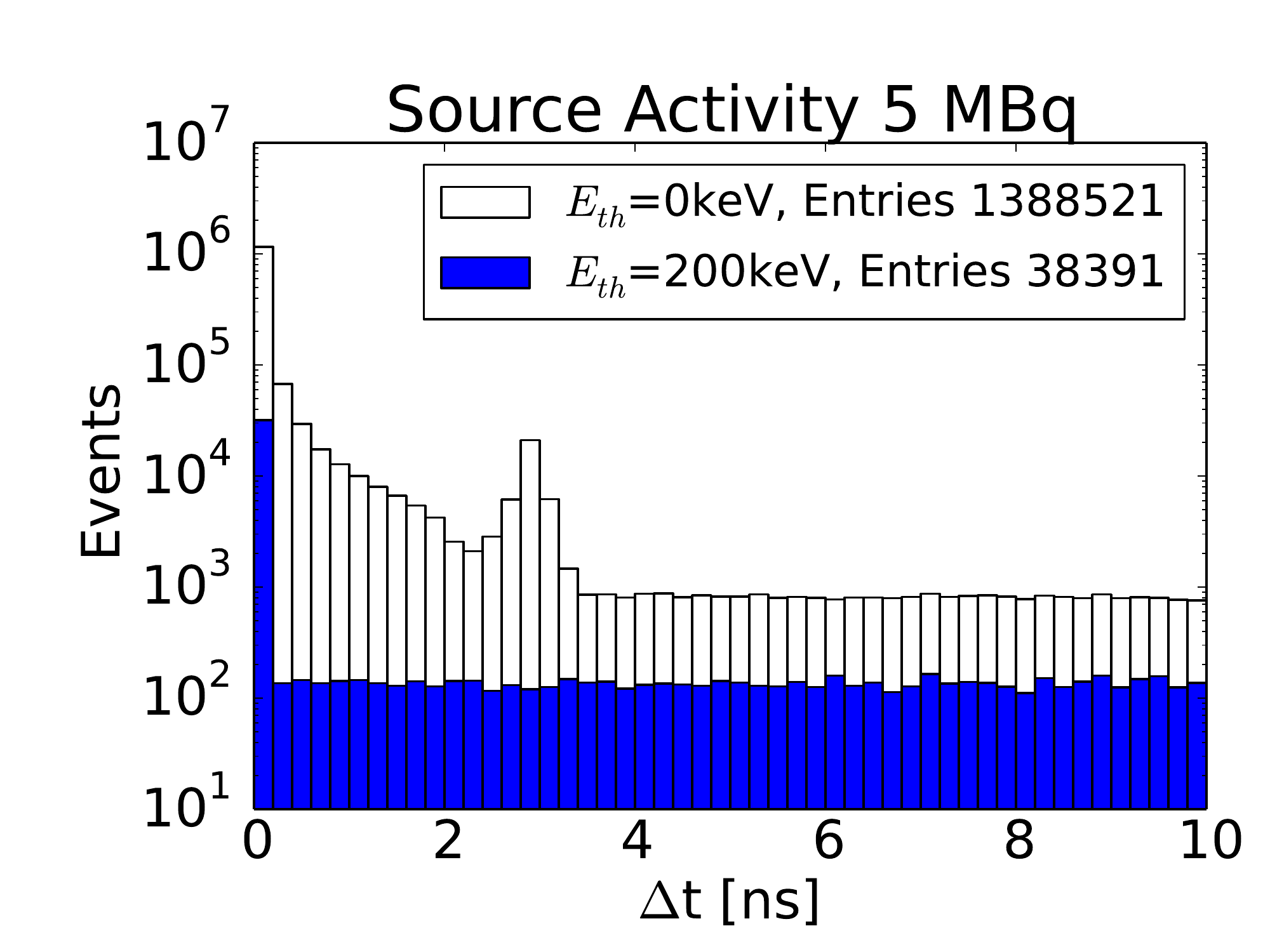}
\end{subfigure}
\begin{subfigure}{0.49\textwidth}
    \centering	\includegraphics[width=\textwidth]{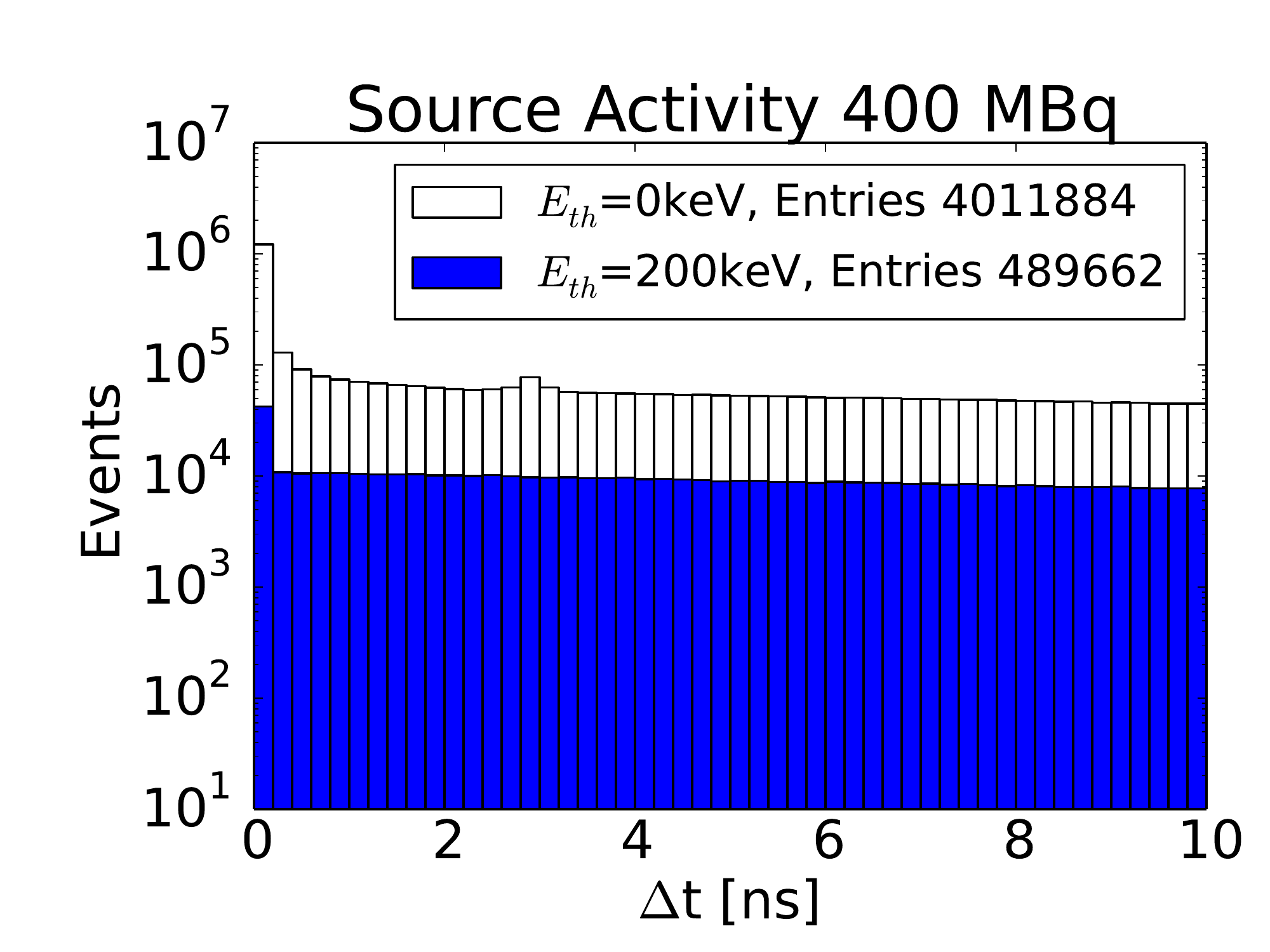}
\end{subfigure}

\begin{subfigure}{0.49\textwidth}
    \centering	\includegraphics[width=\textwidth]{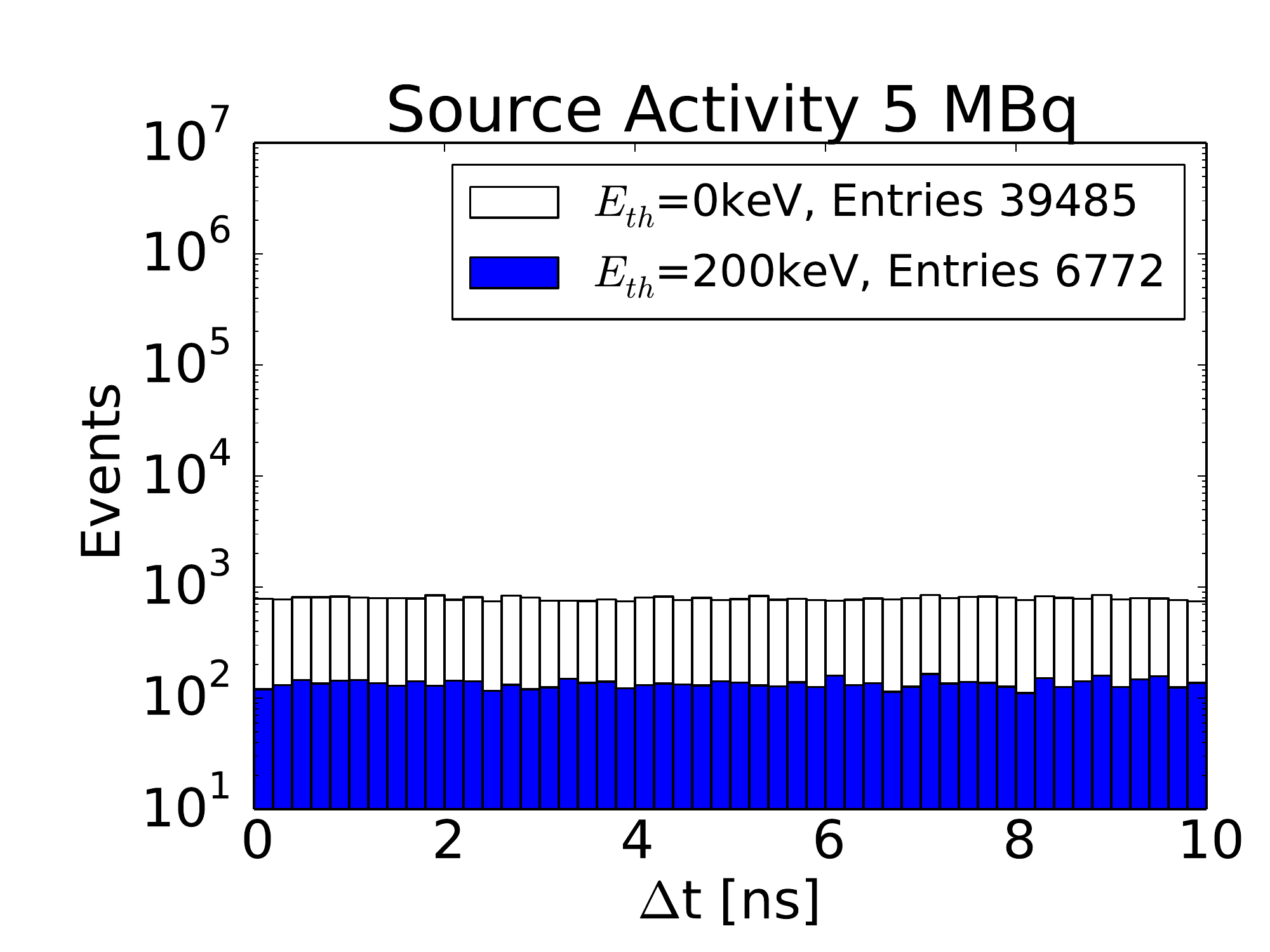}
\end{subfigure}
\begin{subfigure}{0.49\textwidth}
    \centering	\includegraphics[width=\textwidth]{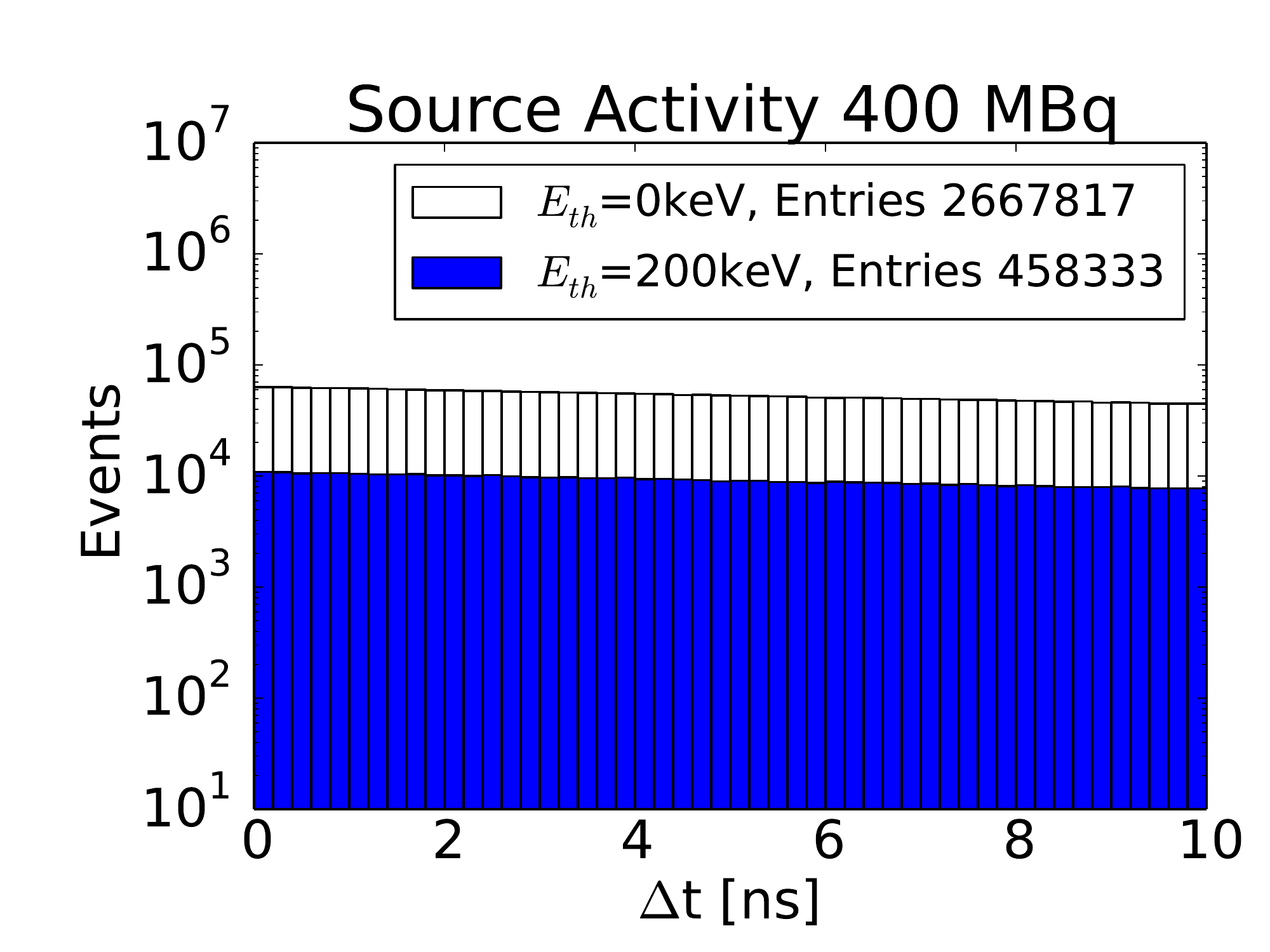}
\end{subfigure}

\caption{Distributions of time differences between hits in the detector  originating from the same and from different annihilations (top panel) and from the different annihilations only (bottom panel). Shown are results of simulations for activities of 5 MBq and 400 MBq and for two different energy threshold as indicated in the figures.}
\label{time_differences_eventID0_10ns}
\end{figure}

Simulations described in this section were performed for other activities of the source: 5~MBq, 10~MBq, 20~MBq, 30~MBq, 50~MBq, 100~MBq, 150~MBq, 200~MBq, 250~MBq, 300~MBq, 350~MBq and 400~MBq. For each of these activities $10^8$ annihilations were simulated.
Results of simulations for the smallest (5~MBq) and the largest (400~MBq) activity are presented in Figs \ref{time_differences_all} and \ref{time_differences_eventID0_10ns}.

In Fig. \ref{time_differences_all} histograms contain all time differences both for hits from the same event and for hits from different events (there is no time window). 
One can see that the first bin is higher than expected from the general exponential dependence. 
This is because  this bin contains both  true and accidental coincidences. The structure is better visible in Fig. \ref{time_differences_eventID0_10ns}.
In the upper panel of this figure, histograms contain time differences between hits from the same and from different annihilations. 
If time differences from the same annihilations were omitted, there would be only accidental coincidences, as it is presented in the bottom panel of the figure.

\clearpage
\subsubsection{Accidental coincidences for time windows 3 ns and 5 ns} 

\begin{figure}[h!]
\centering

\begin{subfigure}{0.49\textwidth}
    \centering	\includegraphics[width=\textwidth]{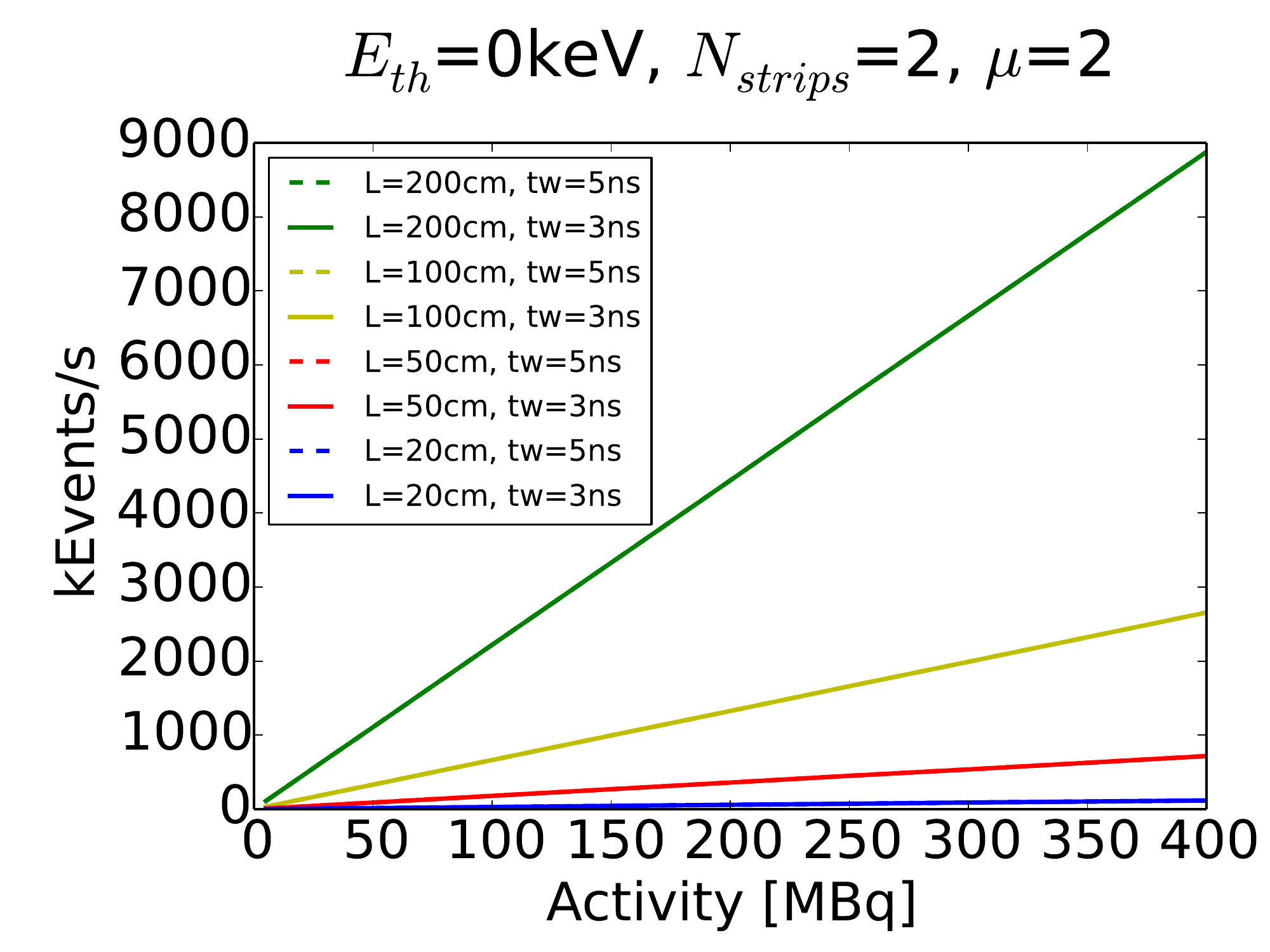}
\end{subfigure}
\begin{subfigure}{0.49\textwidth}
    \centering	\includegraphics[width=\textwidth]{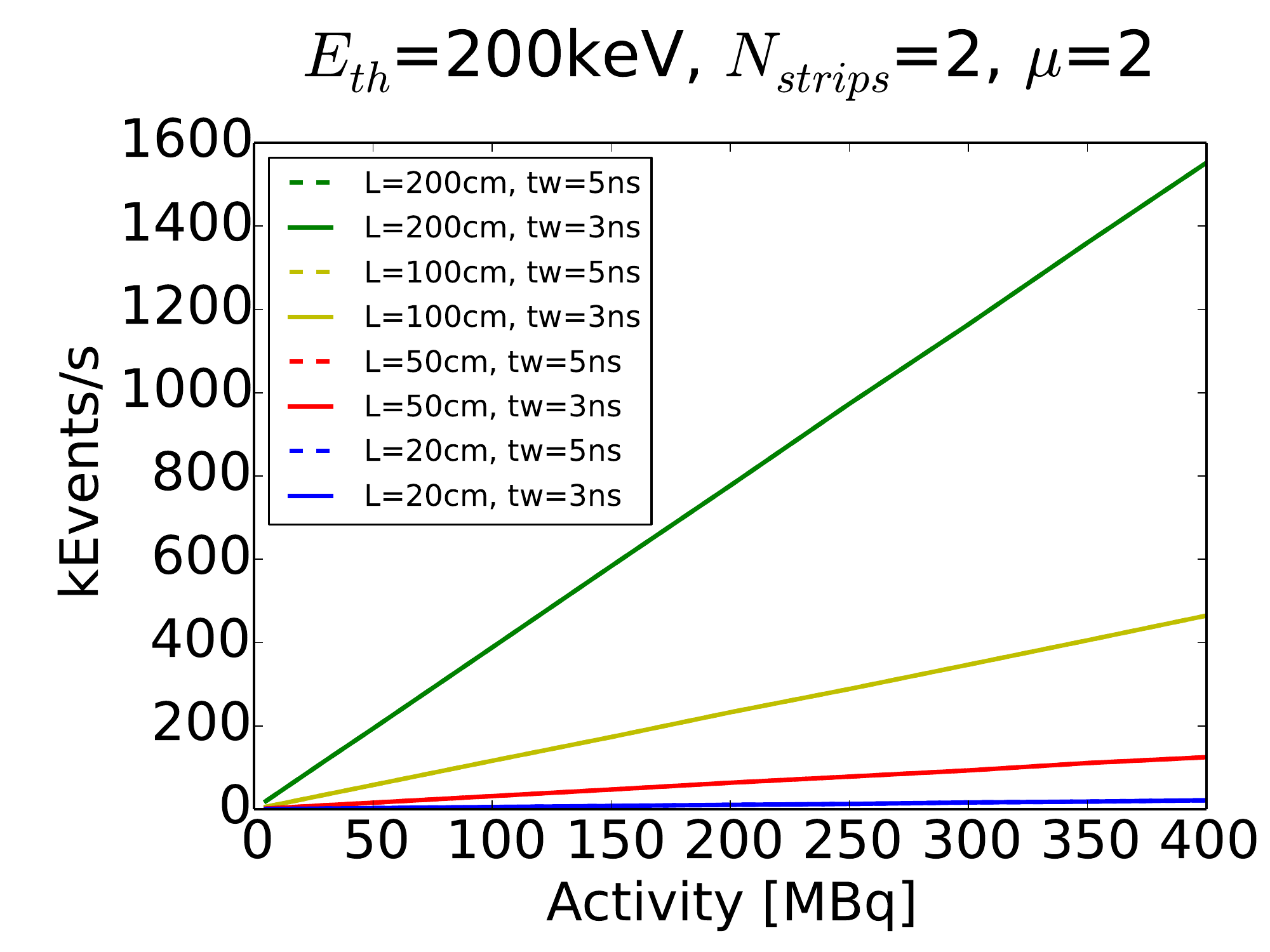}
\end{subfigure}

\begin{subfigure}{0.49\textwidth}
    \centering	\includegraphics[width=\textwidth]{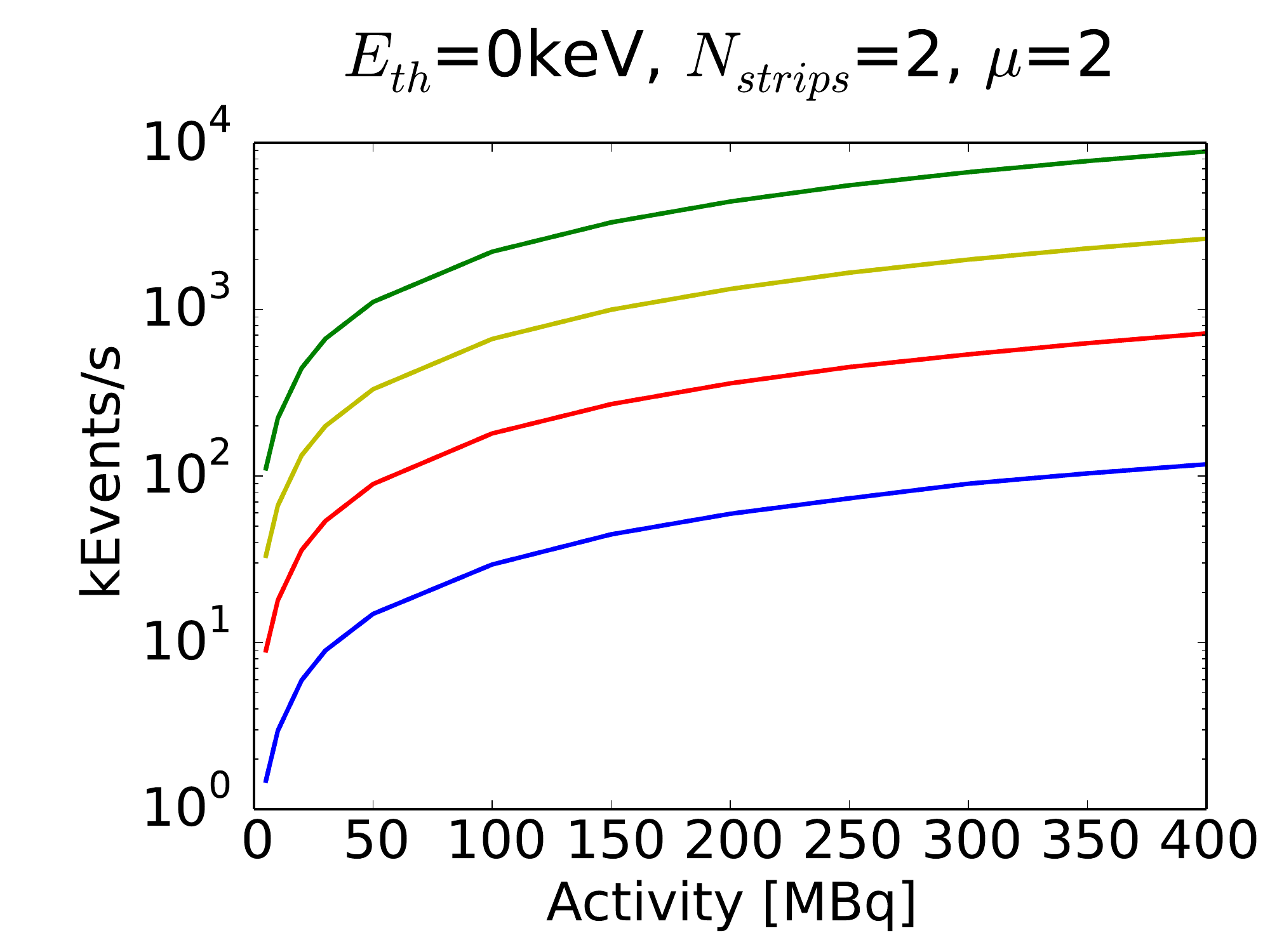}
\end{subfigure}
\begin{subfigure}{0.49\textwidth}
    \centering	\includegraphics[width=\textwidth]{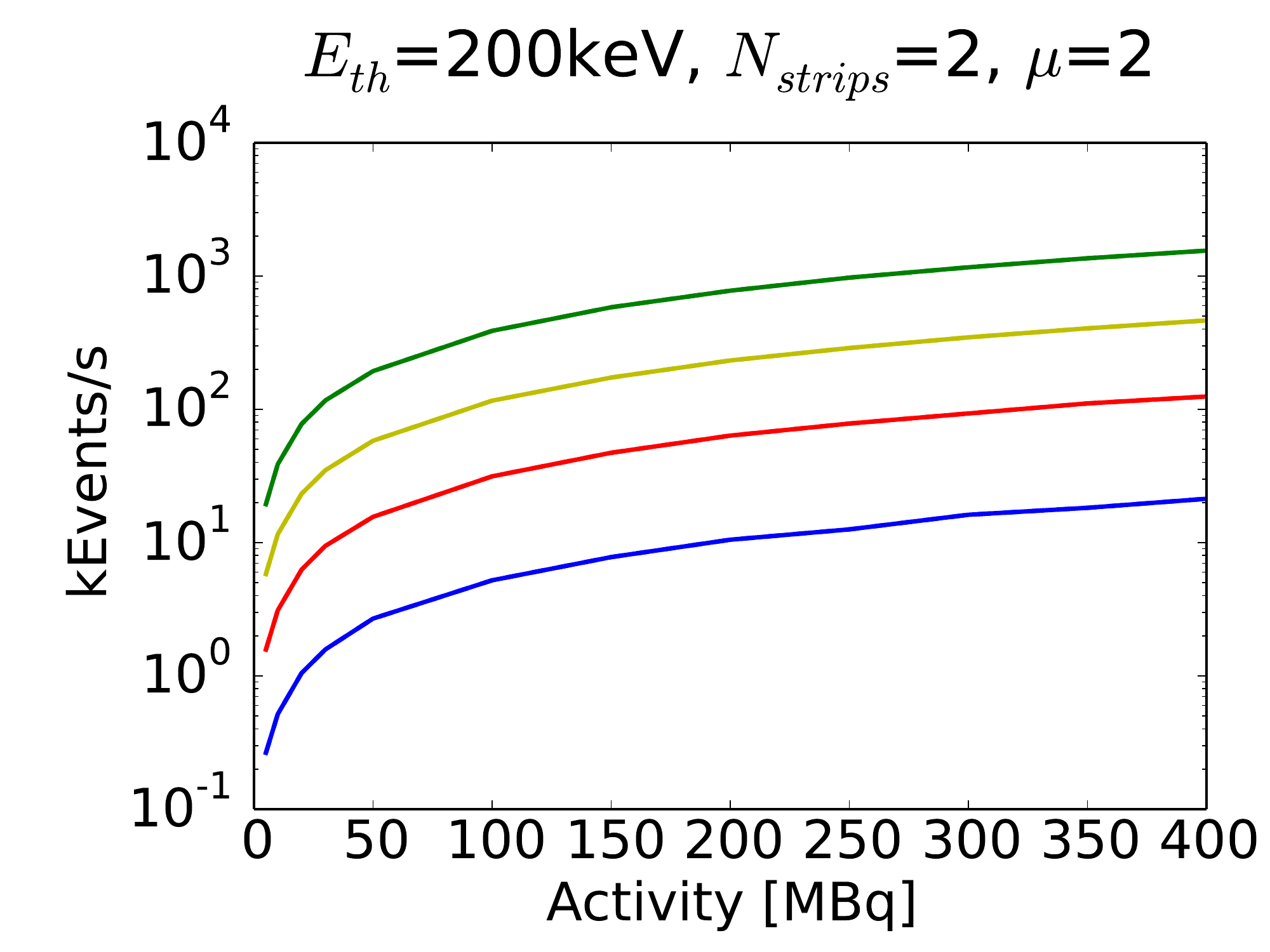}
\end{subfigure}

\caption{Simulated rate of true coincidences as a function of time window, activity and  detector length. The sequence of curves in the figure is the same as in the legend (from top to bottom); bottom pictures present the same data as the top ones but in logarithmic scale. Results for the time window of  3ns (solid lines) are indistinguishable from the results for time window  of 5 ns (dashed lines).}
\label{true_coincidences}
\end{figure}

For simulations described in this article with the virtual linear source of annihilations placed along the main axis of the scanner, true coincidences are defined as two hits from the same annihilation 
having $\Delta ID$ vs $\Delta t$ above the black lines shown in Figs
\ref{Nstrips2} - \ref{histograms_diff_lengths}.
Fig. \ref{true_coincidences} presents rate of such defined true coincidences as a function of annihilation source activity, time window, energy threshold, and detector length L.

\begin{figure}[h!]
\centering

\begin{subfigure}{0.49\textwidth}
    \centering	\includegraphics[width=\textwidth]{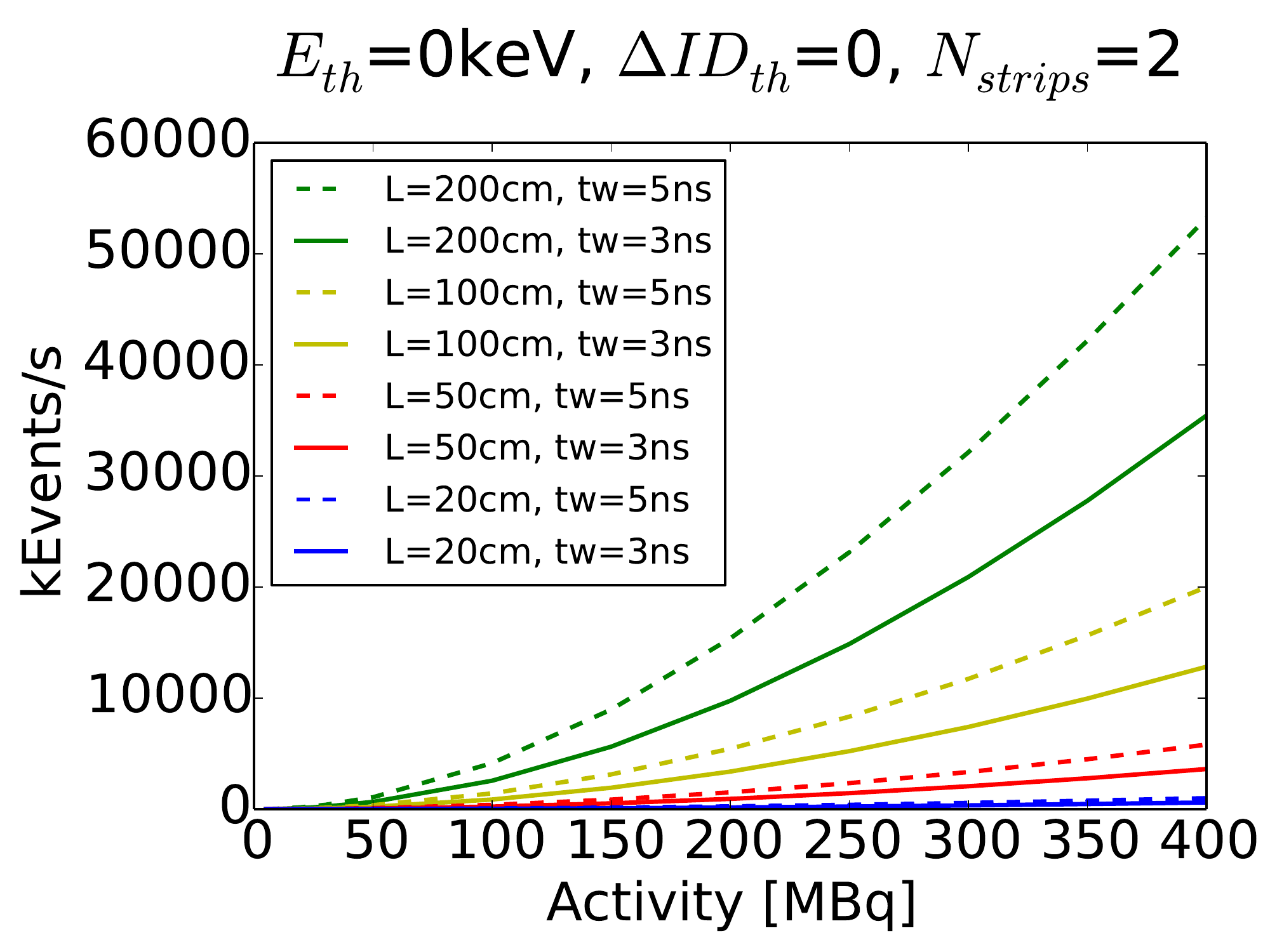}
\end{subfigure}
\begin{subfigure}{0.49\textwidth}
    \centering	\includegraphics[width=\textwidth]{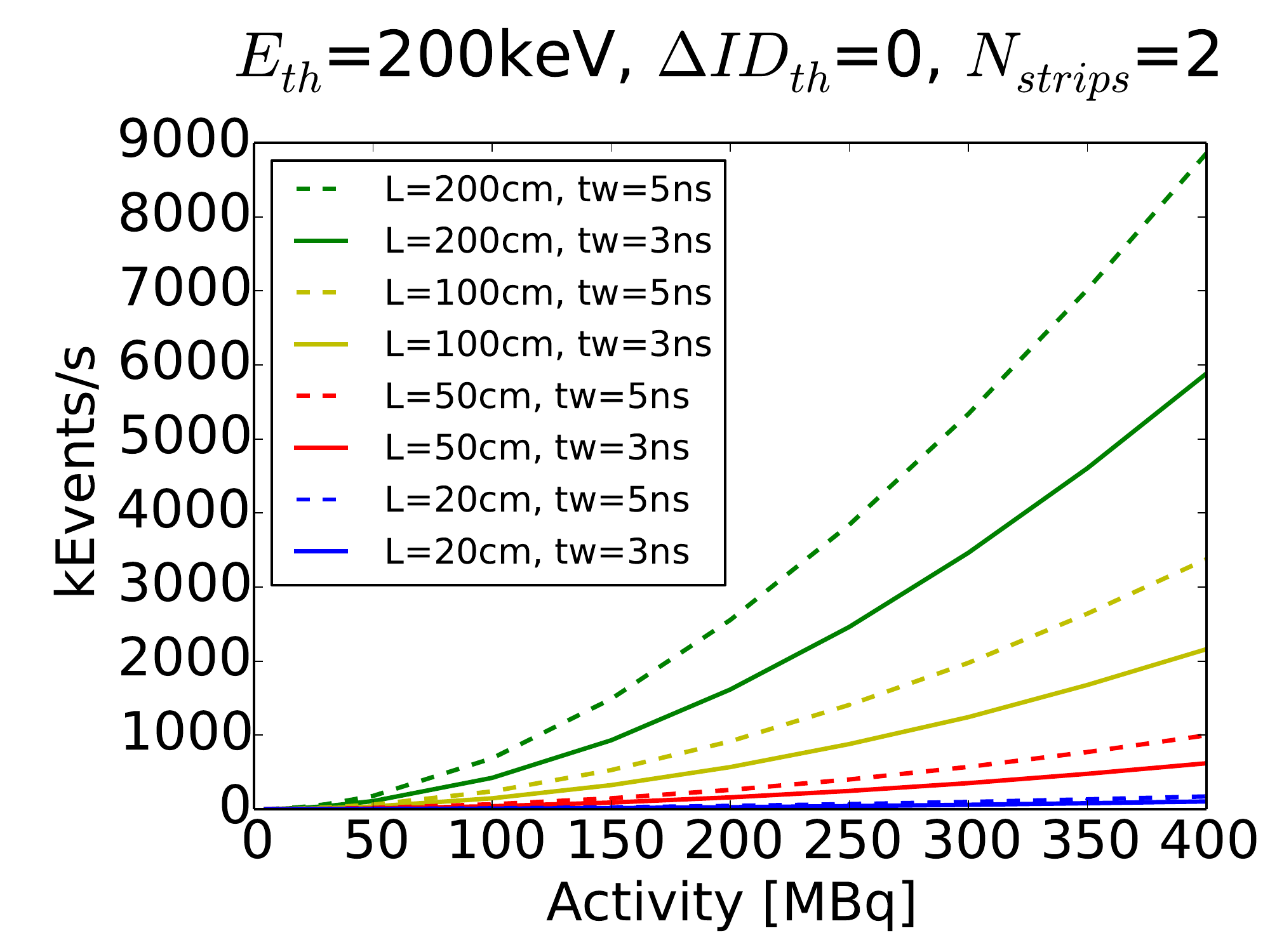}
\end{subfigure}

\begin{subfigure}{0.49\textwidth}
    \centering	\includegraphics[width=\textwidth]{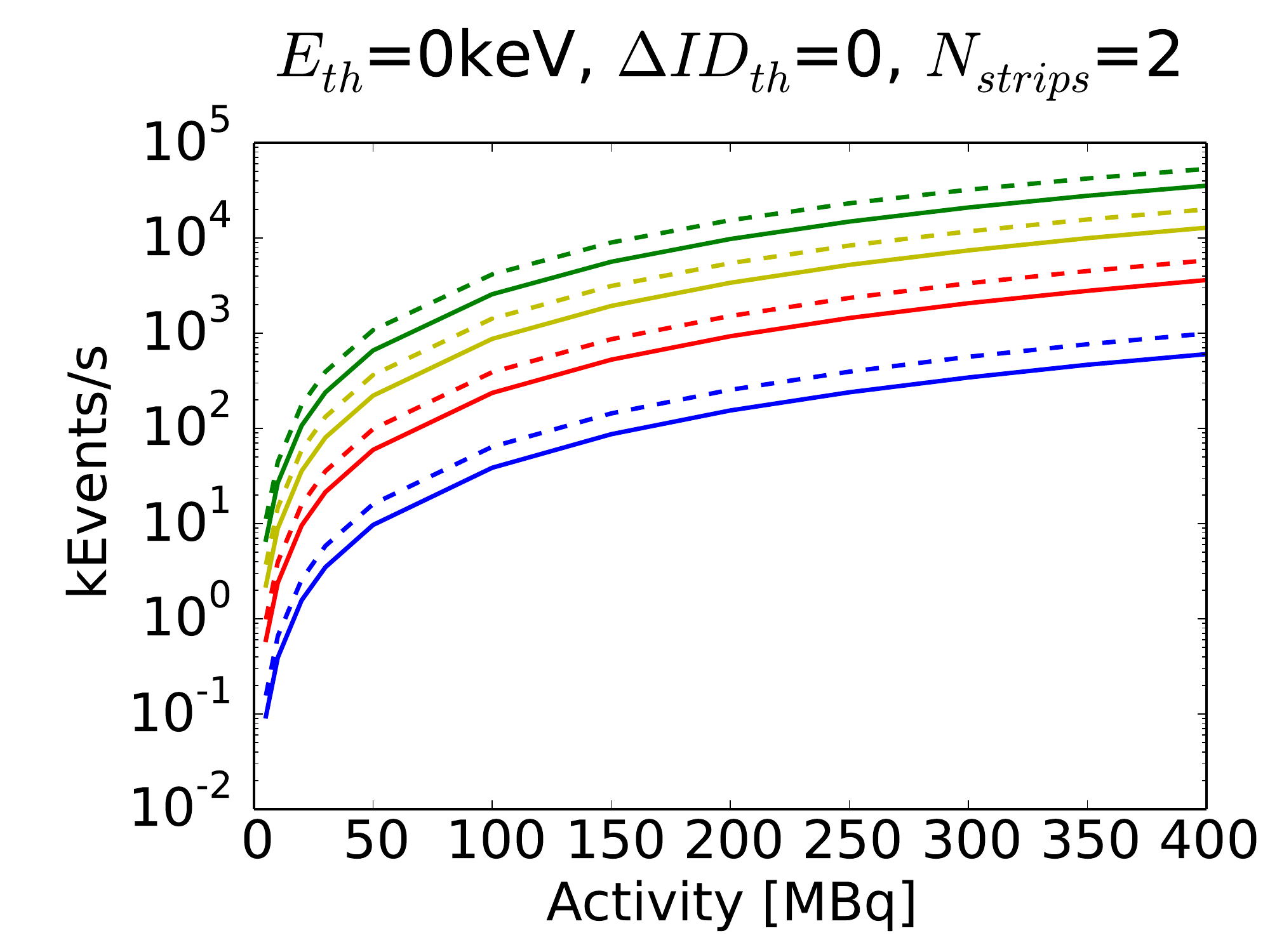}
\end{subfigure}
\begin{subfigure}{0.49\textwidth}
    \centering	\includegraphics[width=\textwidth]{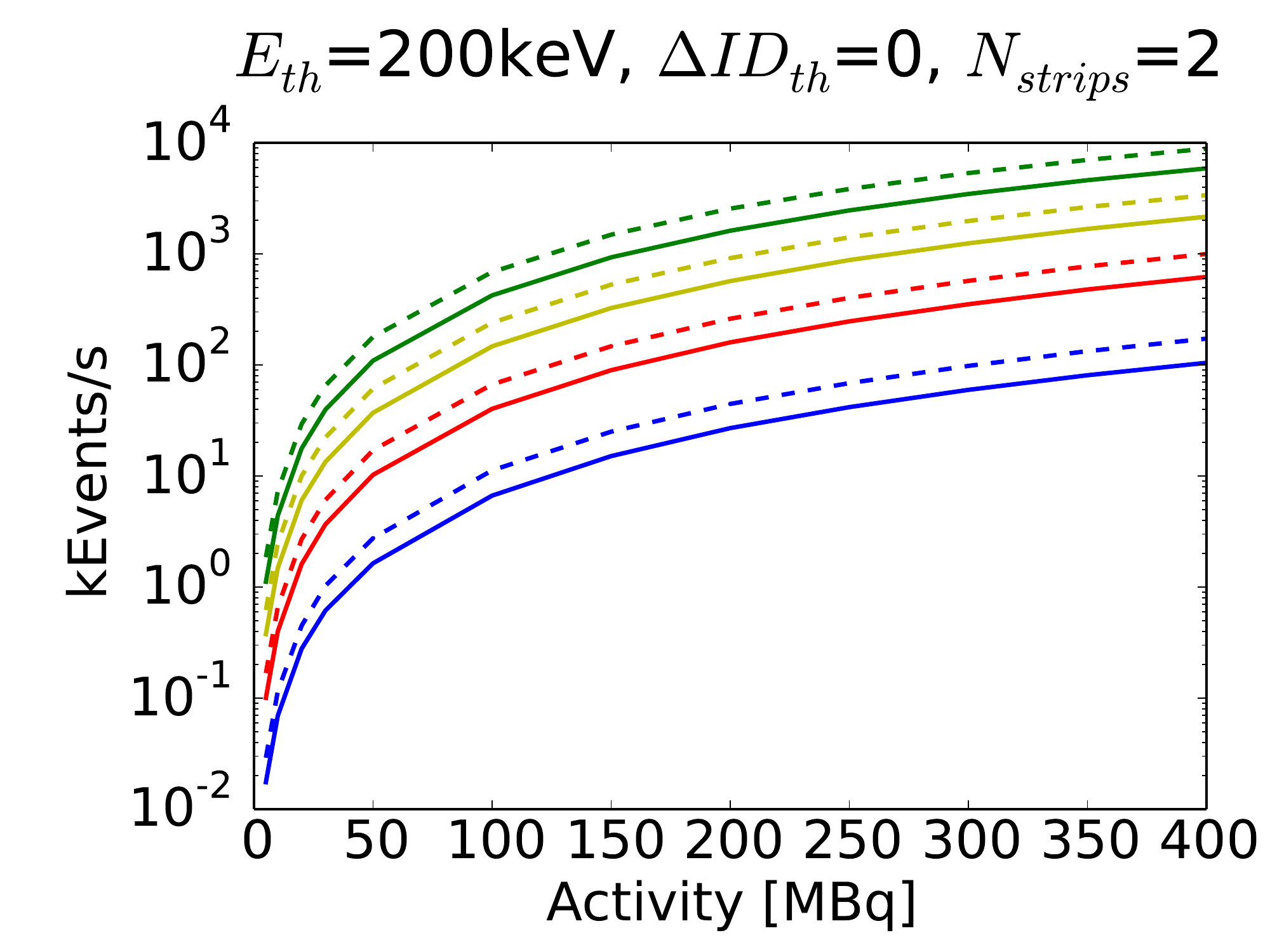}
\end{subfigure}

\caption{Accidental coincidences for time windows 3~ns (solid lines) and 5~ns (dashed lines) for different lengths of the scintillators (as indicated in the legend); the sequence of curves in the figure is the same as in the legend (from top to bottom); bottom pictures present the same data as the top ones but in logarithmic scale.}
\label{acci_coincidences}
\end{figure}

Accidental coincidences for time windows 3 ns and 5 ns and for four lengths of the scintillators are presented in Fig. \ref{acci_coincidences}. One can see that if the energy threshold is 200 keV (right column of the figure), a rate of accidental coincidences is reduced by the factor of about 7 in comparison to situation when there is no energy threshold (left column of the figure). 

\begin{figure}[h!]
\centering

\begin{subfigure}{0.49\textwidth}
    \centering	\includegraphics[width=\textwidth]{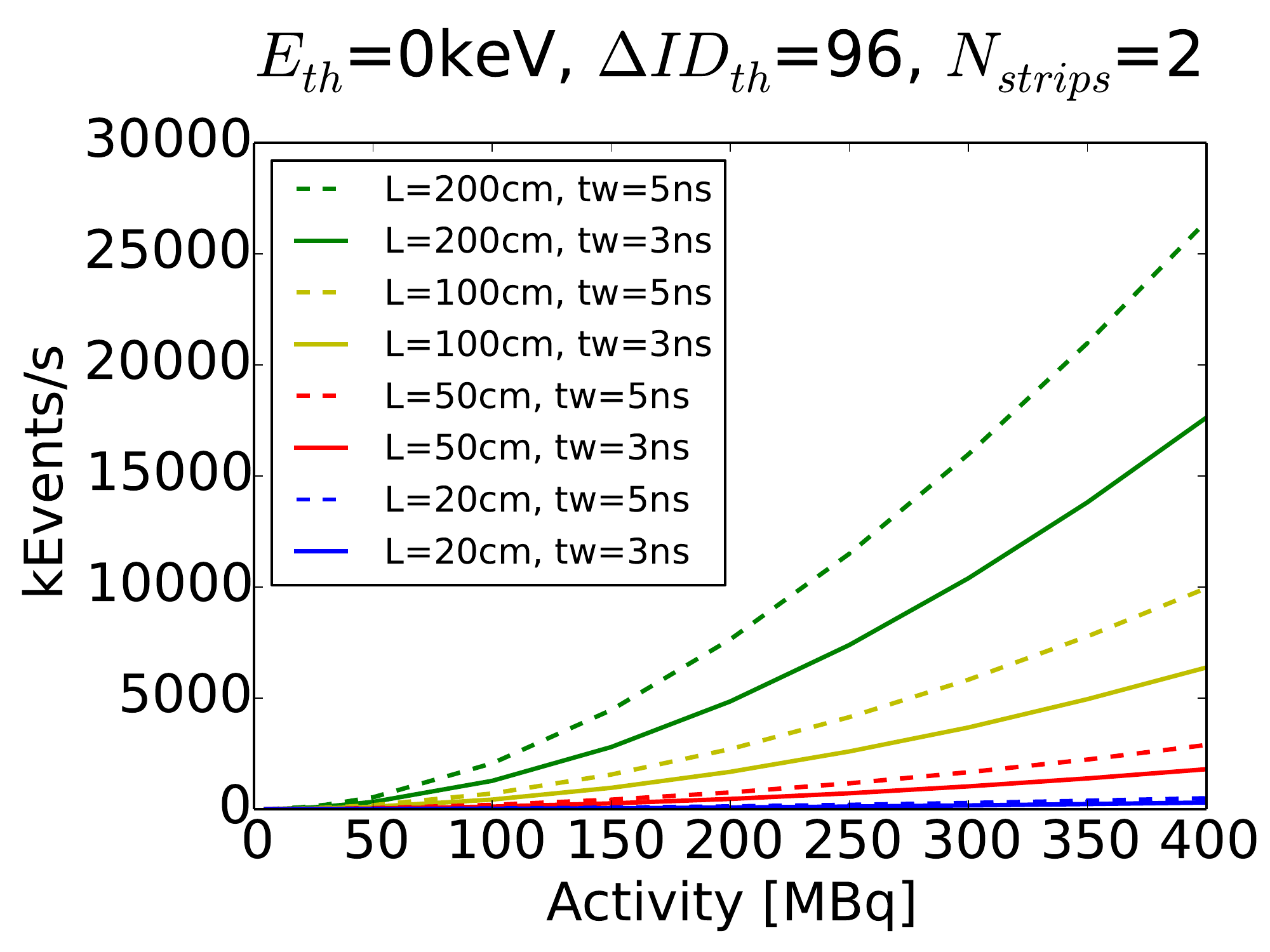}
\end{subfigure}
\begin{subfigure}{0.49\textwidth}
    \centering	\includegraphics[width=\textwidth]{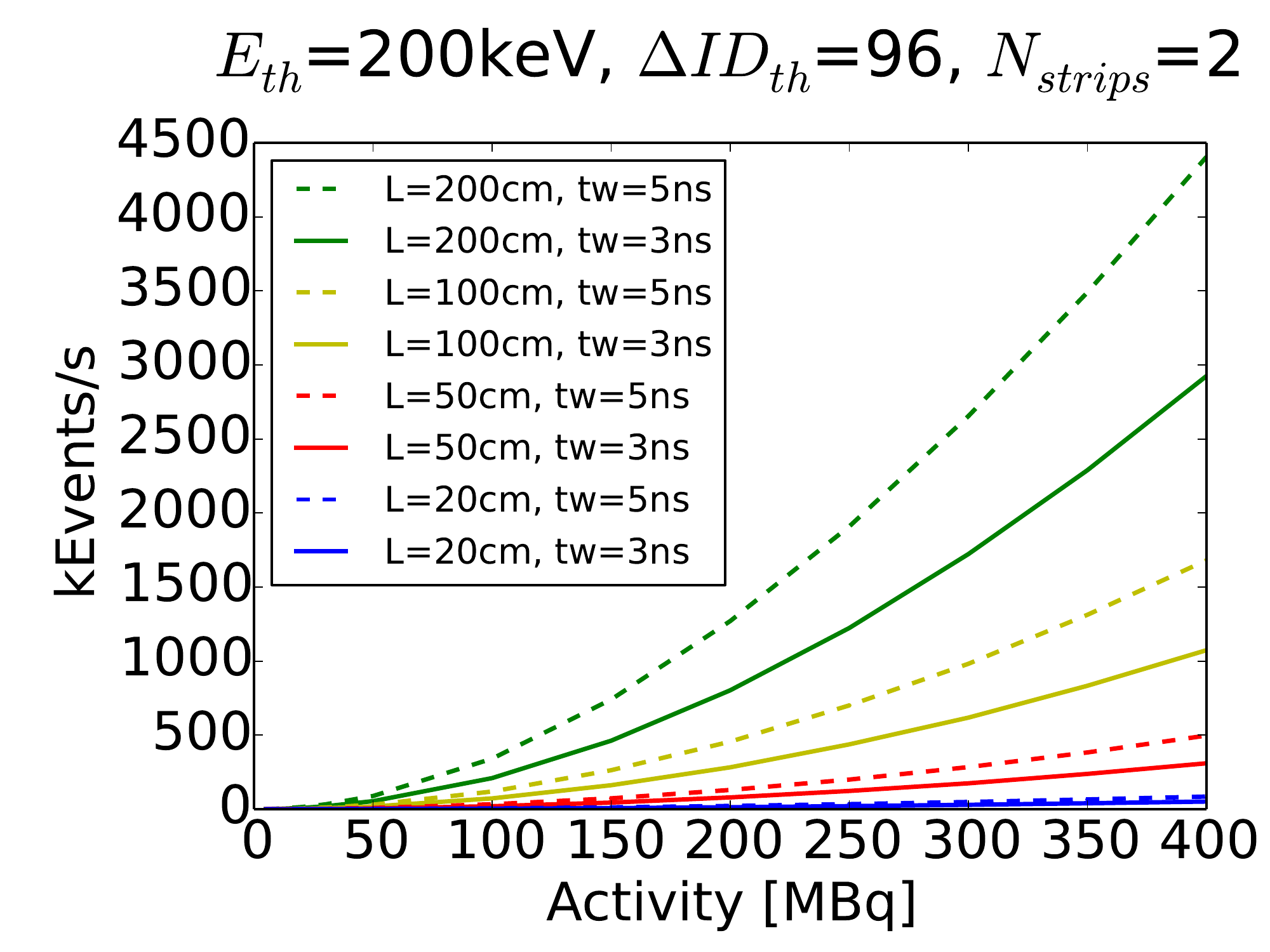}
\end{subfigure}

\begin{subfigure}{0.49\textwidth}
    \centering	\includegraphics[width=\textwidth]{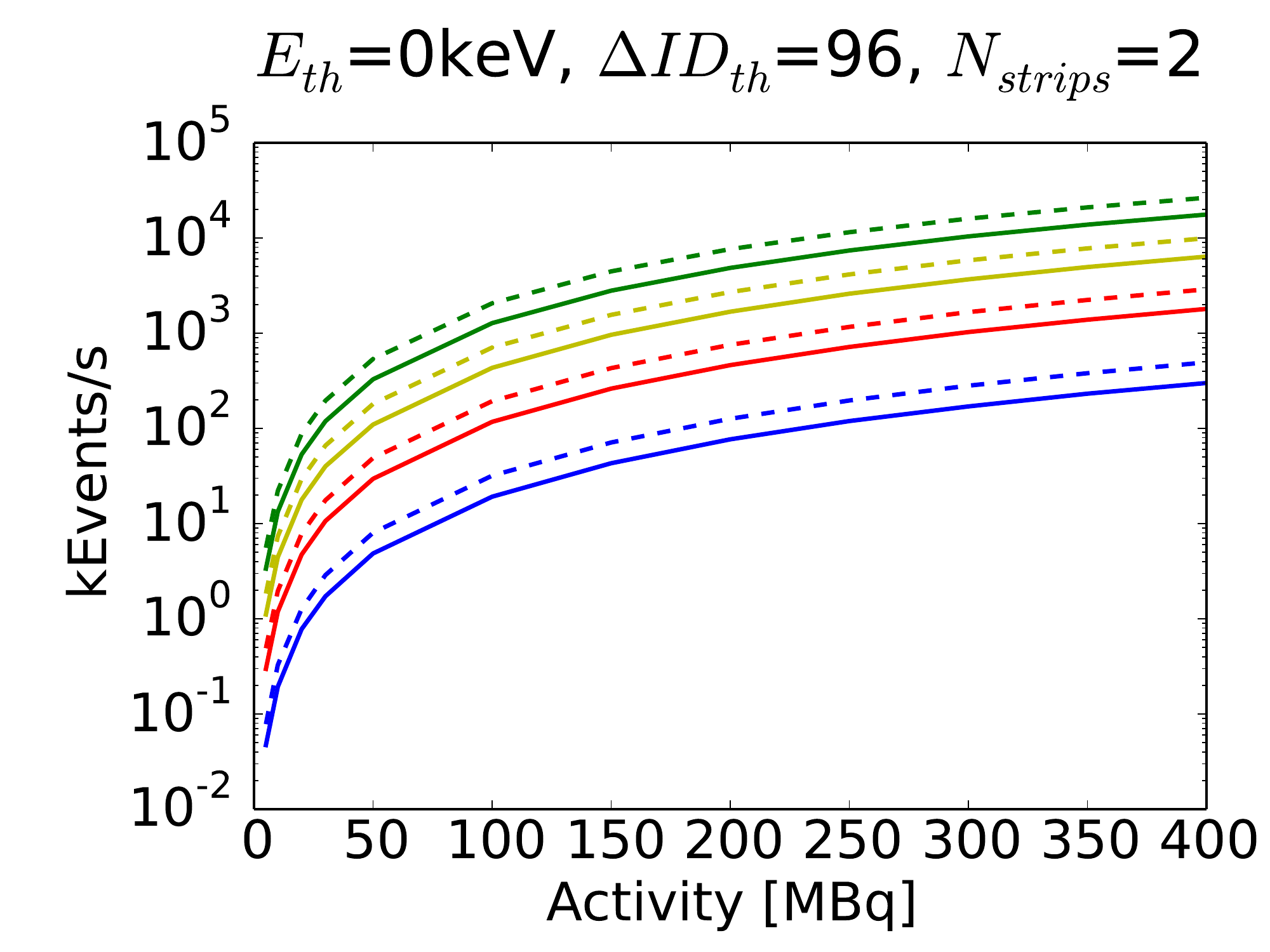}
\end{subfigure}
\begin{subfigure}{0.49\textwidth}
    \centering	\includegraphics[width=\textwidth]{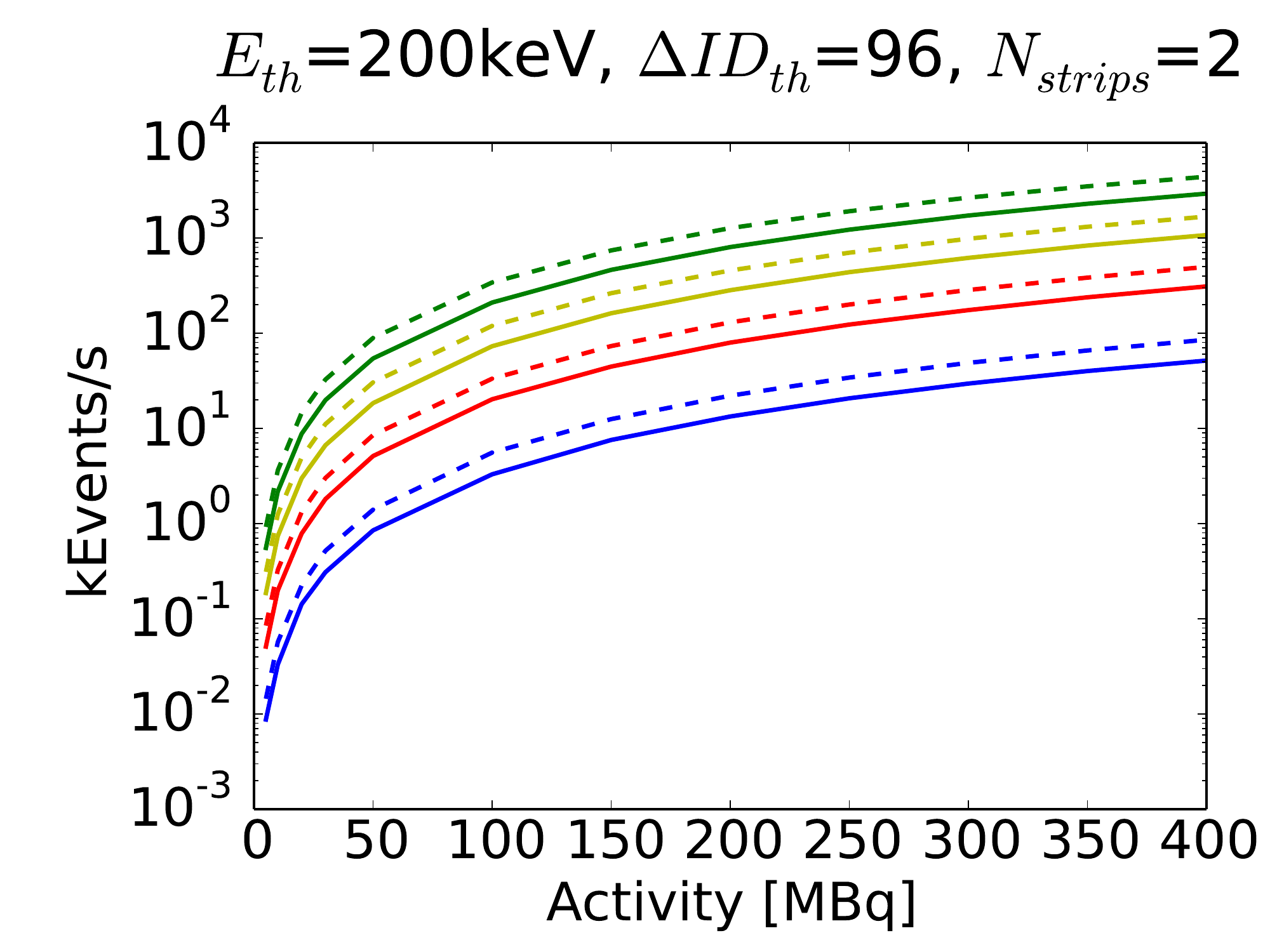}
\end{subfigure}

\caption{Accidental coincidences for time windows 3~ns (solid lines) and 5~ns (dashed lines) for different lengths of the scintillators (as indicated in the legend); minimum difference between IDs of the strips is equal to $\Delta ID_{th} = 96$; the sequence of curves in the figure is the same as in the legend (from top to bottom); bottom pictures present the same data as the top ones but in logarithmic scale.}
\label{acci_coincidences_96}
\end{figure}

Fig. \ref{acci_coincidences_96} shows rate of accidental coincidences under condition that difference $\Delta ID$ is larger than 96. Which means that interactions of gamma quanta occurs in two different  quarters of the cylinder (consisting of 384 scintillator strips). Such condition decrease the field of view of the detector to the cylinder with radius of 30 cm, however this additional condition reduces the number of accidental coincidences by the factor of~2.

\begin{figure}[h!]
\centering

\begin{subfigure}{0.49\textwidth}
    \centering	\includegraphics[width=\textwidth]{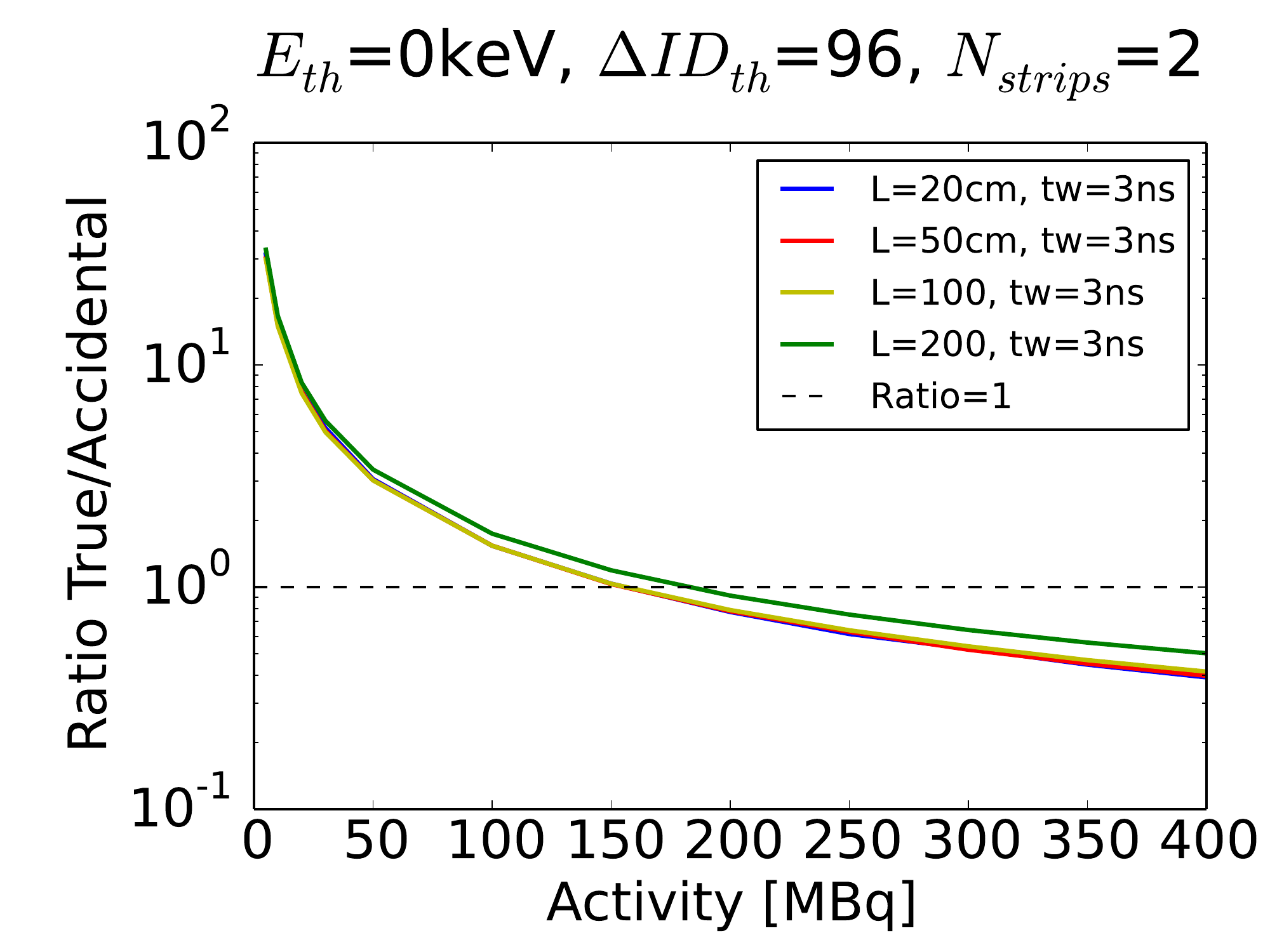}
\end{subfigure}
\begin{subfigure}{0.49\textwidth}
    \centering	\includegraphics[width=\textwidth]{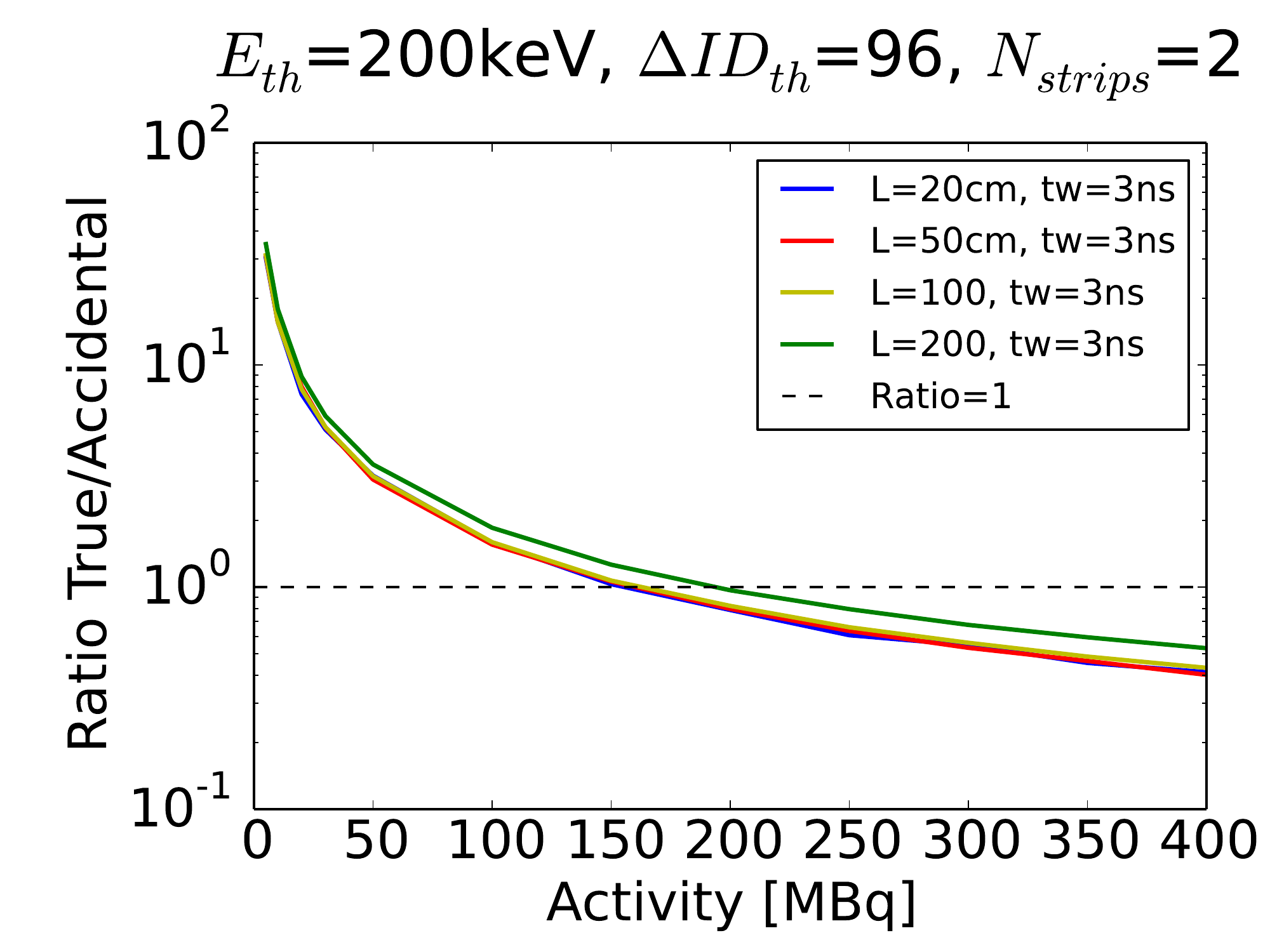}
\end{subfigure}

\caption{Ratios between true and accidental coincidences for time window 3~ns and minimum difference between IDs of the hit scintillator strips equal to 96. Right figure shows results for energy threshold of $E_{th} = 200~keV$ whereas results in left figure were obtained for $E_{th} = 0$.}
\label{ratios_true_acci}
\end{figure}

In Fig. \ref{ratios_true_acci} rates of true and accidental coincidences are presented. The ratio is larger for longer scintillators. It is caused by the fact that for short scintillators there are additional accidental coincidences caused by the gamma quanta from outside of the tomograph.



\section{Summary}

Physical properties of the scintillating material and the photomultiplier used in the J-PET detector were implemented in the GATE software. The simulations procedures were validated by the comparison of simulated and experimental results for the number of photoelectron spectra. 

In previous research, studies of simplified Strip-PET scanner were presented \cite{map_of_efficiency}. Map of efficiency of 2-strip scanner was calculated and compared with the geometrical efficiency of such a device. In present studies, background given by accidental coincidences and multiple scattering of gamma quanta was investigated for single-layer 384-strip J-PET scanner.

In presented simulations, the source of annihilations was assumed to be a 2~m long line placed along the main axis of the scanner. In order to compare precisely obtained results with results for another devices, in the future the source will be simulated in accordance with NEMA-NU-2 standard \cite{nema}. Even so, it is possible to compare orders of magnitudes of calculated parameters. For example, results obtained for 2~m long J-PET scanner for activity  of 200 MBq are similar to these simulated for the same length RPC-PET [10]. 

\section*{Acknowledgements}

We acknowledge technical and administrative support by T.~Gucwa-Ry\'s, A.~Heczko, M.~Kajetanowicz, G.~Konopka-Cupia\l, W. Migda\l, and the financial support by the Polish National Center for Development and Research through grant INNOTECH-K1/IN1/64/159174/NCBR/12, the Foundation for Polish Science through MPD programme, the EU and MSHE Grant No. POIG.02.03.00-161 00-013/09, Doctus – the Lesser Poland PhD Scholarship Fund and Marian Smoluchowski Krak\'ow Research Consortium "Matter-Energy-Future".

%
%

\end{document}